\documentclass[a4paper,11pt]{article}
\pdfoutput=1 
\usepackage{jheppub} 
                     
\usepackage{romannum}
\usepackage{scalerel}
\usepackage[inline]{enumitem}
\usepackage{dcolumn}
\usepackage{placeins}
\usepackage{bm}
\usepackage{dirtytalk}
\usepackage[utf8]{inputenc}
\usepackage{amsmath}
\usepackage[dvipsnames]{xcolor}
\usepackage{graphicx}
\usepackage{fancyhdr}
\usepackage{lastpage}
\usepackage[T1]{fontenc}
\usepackage{physics,slashed}
\usepackage{amssymb}
\usepackage{mathtools}
\usepackage[capitalise]{cleveref}
\usepackage{multirow}
\usepackage[normalem]{ulem}
\usepackage{mathrsfs}  
\usepackage[export]{adjustbox}
\usepackage{cleveref}
\usepackage{orcidlink}
\usepackage{fontawesome5} 
\definecolor{blue-violet}{rgb}{0.33, 0.17, 0.89}
\usepackage{nicefrac}

\renewcommand{\phi}{\varphi}

\newcounter{CommentCount}
\newcolumntype{L}{>{$}l<{$}} 

\definecolor{jhg}{rgb}{0.984314,0.145098,0.462745}

\newcommand{\gitlink}{\href{https://github.com/mhostert/Heavy-Neutrino-Limits}{\textsc{g}it\textsc{h}ub {\large\color{blue-violet}\faGithub}}}

\begin{document}

\preprint{\hfill 
FTUV-23-0303.1224; IFIC/23-09; IFT-UAM/CSIC-23-39}

\title{Effective portals to heavy neutral leptons}

\author[1,2]{Enrique Fern\'andez-Mart\'inez}
\author{\orcidlink{0000-0002-6274-4473}}
\emailAdd{enrique.fernandez-martinez@uam.es}
\author[1,2]{Manuel Gonz\'alez-L\'opez}
\author{\orcidlink{0000-0001-7276-2192}}
\emailAdd{manuel.gonzalezl@uam.es}
\author[3]{\\ Josu Hern\'andez-Garc\'ia}
\author{\orcidlink{0000-0003-0734-0879}}
\emailAdd{josu.hernandez@ttk.elte.hu}
\author[4,5,6]{Matheus Hostert}
\author{\orcidlink{0000-0002-9584-8877}}
\emailAdd{mhostert@pitp.ca}
\author[7]{Jacobo L\'opez-Pav\'on}
\author{\orcidlink{0000-0002-9554-5075}}
\emailAdd{jacobo.lopez@uv.es}

\affiliation[1]{Instituto de F\'{\i}sica Te\'orica, Universidad Aut\'onoma de Madrid \& CSIC, 
Campus de Cantoblanco, 28049 Madrid, Spain}
\affiliation[2]{Departamento de F\'isica Te\'orica, Universidad Aut\'onoma de Madrid, 
Campus de Cantoblanco, 28049 Madrid, Spain}
\affiliation[3]{Institute for Theoretical Physics, ELTE E\"otv\"os Lor\'and University, P\'azm\'any P\'eter s\'et\'any 1/A, H-1117 Budapest, Hungary}
\affiliation[4]{Perimeter Institute for Theoretical Physics, Waterloo, ON N2J 2W9, Canada}
\affiliation[5]{School of Physics and Astronomy, University of Minnesota, Minneapolis, MN 55455, USA}
\affiliation[6]{William I. Fine Theoretical Physics Institute, School of Physics and Astronomy, University of Minnesota, Minneapolis, MN 55455, USA}
\affiliation[7]{Instituto de F\'{\i}sica Corpuscular, Universidad de Valencia and CSIC, Edificio Institutos Investigaci\'on, Catedr\'atico Jos\'e Beltr\'an 2, 46980, Spain}

\date{\today}

\begin{abstract}
{The existence of right-handed neutrinos, or heavy neutral leptons (HNLs), is strongly motivated by the observation of neutrino masses and mixing.
The mass of these new particles could lie below the electroweak scale, making them accessible to low-energy laboratory experiments.
Additional new physics at high energies can mediate new interactions between the Standard Model particles and HNLs, and is most conveniently parametrized by the neutrino Standard Model Effective Field Theory, or $\nu$SMEFT for short.
In this work, we consider the dimension six $\nu$SMEFT operators involving one HNL field in the mass range of $\mathcal{O}(1)$~MeV $<M_N< \mathcal{O}(100)$~GeV.
By recasting existing experimental limits on the production and decay of new light particles, we constrain the Wilson coefficients and new physics scale of each operator as a function of the HNL mass.\footnote{Heavy-Neutrino-Limits: \gitlink}}
\end{abstract}

\maketitle
\setcounter{tocdepth}{1}

\section{Introduction}

The evidence for neutrino masses and mixing, as observed in neutrino oscillations, calls for an extension of the Standard Model (SM) of particle physics. Among the many different options, the completion of the SM particle content with the right-handed counterpart of the left-handed neutrinos is the simplest possibility. Apart from the usual Yukawa couplings with the Higgs, the unique gauge-singlet nature of right-handed neutrinos allows for them also a lepton-number-violating Majorana mass term. This new scale would be the first dimensionful term in the Lagrangian not directly related to the electroweak scale and the Higgs mechanism, and, as such, its value remains utterly unknown.

A common assumption is that the Majorana mass of the right-handed neutrinos is much larger than the electroweak scale. This heavy scale would suppress the induced mass for the mostly-active neutrino mass eigenstates via the $d=5$ Weinberg operator~\cite{Weinberg:1979sa}, providing a rationale for their smallness. 
This is the celebrated high-scale Seesaw mechanism~\cite{Minkowski:1977sc,Mohapatra:1979ia,Yanagida:1979as,Gell-Mann:1979vob} which, as a bonus, may also explain the origin of the matter-antimatter asymmetry of the universe via the leptogenesis mechanism~\cite{Fukugita:1986hr}. The drawback of this scenario is that the same very large mass scale that suppresses the light neutrino masses also renders the model virtually untestable since all phenomenology would be similarly suppressed. Furthermore, the addition of a new mass scale much above the electroweak one worsens the Higgs hierarchy problem~\cite{Vissani:1997ys,Casas:2004gh}.

Nevertheless, the extreme lightness of neutrino masses may also be naturally explained through a symmetry argument~\cite{Branco:1988ex,Kersten:2007vk,Abada:2007ux,Moffat:2017feq}. Indeed, the Weinberg operator violates $B-L$, which is an accidental and non-anomalous symmetry of the SM. This fact is exploited by different low-scale variants of the Seesaw mechanism, such as the inverse~\cite{Mohapatra:1986aw,Mohapatra:1986bd} or linear~\cite{Akhmedov:1995ip,Malinsky:2005bi} Seesaws, to lower the mass scale of right-handed neutrinos. This way, the Higgs hierarchy problem is not worsened, and these new ``Heavy Neutral Leptons'' (HNLs) may be probed at laboratory energies. 
Still, even if light, the interactions of the HNLs with the rest of the SM particles are also suppressed by an unknown but small mixing with active neutrinos.
The HNLs then constitute long-lived particles with \say{weaker-than-weak} interactions.

In this context, if additional new physics is present above the electroweak scale, it can have an important effect not only on SM processes but also on the behaviour of HNLs, possibly dominating over its weaker-than-weak interactions.
To explore the impact of these high-energy new physics scenarios on HNLs, we turn to the framework of effective field theories (EFTs). 

Effective field theories are an extremely powerful and versatile tool that encodes the indirect effects of new physics appearing at energy scales above a given physical process. 
They rely on an expansion in inverse powers of the new physics scale, introducing operators of dimension higher than 4; the higher the dimension, the more suppressed the operator is. In this context, the naive expectation is to see the first evidence of new physics from the Wilson coefficients corresponding to the lowest dimension and least suppressed operators. Thus, it is very suggestive that the only $d=5$ operator that can be built out of the SM particle content is the Weinberg operator, which can account for the evidence of neutrino masses and mixings. EFTs are renormalizable order by order, with a finite number of counterterms required at each order of the EFT expansion. Therefore, all possible operators that can be built with low-energy particle content which are compatible with the symmetries of the theory should be included at each order. In this context, the Standard Model Effective Field Theory (SMEFT), that is, the series of higher dimension operators that can be built from the SM particle content, is a very useful and successful tool to constrain generic new physics (see, for instance, Ref.~\cite{Brivio:2017vri} for a recent review).

Since the evidence for non-zero neutrino masses can be explained by HNLs light enough to be included in the low energy spectrum, it is then interesting to extend the SMEFT description with a basis of new operators also incorporating the HNLs as fundamental building blocks: the $\nu$SMEFT~\cite{Graesser:2007yj,Graesser:2007pc,delAguila:2008ir,Aparici:2009fh,Peressutti:2014lka,Duarte:2014zea,Bhattacharya:2015vja,Duarte:2015iba,Duarte:2016caz,Liao:2016qyd,Duarte:2016miz,Caputo:2017pit,Duarte:2018xst,Alcaide:2019pnf,Chala:2020vqp,Barducci:2020icf,Dekens:2020ttz,Biekotter:2020tbd,Duarte:2020vgj,Barducci:2020ncz,Dekens:2021qch,Cottin:2021lzz,Li:2021tsq,Cirigliano:2021peb,DeVries:2020jbs,Zhou:2021ylt,Zhou:2021lnl,Beltran:2022ast,Delgado:2022fea,Barducci:2022gdv,Talbert:2022unj,Barducci:2022hll,Zapata:2022qwo,Mitra:2022nri,Beltran:2023nli,Dekens:2023iyc,Bischer:2019ttk}.

In this work, we explore the effective operators (up to $d=6$) that form a basis including these new particles and study the leading constraints on them from different sets of existing observables. 
For the first time, we present a comprehensive list of bounds that apply to each operator as a function of the HNL mass. 
As a first approach, we will only consider one operator at a time. We leave a more realistic global analysis for future work. 
Nevertheless, this approach generally leads to conservative constraints, as it does not allow for HNL production and detection through different operators. 
When relevant, we will comment on exceptions to this rule when cancellations among different operators lead to flat directions. 
In the same spirit, we will also assume negligible mixing between the HNLs and the active neutrinos. 
Mixing can lead to new production mechanisms or decay channels that would strengthen the constraints. 

This paper is organized as follows. In \cref{sec:HNL}, we introduce some generalities about HNLs, summarizing the main constraints on their mixing in \cref{sec:HNL_bounds}. Then, in \cref{sec:SMEFT}, we briefly introduce the $\nu$SMEFT and the new operators that extend the SMEFT by including HNLs. 
In \cref{sec:higgs_mix,sec:boson_curr,sec:tensor,sec:four-ferm}, we describe the different types of dimension-6 ($d=6$) operators, the observable effects they can produce, and how current limits can constrain their Wilson coefficients. 
We present our conclusions in \cref{sec:conclus}.

\section{Heavy neutral leptons at the renormalizable level}
\label{sec:HNL}

We will extend the SM particle content by only one~\footnote{At least two HNLs are required to reproduce the measured masses and mixings, but this simplifying assumption encodes the main phenomenology and corresponds to scenarios where the additional HNLs are too heavy to have an impact in low-energy observables.} right-handed neutrino field $N^\prime$, singlet under the SM gauge symmetry group. 
The renormalizable part of the Lagrangian reads
\begin{equation}\label{eq:yukawa_coupl}
    \mathscr{L}_{d=4}=\mathscr{L}_{\rm SM}+\overline{N^\prime}i\slashed{\partial}N^\prime - \left( \frac{M_N}{2} \overline{N^{\prime c}} N^\prime + \sum_{\alpha} y_\alpha \overline{L_\alpha}\tilde{H} N^\prime + \text{ h.c.}\right) \,,
\end{equation}
where $L$ stands for a left-handed lepton doublet of flavour $\alpha$, $M_N$ is the Majorana mass, and $y_\alpha$ are the Yukawa couplings of the HNL to the different flavours of lepton doublets.
The extension of the neutrino sector implies that the flavour eigenstates will generally contain some fraction of the heavy component due to the mixing $U_{\alpha N} = y_\alpha v/\sqrt{2} M_N$:
\begin{equation}\label{eq:mixing}
    \nu_\alpha=\sum_{i=1}^{3}U_{\alpha i}\nu_i+U_{\alpha N}N\,,
\end{equation}
where $\nu_{\alpha(i)}$ denote the interaction (mass) eigenstates and $N$ is the heavy, almost sterile, neutrino.
If the mixing elements $U_{\alpha N} \to 0$ for all $\alpha$, then $N$ can be identified with the flavour state $N^\prime$ and $M_N$ is the physical mass of the heavy state. 
In this work, we assume the mixing elements are small and organize our discussion around the physical field $N$.
As \cref{eq:mixing} shows, HNLs would participate in any process in which active neutrinos appear, ``inheriting" their interactions with an additional suppression from the mixing. 
These elements are typically small, giving rise to weaker-than-weak interactions, long lifetimes, and suppressed production. 
Nevertheless, laboratory experiments can be sensitive to HNLs in a wide mass range. 

Experimental searches have not found any significant evidence for their existence, but have provided strong limits on the mixing elements, $U_{\alpha N}$. 
These bounds are flavour-dependent, as the processes relevant to HNL searches are considerably different depending on the flavour structure of the mixing matrix~\cite{Drewes:2018gkc,SHiP:2018xqw,Bondarenko:2021cpc,Tastet:2021vwp}.
Constraints are typically stronger on the mixing with the electron and muon flavours, and less so on the mixing with tau. 
The standard assumption followed by experimental collaborations when providing limits on either $U_{e N}$, $U_{\mu N}$, or $U_{\tau N}$ (strictly speaking, on their moduli squared, which are the only observable quantities) is to consider a simplified flavour structure. It is assumed that the mixing with one flavour dominates over the other two elements, whose impact on the phenomenology is assumed to be negligibly small. In this work, we will follow this simplified scenario. 
However, these commonly used benchmarks do not reflect the flavour structure typically found in realistic neutrino mass models. As a better approximation to the phenomenology of neutrino mass models, two additional benchmarks have been recently proposed in the context of the CERN’s Physics Beyond Colliders initiative~\cite{Drewes:2022akb}.

A comprehensive collection of experimental limits on HNLs is available on \gitlink. 
This database adds to the existing collection of limits on other light particles, such as the limits on axions and dark photons~\cite{Ilten:2018crw,AxionLimits}, as well as complementary efforts for HNLs~\cite{Bolton:2019pcu}.

\subsection{Constraints on heavy neutrino mixing}
\label{sec:HNL_bounds}

Many bounds on effective operators, studied in the following sections, will be obtained by reinterpreting those existing on HNL mixing. 
Thus, we will first summarize the most relevant and up-to-date constraints existing in the literature.
For previous studies summarizing constraints on HNLs, see Refs.~\cite{Atre:2009rg,Ruchayskiy:2011aa,deGouvea:2015euy,Drewes:2015iva,Antusch:2015mia,Fernandez-Martinez:2016lgt,Bolton:2019pcu,Chrzaszcz:2019inj} as well as recent reviews~\cite{Agrawal:2021dbo,Abdullahi:2022jlv}.
As mentioned, single flavour dominance is assumed, so we will discuss limits that apply separately on either $|U_{e N}|$, $|U_{\mu N}|$, or $|U_{\tau N}|$. 
Many experimental bounds are available, but we will only mention the leading limits in each HNL mass window here. 

\paragraph{Electron flavour dominance ($|U_{e N}|$):} 
At the lowest masses, the best limits are provided by searches for an invisible resonance in $\pi^+\to e^+ N$ decays.
The best constraints are given by these ``peak searches" at TRIUMF~\cite{Britton:1992xv} and PIENU~\cite{Aguilar-Arevalo:2017vlf}, including phenomenological uses of their data~\cite{Bryman:2019bjg}.
Below $16$~MeV, Borexino has searched for the $N\to \nu e^+e^-$ decays of HNLs produced in $^8$B decays in the Sun, providing competitive limits~\cite{Borexino:2013bot}.
Above the pion mass, the dominant constraints come from peak searches in leptonic kaon decays, $K^+\to e^+ N$, in particular, those performed by the NA62 collaboration~\cite{NA62:2020mcv}. 
HNL decay-in-flight signatures at neutrino experiments in a similar mass region provide competitive constraints with a complementary technique.
The T2K collaboration searched for HNL charged-current (CC) decay channels in their near detector, ND280, using HNLs produced in kaon decays at the target~\cite{Abe:2019kgx}.
Above the kaon mass, the CHARM~\cite{Bergsma:1985is} and BEBC~\cite{CooperSarkar:1985nh,Barouki:2022bkt} experiments set the strongest limits. 
Finally, for HNLs with masses above $\sim 2$~GeV, HNLs are no longer copiously produced by meson decays at beam dumps, and colliders provide the best environment for detection. 
One of the most competitive limits was provided by the DELPHI collaboration~\cite{Abreu:1996pa} studying $Z$ decays to produce the HNLs, and searching for their prompt or displaced decay inside the detector.
Because the production proceeds through the neutral-current (NC) coupling, this bound is largely insensitive to the flavour structure of the HNL mixing.
At the LHC, the ATLAS~\cite{ATLAS:2019kpx, ATLAS:2022atq} and CMS~\cite{CMS:2022fut} collaborations have set some of the strongest limits by using both $W$ or $Z$ boson decays and searching for the subsequent HNL decays into dilepton final states. Finally, for masses beyond the reach of collider energies, the HNL mixing can only be probed indirectly by testing the unitarity of the $3 \times 3$ PMNS matrix, which involves only the 3 light mass eigenstates and active neutrino flavours, through flavour and electroweak precision data~\cite{Fernandez-Martinez:2016lgt,dani_in_progress}.

\paragraph{Muon flavour dominance ($|U_{\mu N}|$):} 
This case follows a very similar pattern to the one above. As the muon mass is close to that of the pion, HNLs can only be produced in pion decays if they are lighter than about $30$ MeV.
Just as before, peak searches at PSI~\cite{Daum:1987bg} and the PIENU experiment~\cite{Aguilar-Arevalo:2019owf} provide the best limits. 
Above the pion mass, the decay-in-flight searches at the PS191 and T2K experiments provide the strongest bounds~\cite{Bernardi:1985ny,Bernardi:1987ek,Abe:2019kgx}~\footnote{We omit the PS191 limits from our plots, as these have been shown to have been overestimated and are subdominant to the constraints shown here~\cite{Arguelles:2021dqn,Gorbunov:2021wua}.}.  
The region in between $m_\pi - m_\mu \sim 34$~MeV and $m_\pi$ was not targeted by any dedicated experimental search, but phenomenological studies have recasted limits from MicroBooNE~\cite{Kelly:2021xbv} and T2K~\cite{Arguelles:2021dqn}, as well as from measured kaon decays (KEK)~\cite{Hayano:1982wu,Yamazaki:1984sj}.
Above $M_N \sim 180$~MeV, the E949 and NA62 kaon factories provide the strongest constraints using peak searches in leptonic kaon decays~\cite{Artamonov:2009sz,NA62:2021bji}.
Above the kaon mass, low-energy neutrino experiments and kaon factories become insensitive, and high-energy beam dumps dominate by looking for the decays of HNLs produced in $D$ meson decays and muon-neutrino upscattering in the detector.
The leading limits are provided by BEBC~\cite{WA66:1985mfx}, NuTeV~\cite{Vaitaitis:1999wq}, and CHARM~\cite{Bergsma:1985is,CooperSarkar:1985nh}. 
Finally, for HNLs above $\sim 2$~GeV, high-energy colliders stand out.
The ATLAS~\cite{ATLAS:2019kpx, ATLAS:2022atq} and CMS~\cite{CMS:2022fut} collaborations provide the most stringent constraints using displaced decays of HNLs.
Finally, DELPHI~\cite{Abreu:1996pa} also provides competitive limits using prompt or displaced decays from HNLs produced in $Z$ decays. As for the electron case, for the highest HNL masses constraints can be set through flavour and electroweak precision data probing the leptonic mixing matrix unitarity~\cite{Fernandez-Martinez:2016lgt,dani_in_progress}.

\paragraph{Tau flavour dominance ($|U_{\tau N}|$):} Finally, the mixing with the tau flavour is the less probed. 
Solar neutrinos can produce light HNLs ($M_N < 16$~MeV) inside the Earth, and their subsequent decays in large-volume neutrino detectors, such as Borexino, provide the best limits~\cite{Plestid:2020ssy}.
A similar strategy was pursued by Ref.~\cite{Gustafson:2022rsz} to obtain bounds from Super-Kamiokande on HNLs produced by atmospheric neutrinos scattering in the Earth. 
Atmospheric neutrino data~\cite{Dentler:2018sju} provide a limit of $|U_{\tau N}|^2<0.13$ (90\% C.L.)~\cite{Arguelles:2022tki} for HNL masses in the $10\, \mathrm{eV}\lesssim M_N \lesssim 10\,\rm{MeV}$ range~\cite{Blennow:2016jkn}.
For larger HNL masses, only high-energy beam dumps can produce them, through $D$ and $\tau$ decays, so the best limits are those obtained in phenomenological reanalyses of CHARM~\cite{Orloff:2002de,Boiarska:2021yho} and BEBC~\cite{CooperSarkar:1985nh,Barouki:2022bkt} data.
A dedicated experimental search was also performed by the ArgoNeuT collaboration~\cite{ArgoNeuT:2021clc}.
While peak searches in $\pi$ and $K$ decays are not possible in this case, a peak search in $\tau\to N 3\pi$ decays was performed by the BaBar $e^+e^-$ collider~\cite{BaBar:2022cqj}.
As usual, for HNLs above the GeV scale (roughly $1.5$~GeV), the most competitive limits are given by colliders.
As explained above, the production of HNLs at DELPHI was mostly insensitive to the flavour structure of mixing, so DELPHI provides the strongest constraints~\cite{Abreu:1996pa}. As for the other flavours, the constraints for the highest HNL masses stem from unitarity bounds~\cite{Fernandez-Martinez:2016lgt,dani_in_progress}.

\begin{figure}[t]
\centering
\includegraphics[width=\columnwidth]{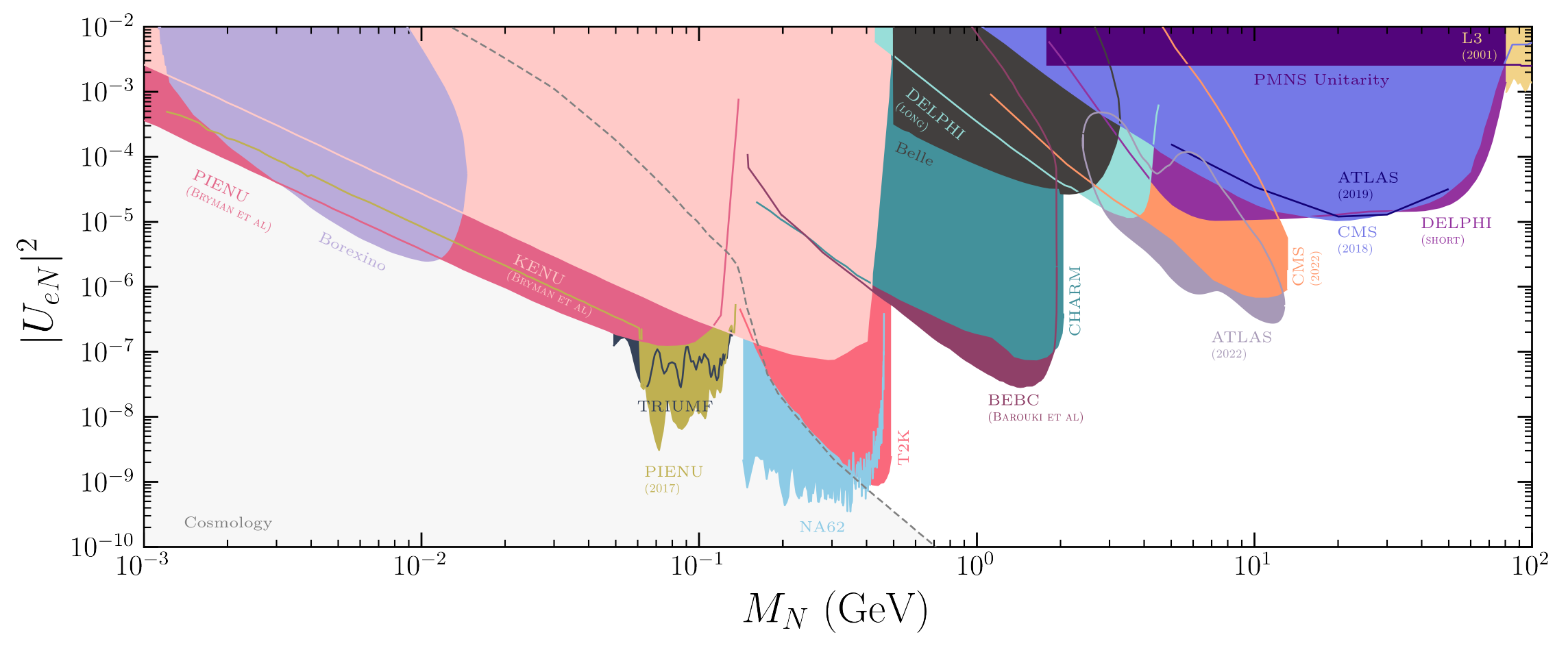}
\includegraphics[width=\columnwidth]{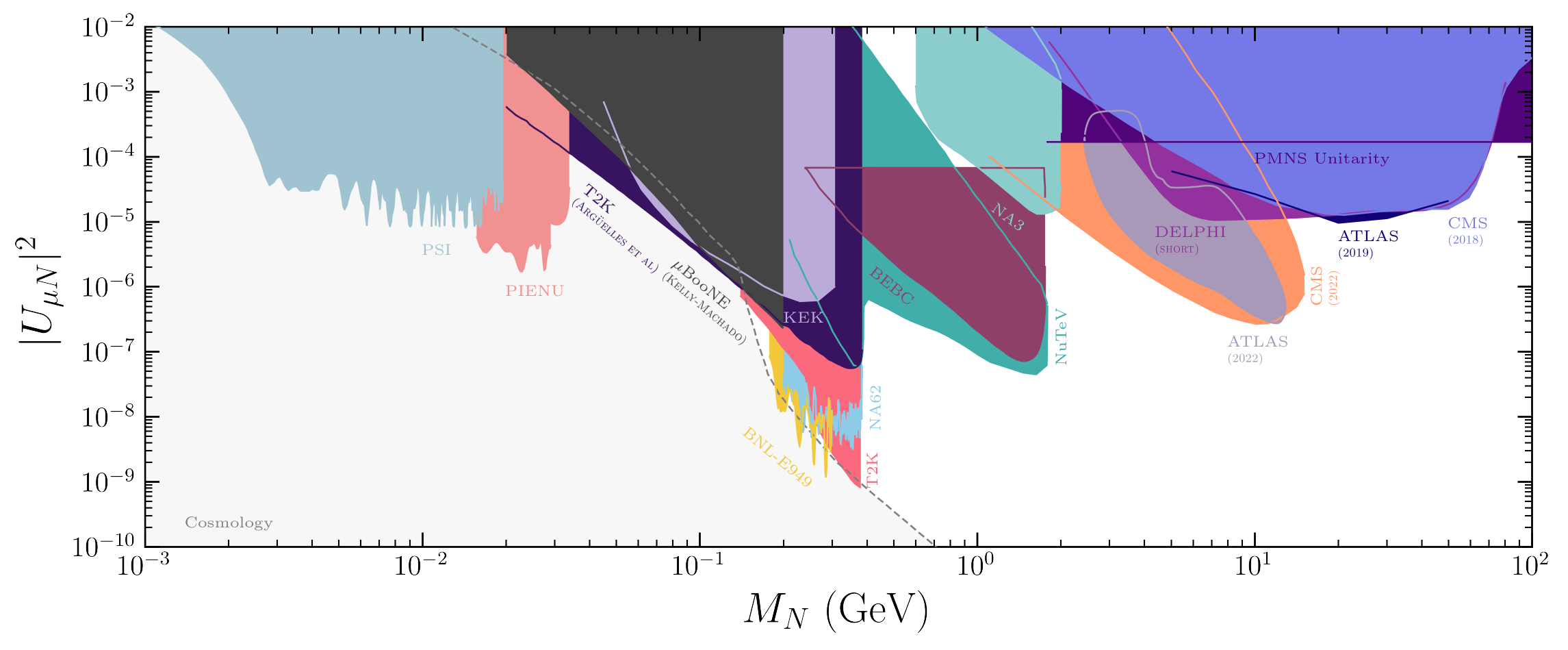}
\includegraphics[width=\columnwidth]{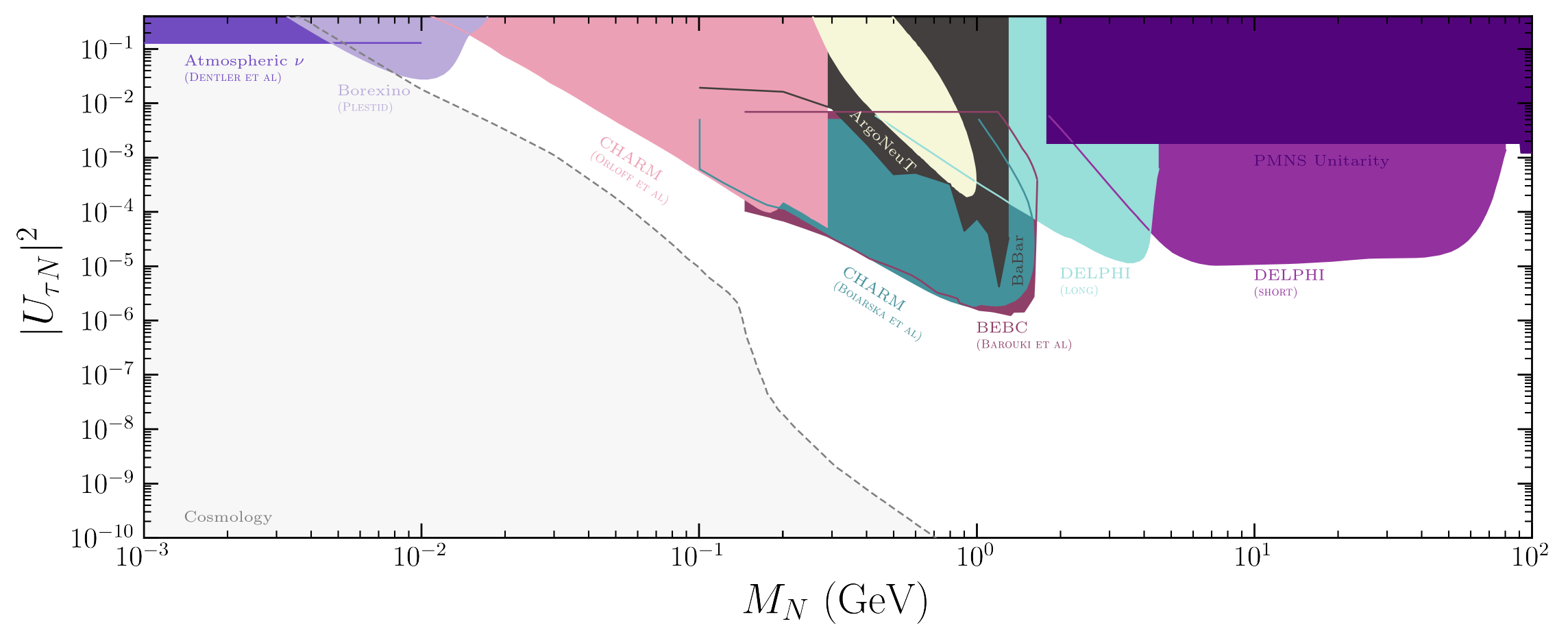}
\caption{Present constraints on heavy neutrino mixing at 90\% C.L., as a function of the HNL mass. The bounds set by the different experiments are displayed separately, but we only show those more relevant in each mass window. Single flavour dominance is assumed, with the limits on $|U_{e N}|^2$, $|U_{\mu N}|^2$, and $|U_{\tau N}|^2$ depicted on the upper, middle, and lower panels, respectively.}
\label{fig:mixing_bounds_dom}
\end{figure}

\Cref{fig:mixing_bounds_dom} displays all the bounds discussed above, as a function of the heavy neutrino mass, for each lepton flavour dominance case.
The constraints set by each experiment are displayed independently. 
It is worth mentioning that, besides these laboratory constraints, HNLs mixing with active neutrinos would have a considerable impact on the cosmological evolution of the Universe. 
Indeed, even though their mixing may not be large enough to thermalize HNLs with the SM plasma, they would still be produced, and, upon becoming non-relativistic, their contribution to the energy density of the universe could be sizable. 
For small masses and mixings, the lifetime of the HNLs could be larger than the age of the Universe, and their predicted population could exceed the measured dark matter component. 
Conversely, for heavier masses, the HNL decays could lead to an unacceptably large contribution to the energy density in radiation, $N_{\rm eff}$. Their decay products to charged particles could also raise the ionization floor after recombination and alter the CMB temperature and polarization power spectra. 
The combination of these effects allows setting very stringent bounds, stronger than the laboratory constraints summarized in \cref{fig:mixing_bounds_dom} in some parts of the parameter space, as shown in Refs.~\cite{Vincent:2014rja,Langhoff:2022bij}. 
Similarly, if HNLs decay during BBN, they might alter its predictions for the primordial abundance of light elements, leading to stringent constraints~\cite{Dolgov:2000pj,Ruchayskiy:2012si,Gelmini:2020ekg,Sabti:2020yrt,Boyarsky:2020dzc}. 
For larger mixing angles, however, HNLs decay before the BBN epoch, and cosmological constraints are no longer effective.
In that sense, they provide an upper as well as a lower limit for the mixing of HNLs, with a direct complementarity to laboratory-based bounds.

The aforementioned constraints also rely on a given cosmological model and its extrapolation to the early universe, and it is fair to treat them separately compared to the more direct laboratory limits. 
The two sets of bounds should therefore be regarded as complementary since a disagreement between them could signal a deviation from the standard cosmological model. 
Thus, in this work, we will focus only on the more direct laboratory and astrophysical constraints, but the existence of these complementary cosmological bounds should also be acknowledged.  

\section{Including HNLs in SMEFT}
\label{sec:SMEFT}

The Standard Model Effective Field Theory (SMEFT) is a very powerful tool to parametrize the effects on low-energy observables of new physics lying at high energies. The possible BSM effects are encoded in effective, non-renormalizable operators of dimension 5 and higher, which are built from the SM fields and respect its fundamental symmetries. Thus, the SMEFT Lagrangian consists of an infinite tower of operators:
\begin{equation}
    \mathscr{L} =  \mathscr{L}_{d=4} +\mathscr{L}_{d=5} +  \mathscr{L}_{d=6}+...\,,
\end{equation}
with
\begin{equation}
    \mathscr{L}_{d=N} = \sum_i \mathcal{O}_i = \frac{1}{\Lambda^{N-4}}\sum_i C_{i} \widetilde{\mathcal{O}_i}\,,
\end{equation}
where $\Lambda$ is the scale of new physics, up to which the effective theory is applicable, and $\widetilde{\mathcal{O}}$ is the dimension $d=N$ effective operator. 
Here, $C_i$ are the Wilson coefficients of each operator: dimensionless parameters that control their respective coupling strength. 
The higher the dimension, the greater the number of possible operators, but the smaller their effects on low-energy observables, as they are increasingly suppressed by larger powers of the new physics scale. 

The effective field theory framework is not suitable if the new particles are lighter than the energy scales of the observables.
In that case, the new propagating degrees of freedom have to be added to the theory.
This is the case for the light HNLs that we consider since $M_N$ can be much smaller than the electroweak scale.
With the addition of the new field $N$, a singlet under the SM symmetry group, the SMEFT framework needs to be extended.
This new framework will contain new operators involving the field $N$ and other SM particles.
This is a compelling approach to thoroughly explore the potential effects of new light particles and takes advantage of the power of EFTs to describe generic and indirect new physics effects.

The operators of this extended theory, which we refer to as $\nu$SMEFT, were first systematically considered in Ref.~\cite{delAguila:2008ir} (see also~\cite{Graesser:2007yj,Graesser:2007pc}), and then in Ref.~\cite{Liao:2016qyd}, where redundant operators were removed, and new dim-7 operators were included. 
Many of these non-renormalizable operators provide a new way for HNLs to interact with other SM particles. 
In addition, many processes mediated by the effective operators are also generated by the renormalizable interactions described in \cref{sec:HNL_bounds}, namely, the mixing between active and heavy neutrinos.
Thus, the bounds arising from this kind of observable may also provide limits on the corresponding Wilson coefficients. 
Whether the existing limits apply and in which way depends on the particular operator considered.

Our main focus will be on $d=6$ operators. 
At higher dimensions, the operators are further suppressed and their number quickly gets out of hand, hindering the extraction of conclusions from an EFT. Let us briefly comment on the only two operators that exist at dimension 5 (see Ref~\cite{Caputo:2017pit} for a detailed discussion), 
\begin{equation}\label{eq:dim5op}
    \mathcal{O}^{d=5}_{\rm Higgs} =\frac{C^{d=5}_{\rm Higgs}}{\Lambda} \overline{N^c} N |H|^2\,, \qquad \qquad
    \mathcal{O}^{d=5}_{\rm dipole} =\frac{C^{d=5}_{\rm dipole}}{\Lambda} \overline{N^c} \sigma_{\mu\nu} N B^{\mu\nu}\,.
\end{equation}
The first one contributes to the Majorana mass term and to the invisible decays of the Higgs boson into two HNLs. 
The associated phenomenology was studied in \cite{Graesser:2007pc,Caputo:2017pit,Barducci:2020icf,Barducci:2020ncz,Barducci:2022hll}.
The focus of these works was on the sensitivity of collider experiments to
prompt or displaced vertex signals originated by HNL visible decays, which depend on the value of the mixing $U_{\alpha N}$ and the HNL mass. Additionally, the ATLAS~\cite{ATLAS:2022vkf} and CMS~\cite{CMS:2022dwd} measurements of the Higgs boson signal strength, $\mu$, provide an important constraint on $C^{d=5}_{\rm Higgs}/\Lambda$, since the exotic Higgs decay $H\rightarrow NN$ would affect this quantity. Following~\cite{Fernandez-Martinez:2022stj} and using the combined bound $\mu \geq 0.94$ ($95\%$~C.L.), we find 
\begin{equation}\label{eq:dim5bound}
    \frac{C^{d=5}_{\rm Higgs}}{\Lambda}<3 \times 10^{-5}~\mathrm{GeV}^{-1} \quad\mathrm{for}\quad M_N< 40~
    \rm{GeV}\,.
\end{equation}
This constraint is comparable with the expectation for the LHC sensitivity from direct searches~\cite{Caputo:2017pit} or even stronger. It should be noted that this current bound is independent of the HNL mixing and mass as long as $M_N< 40$ GeV. The bound becomes increasingly weaker for heavier masses up to $M_H/2$.

The second operator requires at least two different HNL fields to not vanish and amounts to a transition magnetic moment for the HNLs involved. It has been discussed, for instance, in Refs.~\cite{Aparici:2009fh,Barducci:2022gdv}.

In the following sections, we present all the $d=6$ operators involving at least one $N$ field. We will discuss them one at a time, summarizing the processes they mediate and how they are constrained to finally display the bounds on the corresponding Wilson coefficients as a function of the HNL mass. Barring flat directions, which we will try to comment on, this approach is generally conservative and provides the first step towards a more general and ambitious global fit in the future. Details on the procedures followed to derive and recast the constraints are given in~\cref{app:recasting,app:decay_widths}.
A list of the operators and the corresponding limits is shown in \cref{tab:main_table}.

\renewcommand{\arraystretch}{1.7}
\begin{table}[t]
\begin{adjustbox}{width=\textwidth,center} 
\centering
\begin{tabular}{|>{\arraybackslash}p{3 cm}|ll|>{\arraybackslash}p{6 cm}|l|}
\hline
\textbf{Type}                     &                                       \textbf{Operator}       &                                                                                     & \textbf{Leading bounds}                                                                                    & \textbf{Ref.}                     \\ \hline\hline
$H$-dressed mass & $\mathcal{O}^{d=5}_{\rm Higgs}$           & $\overline{N^c}N |H|^2$                                                                               & Higgs signal strength.                                                                             & Eq.~(\ref{eq:dim5bound})                       \\ \hline\hline
$H$-dressed mixing       & $\mathcal{O}_{\rm LNH}^\alpha$            & $\overline{L_\alpha}\widetilde{H}N (H^\dagger H)$                                                                & Standard mixing, invisible $H$ decays.                                            & \cref{fig:higgsdressed}  \\ \hline
\multirow{2}{*}{Bosonic currents} & $\mathcal{O}_{\rm HN}$                    & $\overline{N}\gamma^\mu N (H^\dagger i \overleftrightarrow{D}_\mu H)$                                 & Invisible $Z$ decays, monophoton searches, SN1987A.                                                & \cref{fig:neutbosoncurr} \\ \cline{2-5}
                         & $\mathcal{O}_{\rm HN\ell}^{\alpha}$       & $\overline{N}\gamma^\mu \ell_\alpha (\widetilde{H}^\dagger i \overleftrightarrow{D}_\mu H)$               & Decay-in-flight and peak searches for $e$ and $\mu$. PMNS unitarity and peak searches for $\tau$. & \cref{fig:charbosoncurr} \\ \hline
\multirow{2}{*}{Moments} & $\mathcal{O}_{\rm NB}^\alpha$             & $\left(\overline{L_\alpha} \sigma_{\mu \nu} N\right) \widetilde{H} B^{\mu\nu}$                        & \multirow{2}{6 cm}{Neutrino upscattering, monophoton searches.}                                   & \cref{fig:ONB}           \\ \cline{2-3}\cline{5-5}
                         & $\mathcal{O}_{\rm NW}^\alpha$             & $\left(\overline{L_\alpha}\sigma_{\mu \nu} N\right) \tau^a \widetilde{H} W^{\mu\nu}_a$               &                                                                                                   & \cref{fig:ONW}           \\ \hline
\multirow{3}{*}{4-fermion NC}      & $\mathcal{O}_{\rm ff}$                    & $(\overline{f} \gamma^\mu f) (\overline{N} \gamma_\mu N)$                                             & \multirow{3}{6 cm}{Monophoton and monojet searches, SN1987A.}                                     & \multirow{2}{0cm}{\cref{fig:4ferm_neut_ff}}    \\ \cline{2-3}
                         & $\mathcal{O}_{\rm LN}^\alpha$             & $(\overline{L_\alpha} \gamma^\mu L_\alpha) (\overline{N} \gamma_\mu N)$                               &                                                                                                   &    \\ \cline{2-3}\cline{5-5}
                         & $\mathcal{O}_{\rm QN}$                    & $(\overline{Q_i} \gamma^\mu Q_i) (\overline{N} \gamma_\mu N)$                &                                                                                                   &  \cref{fig:4ferm_neut_left}   \\ \hline
\multirow{4}{*}{4-fermion CC}      & $\mathcal{O}_{\rm LNL\ell}^{\alpha\beta}$ & $(\overline{L_\alpha} N)\epsilon (\overline{L_\alpha} \ell_\beta)$                                    & \multirow{4}{6 cm}{Monolepton searches, decay-in-flight and peak searches.}                      & \cref{fig:4ferm_neut_scalar}              \\ \cline{2-3}\cline{5-5}
                         & $\mathcal{O}_{\rm duN\ell}^{\alpha}$      & $\mathcal{Z}_{ij}^{\rm duN\ell}(\overline{d_i} \gamma^\mu u_j) (\overline{N} \gamma_\mu \ell_\alpha)$ &                                                                                                   & \cref{fig:4ferm_dune}    \\ \cline{2-3}\cline{5-5}
                         & $\mathcal{O}_{\rm LNQd}^\alpha$           & $\mathcal{Z}^{\rm LNQd}_{ij} (\overline{L_\alpha} N)\epsilon (\overline{Q_i} d_j)$                    &                                                                                                   & \cref{fig:4ferm_lnqd}    \\ \cline{2-3}\cline{5-5}
                         & $\mathcal{O}_{\rm QuNL}^\alpha$           & $\mathcal{Z}^{\rm QuNL}_{ij}(\overline{Q_i} u_j)(\overline{N} L_\alpha)$                              &                                                                                                   & \cref{fig:4ferm_qunl}    \\ \hline
\end{tabular}
\end{adjustbox}
\caption{The dimension-six operators of the $\nu$SMEFT involving at least one $N$ field. We list the leading constraints on each one of the operators in the range $1$~MeV $<M_N< 100$~GeV. We also include the bound on the dimension-five operator involving HNLs and the Higgs derived in this work.\label{tab:main_table}}
\end{table}

\section{Higgs-dressed mixing}
\label{sec:higgs_mix}

\begin{figure}[t]
\centering
\includegraphics[width=\columnwidth]{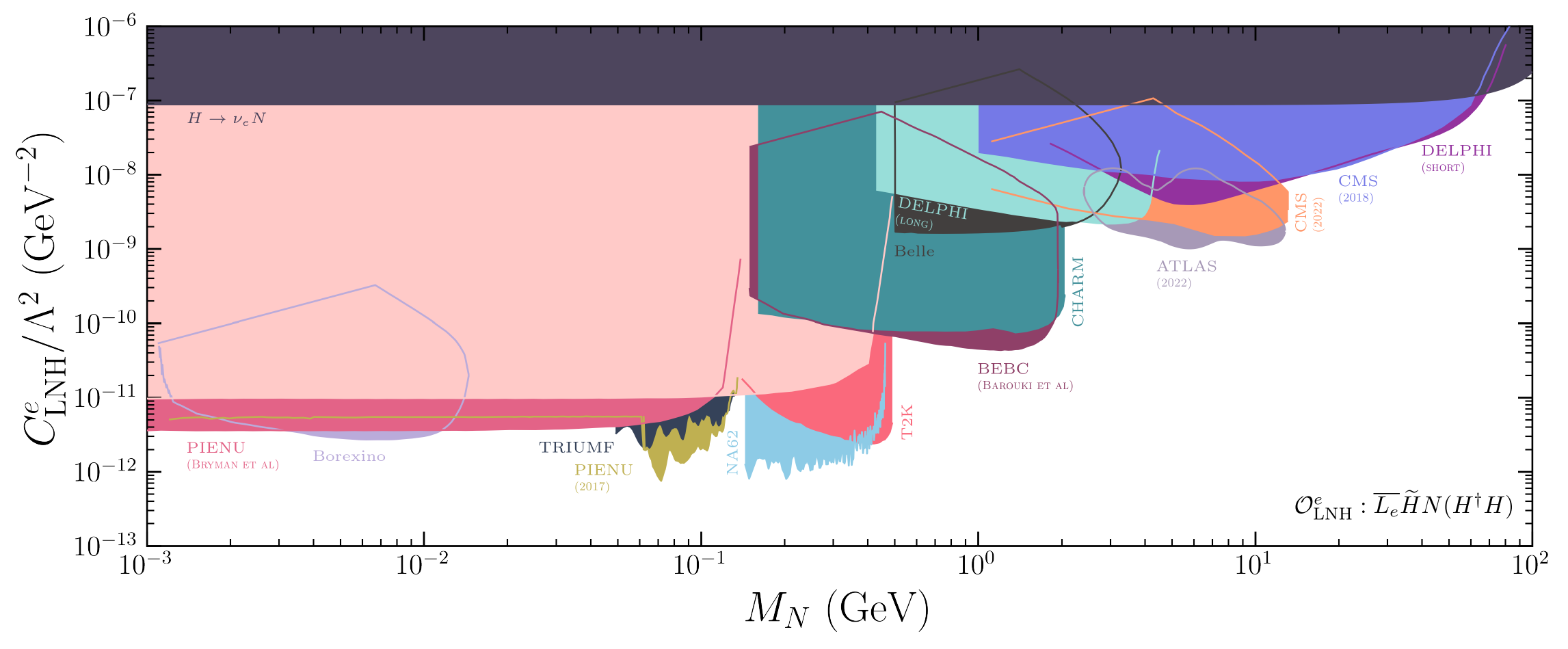}
\includegraphics[width=\columnwidth]{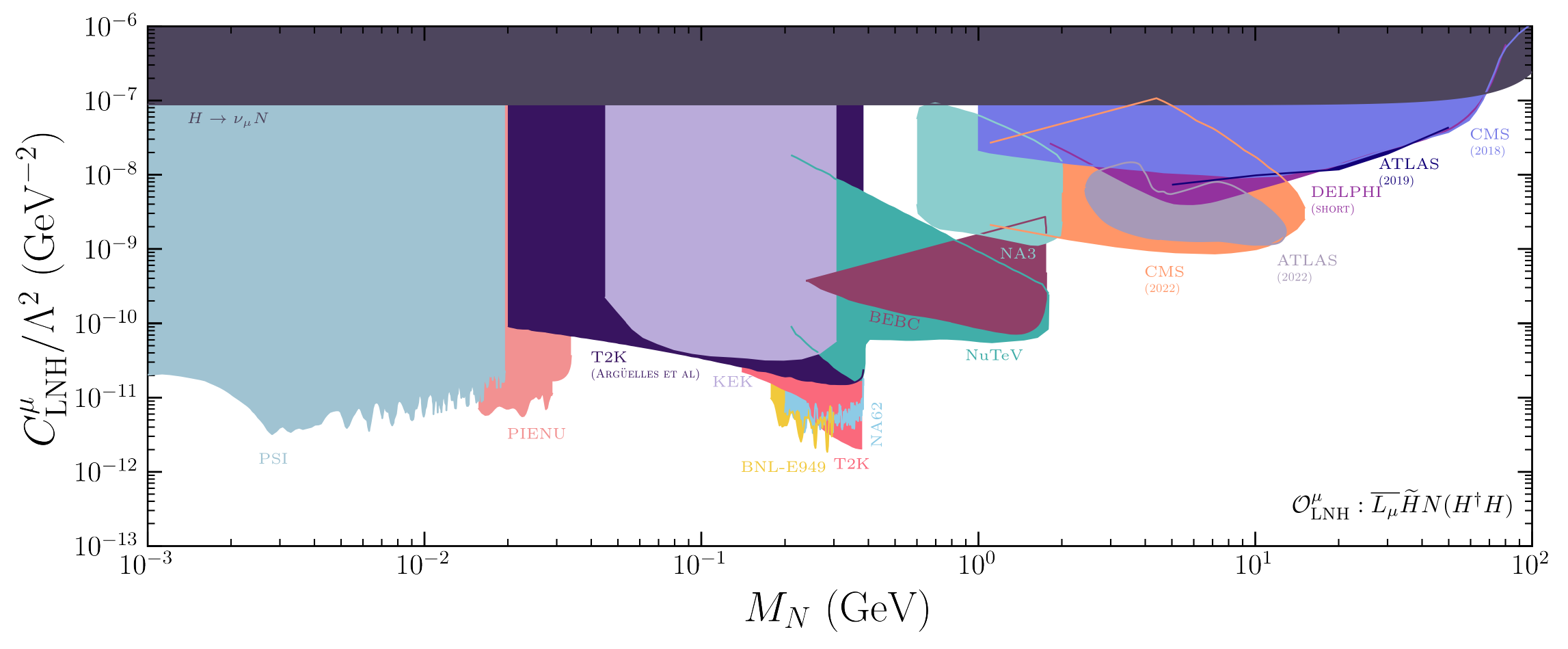}
\includegraphics[width=\columnwidth]{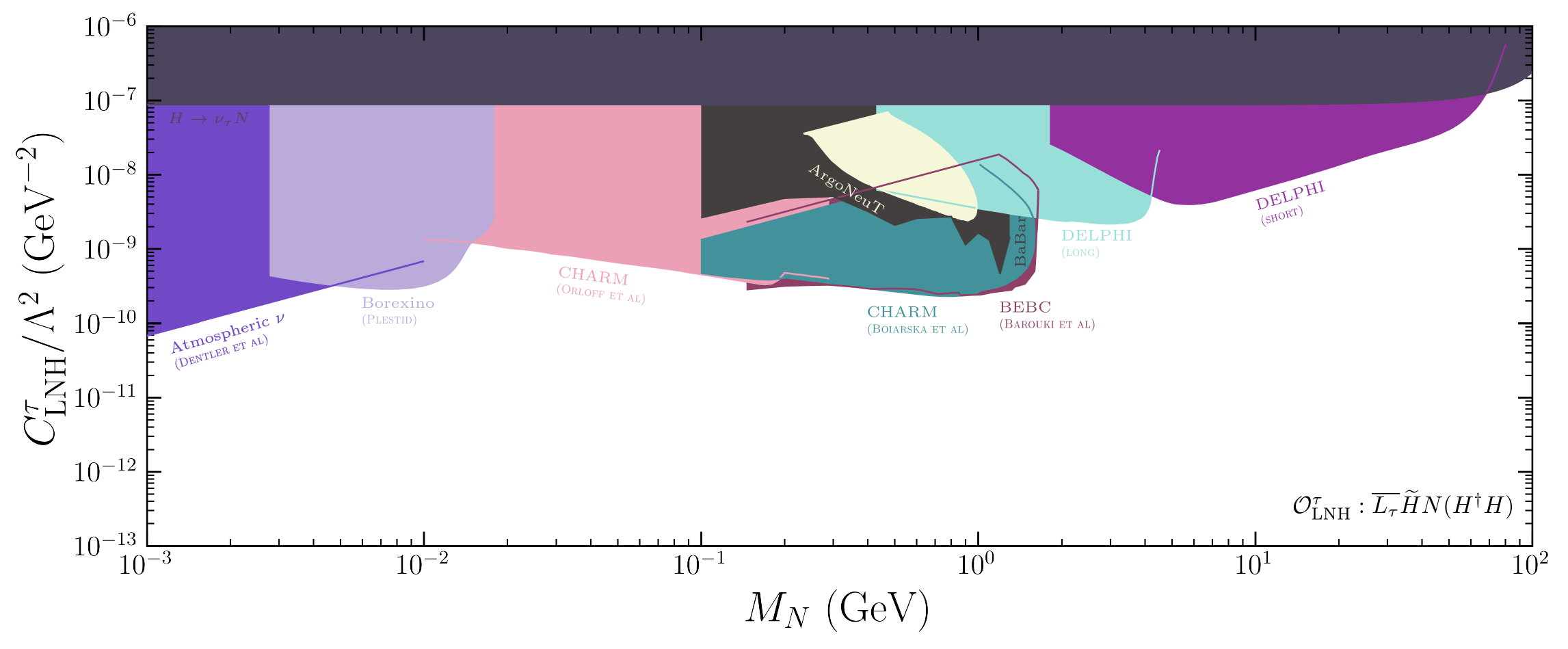}
\caption{The 90\% C.L. constraints on the Wilson coefficient of the Higgs-dressed mixing operator in \cref{eq:higgsdressed}, as a function of the heavy neutrino mass. We display in separate panels the constraints relevant for each lepton flavour, detailing the most dominant sources of limits. See text for details.}
\label{fig:higgsdressed}
\end{figure}

The first and simplest operator we consider is the addition of a Higgs bilinear to the Yukawa coupling of \cref{eq:yukawa_coupl},
\begin{align}\label{eq:higgsdressed}
    \mathcal{O}_{\rm LNH}^\alpha = \frac{C_{\rm LNH}^\alpha}{\Lambda^2}(H^\dagger H) \overline{L_\alpha}\widetilde{H}N  
\end{align}
This operator is a neutrino Yukawa coupling with an extra pair of Higgs doublets. Once they acquire vevs, this operator induces extra contributions at tree level to light neutrino masses and their mixing with the HNLs. 
In the limit where $M_N \ll v \ll \Lambda$, the generated mass and mixing read
\begin{align}
    U_{\alpha N} = \frac{C_{\rm LHN}^\alpha}{2\sqrt{2}} \frac{v}{M_N}\left(\frac{v}{\Lambda}\right)^2, 
    \qquad
    m_{\nu} =
    \frac{(C_{\rm LHN}^\alpha)^2}{8} \frac{v^2}{M_N}\left(\frac{v}{\Lambda}\right)^4\,.
    \label{eq:higgs_mix}
\end{align}
Note that there are three copies of this operator, one for each lepton flavour, whose Wilson coefficients are, in principle, independent.

At low energies, the mixing induced by this operator gives rise to an identical phenomenology to that already studied in the case of the Yukawa coupling of \cref{eq:yukawa_coupl}; thus, all bounds on mixings from direct constraints apply directly, yielding limits on the Wilson coefficient. 
There could be possible cancellations between this contribution and the tree level $d=4$ Yukawa couplings, leading to possible flat directions; as we work under the assumption of negligible standard mixing, we will not take them into account. 
In principle, it could be possible to lift this flat direction, as this effective operator also mediates processes involving two HNLs and several Higgs bosons, and the $d=4$ Yukawa does not. 
However, these processes, such as $HH(H)\to NN$, require multi-Higgs production and thus do not offer promising experimental prospects.

The existing limits on the absolute neutrino mass scale would also allow setting extremely stringent constraints on the Wilson coefficient. However, the expression shown in \cref{eq:higgs_mix} corresponds to the inclusion of a single HNL. When several of these new particles are considered, their contributions to  neutrino masses could cancel each other, while this is not a possibility for their mixings. Indeed, this is exactly what would be expected in low-scale realizations of the Seesaw mechanism. In those scenarios, the contribution to the neutrino mass cancels, due to its protection from the approximate $B-L$ symmetry, while the mixing does not. Such a protection mechanism needs to be in place to have such light HNLs in the first place. Thus, we do not include bounds on the Wilson coefficient arising from the current information on the absolute value of neutrino masses.

Apart from contributing to the neutrino masses and mixing, this operator induces an invisible Higgs decay channel, into a light and a heavy neutrino.
The direct limits on the invisible branching ratio of the Higgs are
$\mathcal{B}(h\to N\nu_{\alpha}) < 0.13$ , as reported by ATLAS~\cite{ATLAS:2022vkf}, and $\mathcal{B}(h\to N\nu_{\alpha}) < 0.16$, as reported by CMS~\cite{CMS:2022dwd}, both at 95\% C.L.
In our work, we will combine the most stringent upper limits based on the Higgs signal strength by ATLAS~\cite{ATLAS:2022vkf} and CMS~\cite{CMS:2022dwd}, yielding $\mathcal{B}(h\to N\nu_{\alpha}) < 0.06$ at 95\% C.L. following the Feldman-Cousins~\cite{Feldman:1997qc} prescription.

In \cref{fig:higgsdressed}, we show the bounds on the Wilson coefficients of this operator under the three lepton-flavour dominance scenarios, shown in three separate panels. 
Most shaded regions (except the dark grey bands) correspond to reinterpreted mixing bounds, and, according to the assumption of single flavour dominance, are different for each lepton flavour. The limits from Higgs invisible decay (dark grey region) apply to the dominant Wilson coefficient, independently of its flavour.

\section{Bosonic currents}
\label{sec:boson_curr}

It is possible to couple bosonic currents to the $V+A$ currents constructed out of heavy neutrinos. The resulting operators read
\begin{align}
\label{eq:OHN}
    \mathcal{O}_{\rm HN} &=\frac{C_{\rm HN}}{\Lambda^2} \overline{N}\gamma^\mu N (H^\dagger i \overleftrightarrow{D}_\mu H) \,,
\\
\label{eq:OHNe}
    \mathcal{O}_{\rm HN\ell}^{\alpha} &= \frac{C_{ \rm HN\ell}^{\alpha}}{\Lambda^2} \overline{N}\gamma^\mu \ell_\alpha (\widetilde{H}^\dagger i \overleftrightarrow{D}_\mu H) \,,
\end{align}
where $\ell_\alpha$ is a right-handed charged lepton of flavour $\alpha$. After electroweak symmetry breaking, the first operator will yield a coupling of the $Z$ boson to two HNLs, while the second operator will produce a coupling of the $W$ boson to one HNL and a charged lepton. The latter will exhibit three copies, depending on the flavour of the charged lepton. The phenomenological consequences of these two operators are quite distinct, and so are the constraints placed upon their corresponding Wilson coefficients. 
They are discussed separately below.

\subsection{Neutral bosonic currents}
\label{sec:neut_bos_curr}
The operator $\mathcal{O}_{\rm HN}$ yields a vertex between two HNLs and a $Z$, namely $\left(\frac{C_{\rm HN}}{\Lambda^2}\frac{g v^2}{2 c_{\rm W}}\right) \overline{N}\gamma^\mu N Z_{\mu}$, opening a new decay channel for this boson if kinematically allowed by the masses of the HNLs. This decay, controlled by the Wilson coefficient $C_{\rm HN}$, constitutes an extra contribution to the invisible decay of the $Z$ boson, which was measured with high accuracy at LEP~\cite{ParticleDataGroup:2022pth,Janot:2019oyi}. We have performed a $\chi^2$ fit to this observable, obtaining upper limits on $C_{\rm HN}/\Lambda^2$.

Similarly, this neutral bosonic current would mediate decays of neutral mesons ($\pi^0$, $\eta$, $\eta'$, $J/\psi$, $\rho^0$, $\Upsilon$(1S)) into two HNLs, if light enough. Analogous invisible decays are also present in the SM, with light neutrinos in the final state~\cite{Marciano:1996wy,Chang:1997tq}. 
For pseudoscalar mesons, these processes require a chirality flip, so in the SM they are suppressed by the small neutrino masses; for vectors, they are very subdominant with respect to hadronic and electromagnetic channels.
Thus, this kind of decay has not been observed so far, and only upper bounds are available.  
Imposing the decays into HNLs to respect the experimental limits provides another method to constrain the Wilson coefficient. We will only display the upper bound on the neutral pion and the $\Upsilon$(1S) invisible decay, as they are the most stringent, depending on the HNL mass. In particular, the NA62 collaboration determined the corresponding branching ratio of the pion to be smaller than $4.4 \times 10^{-9}$, at 90\% C.L.~\cite{CortinaGil:2020zwa}. For the $\Upsilon$(1S), we employ the limit of $3 \times 10^{-4}$, at 90\% C.L.~\cite{BaBar:2009gco}.
Stronger limits on this invisible branching ratio can be obtained using $\gamma \gamma \to \pi^0 \to N N$ in supernovae (see also the supernova discussion below).
The limit is $3.2 \times 10^{-13}$ for the BR~\cite{Natale:1990yx}.

Another probe of this effective operator comes from collider physics, in particular, monophoton searches performed at LEP~\cite{DELPHI:2003dlq}. This channel, intended to look for processes in which one or more invisible particles were produced in the $e^+e^-$ collision, has the photon as the only detectable signal. These searches are blind to the nature of the invisible particle, dark matter being one of the most recurrent candidates. We follow Ref.~\cite{Fox:2011fx}, where an EFT framework was also adopted, with dark matter fermion singlets playing a similar role as our HNLs. We translate this information into constraints on $C_{\rm HN}/\Lambda^2$. Note that Ref.~\cite{Fox:2011fx} employs a basis of effective operators different from the one considered here. To derive our bounds, we compute the monophoton production cross section for the operators of our basis, applying the same cuts in the photon energy and angle, and rescale the constraints accordingly to obtain the same cross section in the signal region as for the operators in Ref.~\cite{Fox:2011fx}. 
Details on this rescaling procedure can be found in \cref{app:monophoton}.

A very similar process could be mediated by this operator replacing electrons with quarks and the photon with a gluon, yielding experimental signals composed of a single jet and missing energy. Thus, monojet searches at the LHC pose an additional constraint on the corresponding Wilson coefficient. 
Limits from these processes are not available for the operator under consideration, and a proper rescaling would involve a detailed simulation, which is out of the scope of this work. Nevertheless, a crude estimate, obtained by rescaling parton-level cross sections, yields bounds similar to those from monophoton searches. These searches will also be relevant for the observables discussed in \cref{sec:four-ferm}.

Supernovae also offer competitive probes of particles that couple very feebly with the SM. 
Due to their weakly interacting nature, neutrinos are the dominant cooling mechanism of core-collapse supernovae.
This mechanism is compatible with the observation of neutrinos from SN1987A by the Kamiokande-II~\cite{Hirata:1987hu} and IMB~\cite{Bionta:1987qt} neutrino detectors.
Thus, a BSM particle would be incompatible with observations if it exhibited a mass and couplings such that it extracted energy from the supernova at a faster rate than SM neutrinos~\cite{Raffelt:1987yt}. 
This usually provides upper and lower limits on the couplings: if the new particle couples too strongly with the SM, it would be trapped inside the supernova, whereas, if the coupling is too small, the new particle is produced much less often than neutrinos, cooling the supernova less. We follow Ref.~\cite{DeRocco:2019jti}, where these arguments are used to constrain the parameter space of a fermion singlet, also working in an EFT framework. 
We again rescale their results so that the production and scattering cross sections of the HNLs coincide with those for the operators in Ref.~\cite{DeRocco:2019jti} along the lines defining the lower and upper bounds respectively (see \cref{app:supernova}). 
Note that, depending on the HNL mass, the most stringent upper bound may be given by HNL production in $\pi^0$ decay in supernovae, as discussed above.

Finally, it should be noted that for this operator, or, more generally, for those involving two HNL fields, and in absence of mixing or other interactions, HNLs would be completely stable. Thus, their production in the early Universe could lead to an overabundant dark matter component and very stringent constraints. As discussed at the end of \cref{sec:HNL}, we regard these important cosmological constraints as complementary to the direct bounds discussed in this work. Indeed, their comparison provides a test of the necessary assumptions on the cosmological model, which are not present in the laboratory or astrophysical observations we consider here. For instance, if the reheating temperature were below the HNL mass, the overproduction constraint would not apply. 

\begin{figure}[t]
\includegraphics[width=\columnwidth]{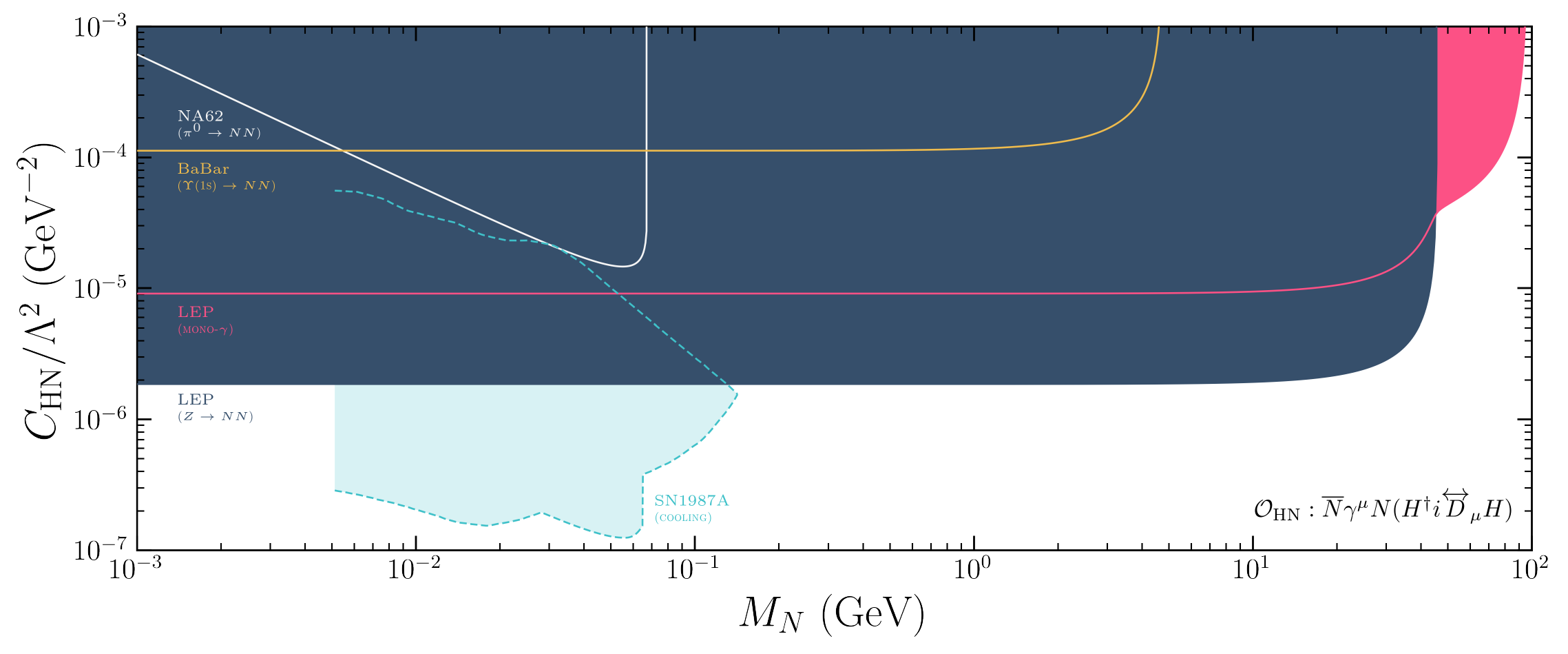}
\caption{The 90\% C.L. constraints on the Wilson coefficient of the neutral bosonic current operator of \cref{eq:OHN}, as a function of the heavy neutrino mass. See text for details. 
}
\label{fig:neutbosoncurr}
\end{figure}

\Cref{fig:neutbosoncurr} collects the bounds on $C_{\rm HN}/\Lambda^2$ derived from the different observables discussed above: invisible $\pi^0$, $\Upsilon$(1S), and $Z$ decays (white, yellow and dark blue regions respectively), monophoton searches (pink area) and supernova cooling arguments (light blue region). Note that the bounds on the standard HNL mixing do not apply, as they typically involve CC interactions (mostly charged mesons decay), which are not mediated by this operator.

\subsection{Charged bosonic current}
After EW symmetry breaking, the operator $\mathcal{O}_{\rm HN\ell}^{\alpha}$ leads to a vertex between an HNL, a charged lepton, and a $W$, with a similar chiral structure to the SM charged currents: $\left( \frac{C_{ \rm HN\ell}^{\alpha}}{\Lambda^2}\frac{g v^2}{2}\right) \overline{N} \gamma^\mu \ell_\alpha W^-_\mu$. Thus, this operator yields a coupling to the $W$ to a right-handed current, analogous to the left-handed counterpart inherited through mixing, controlled by the Wilson coefficient:
\begin{equation}\label{eq:effectiveCCmixing}
U_{\alpha N}^{\rm CC} \equiv \frac{C_{\rm HN\ell}^{\alpha} v^2}{\sqrt{2}\Lambda^2}\,.
\end{equation}
We then reinterpret the current bounds on HNL mixing as limits on $C_{\rm HN\ell}^{\alpha}/\Lambda^2$. 
Most of the bounds only involve processes mediated by the SM charged currents, and thus translate directly by means of $U_{\alpha N}^{\rm CC} = U_{\alpha N}$. 
However, some of the experimental searches also consider NC-mediated processes, absent for the effective operator under consideration. 
The limits provided by those searches need to be rescaled, to remove the neutral contributions while keeping constant the expected number of events at the detector.
We describe our rescaling procedure in \cref{app:rescaling}.

Supernova cooling and LEP monophoton searches are also able to probe charged bosonic currents. There are no basic conceptual differences with the neutral case; the main distinctions appear at a diagrammatic level, as the HNLs are now produced via the exchange of a $W$ in a $t$ channel. 
However, the associated bounds are not competitive with those arising from the mixing, as the corresponding processes are NC-like. Thus, we will not display them.

Finally, this effective operator mediates leptonic decays of muon and tau, $\ell_\alpha\to\ell_\beta\nu_\beta N$. We have performed a $\chi^2$ fit of the corresponding decay widths of the muon and the tau to the experimental measurements~\cite{ParticleDataGroup:2022pth}, finding limits on the Wilson coefficient that dominate at low masses. Note that, in the channel $\tau\to\mu\nu N$, the experimental determination exhibits a slight tension, of roughly 2$\sigma$, with the SM prediction. This means that, at this confidence level, the Wilson coefficient in the tau flavour dominance scenario is constrained to a band instead of only an upper bound.
We display the corresponding lines in a dashed style, as they are not completely equivalent to those arising from other observables.

\begin{figure}[t]
\includegraphics[width=\columnwidth]{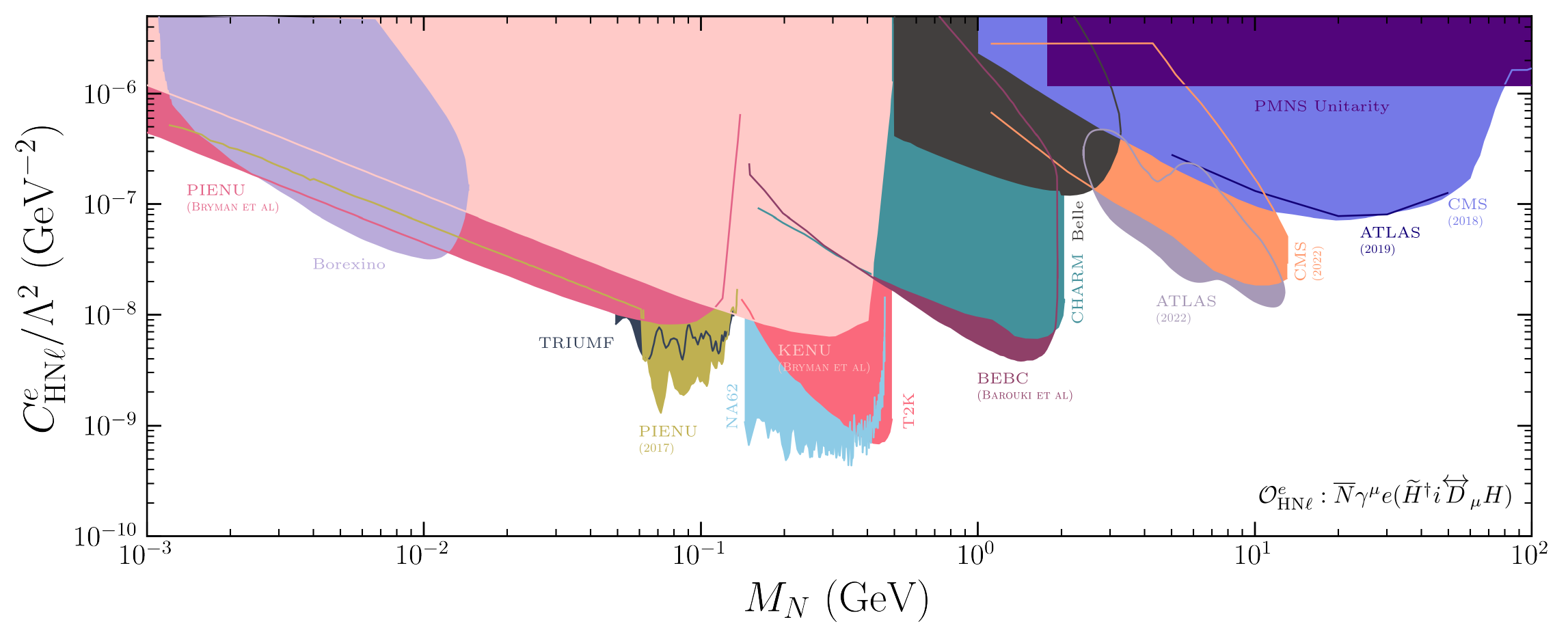}
\includegraphics[width=\columnwidth]{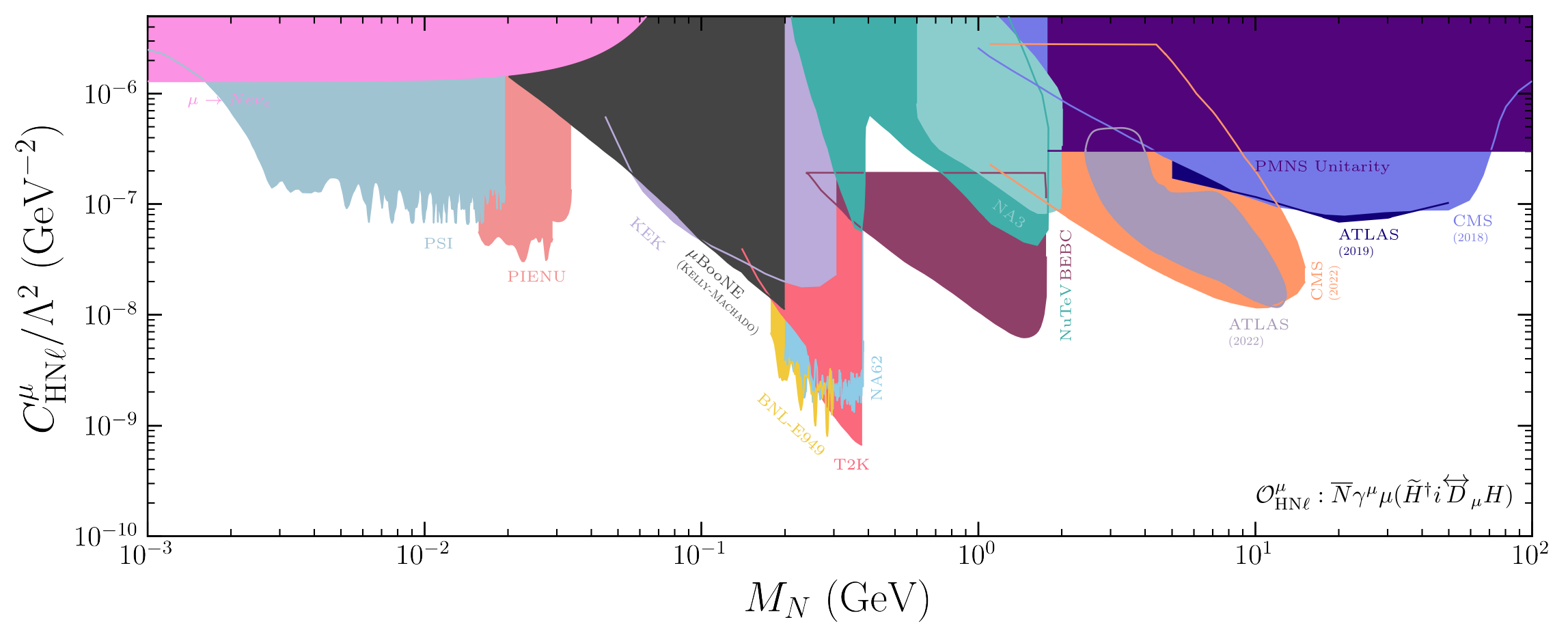}
\includegraphics[width=\columnwidth]{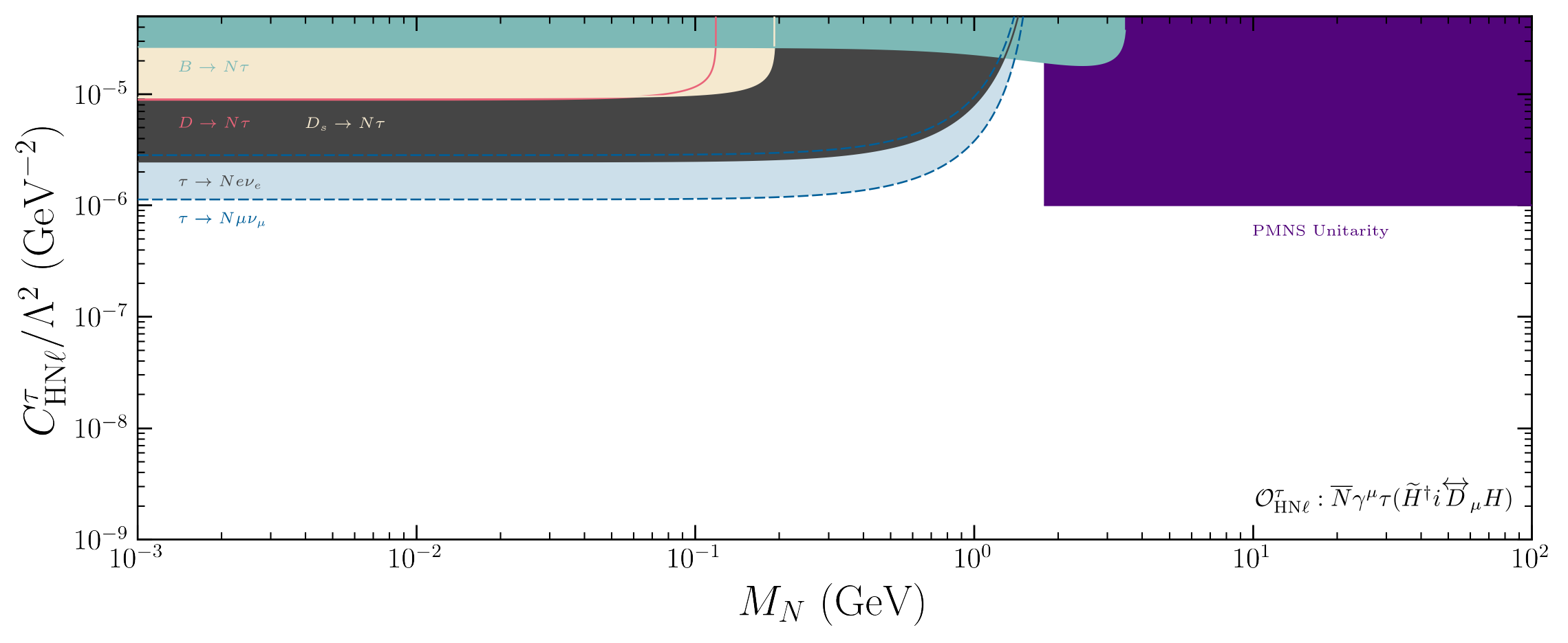}
\caption{The 90\% C.L. constraints on the Wilson coefficient of the charged bosonic current operator in \cref{eq:OHNe}, as a function of the heavy neutrino mass. We display in separate panels the constraints relevant to each lepton flavour, specifying the most relevant experiments. See text for details.}
\label{fig:charbosoncurr}
\end{figure}

\Cref{fig:charbosoncurr} contains the bounds on the Wilson coefficients of this operator, coming from reinterpretations of limits on heavy neutrino mixing. 
Once again, we assume single flavour dominance, displaying on different panels the relevant constraints for each lepton flavour.

\section{Tensor current}
\label{sec:tensor}

Similar to the $d=5$ operator in \cref{eq:dim5op}, dipole couplings between the HNL and the lepton doublet can be considered. 
Two different operators arise:
\begin{align}\label{eq:neut_tensor}
    \mathcal{O}_{\rm NB}^\alpha &= \dfrac{C_{\rm NB}^\alpha}{\Lambda^2}\left(\overline{L_\alpha} \sigma_{\mu \nu} N\right) \widetilde{H} B^{\mu\nu}+ \rm{h.c.}\,,
    \\\label{eq:char_tensor}
    \mathcal{O}_{\rm NW}^\alpha &= \dfrac{C_{\rm NW}^\alpha}{\Lambda^2}\left(\overline{L _\alpha}\sigma_{\mu \nu} N\right) \tau^a \widetilde{H} W^{\mu\nu}_a + \rm{h.c.}\,.
\end{align}
Three copies of these two operators exist, one for each lepton flavour. Once again, their Wilson coefficients are, in principle, independent.
After the Higgs develops a vacuum expectation value, these operators are translated into three dipoles for the HNL, coupling it to the photon, the $W$, and the $Z$:
\begin{eqnarray}
    \label{eq_dipole_coefficient}
    d_\gamma^\alpha &\equiv& \dfrac{\mu_{\nu}^{\alpha}}{2}=\dfrac{v \cos{\theta_{\rm W}}}{\sqrt{2} \Lambda^2}\left(C_{\rm NB}^\alpha+\tan{\theta_{\rm W}} C_{\rm NW}^\alpha\right)\,,
    \\
    d_Z^\alpha &=&\dfrac{v \cos{\theta_{\rm W}}}{\sqrt{2} \Lambda^2}\left(C_{\rm NW}^\alpha-\tan{\theta_{\rm W}}C_{\rm NB}^\alpha\right)\,,
    \\
    d_W^\alpha &=& \dfrac{v}{\Lambda^2} C_{\rm NW}^\alpha\,.
\end{eqnarray}

\begin{figure}[t]
\centering
\includegraphics[width=\columnwidth]{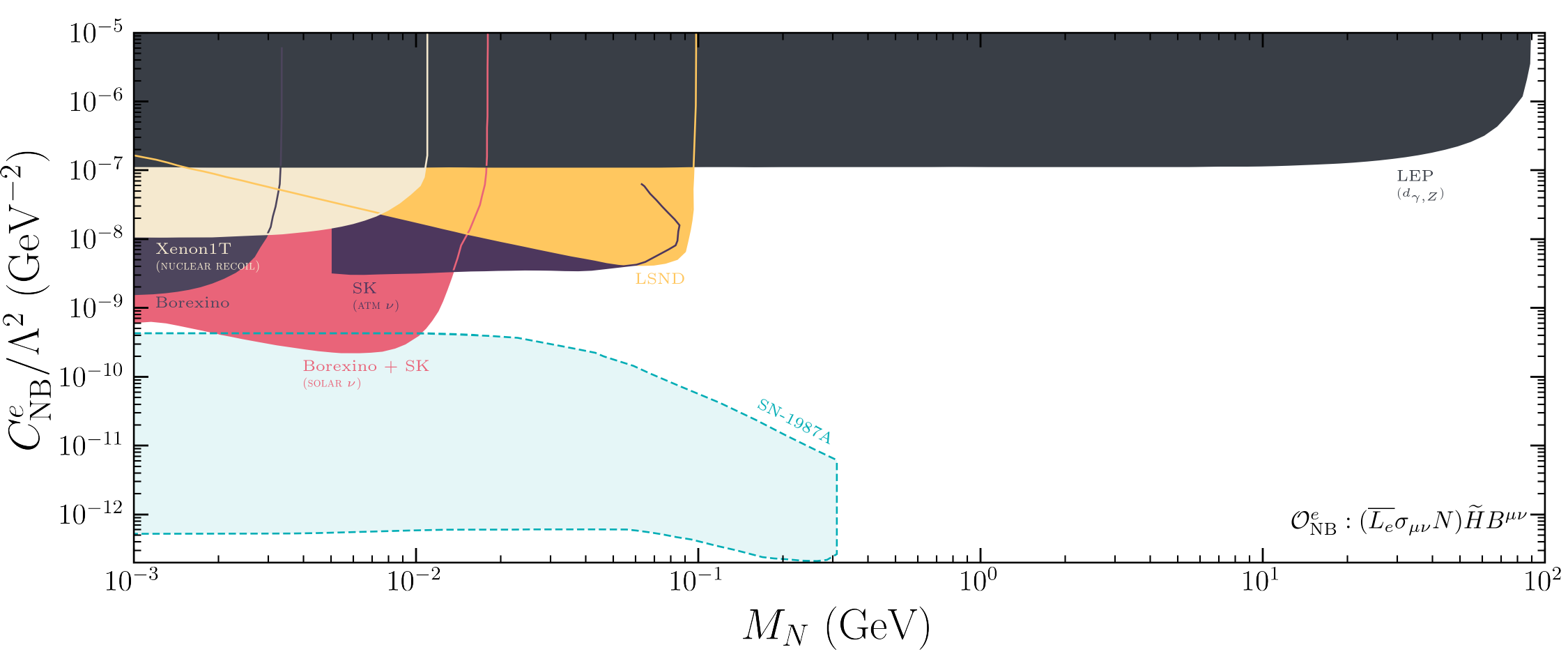}
\includegraphics[width=\columnwidth]{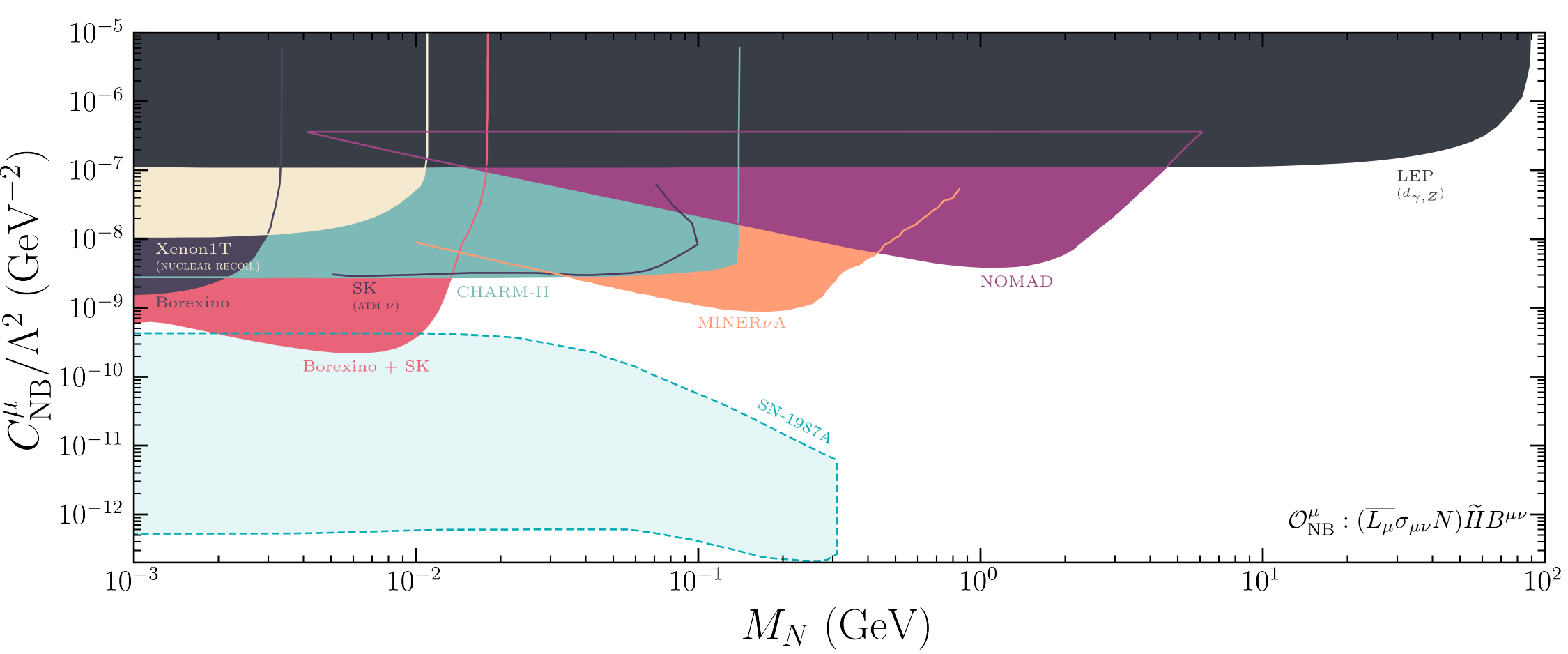}
\includegraphics[width=\columnwidth]{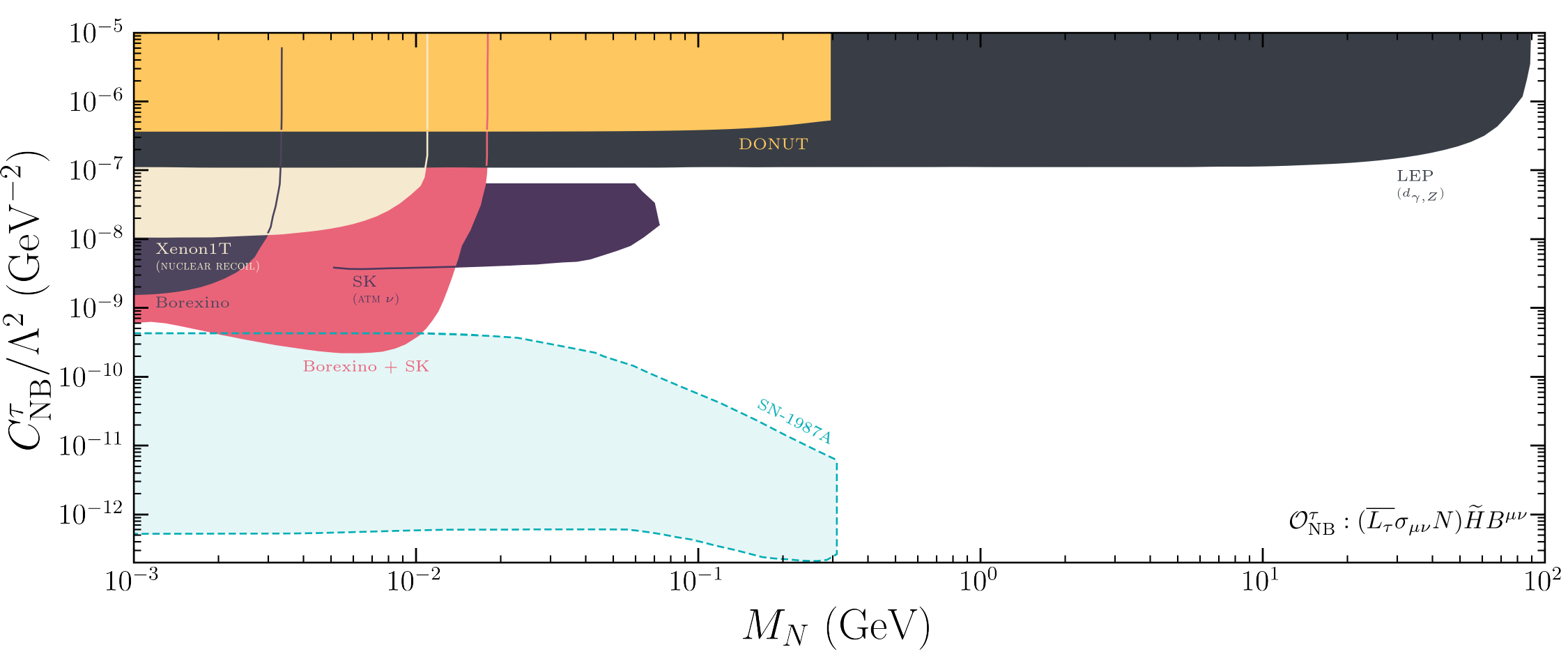}
\caption{The 90\% C.L. limits on the Wilson coefficient of the tensor current operator in ~\cref{eq:neut_tensor}, as a function of the heavy neutrino mass. We display in separate panels the constraints relevant to each lepton flavour, detailing the most dominant sources of limits. See text for details.
\label{fig:ONB}
}
\end{figure}

\begin{figure}[t]
\centering
\includegraphics[width=\columnwidth]{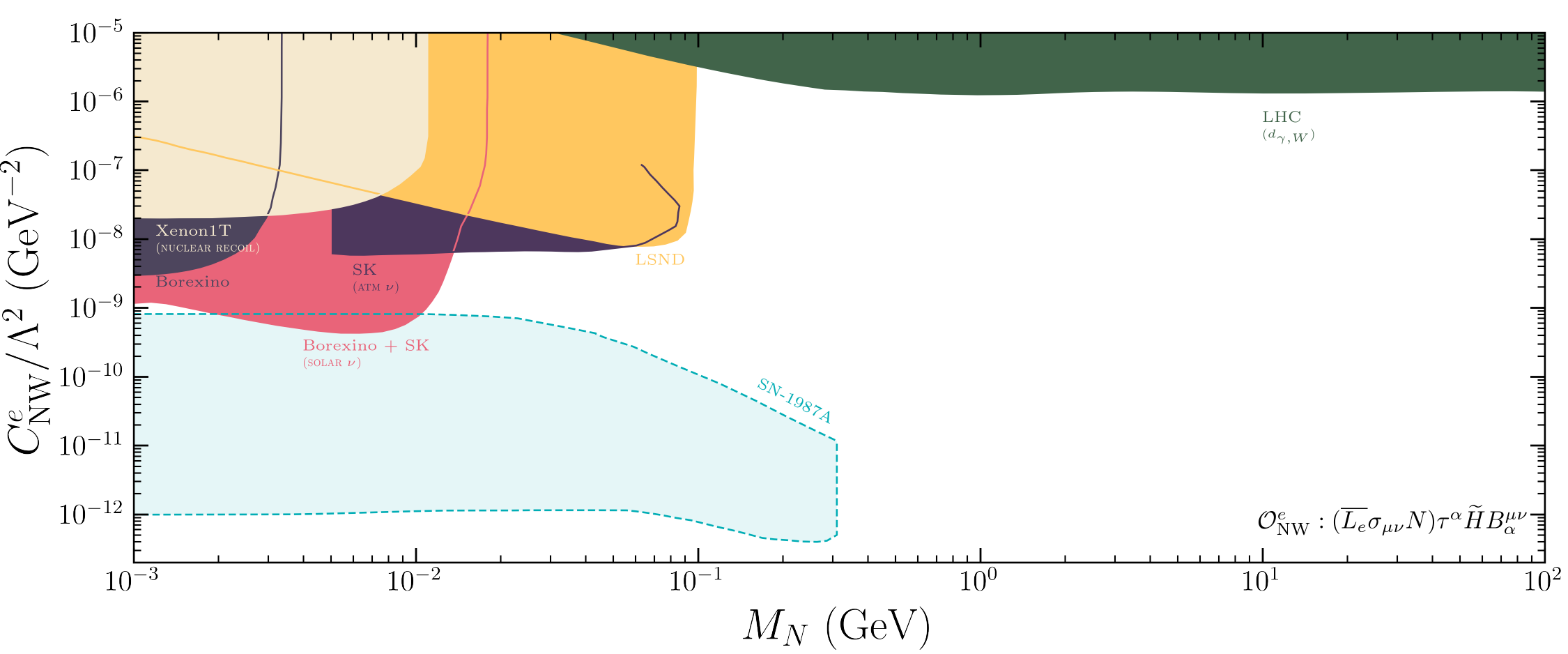}
\includegraphics[width=\columnwidth]{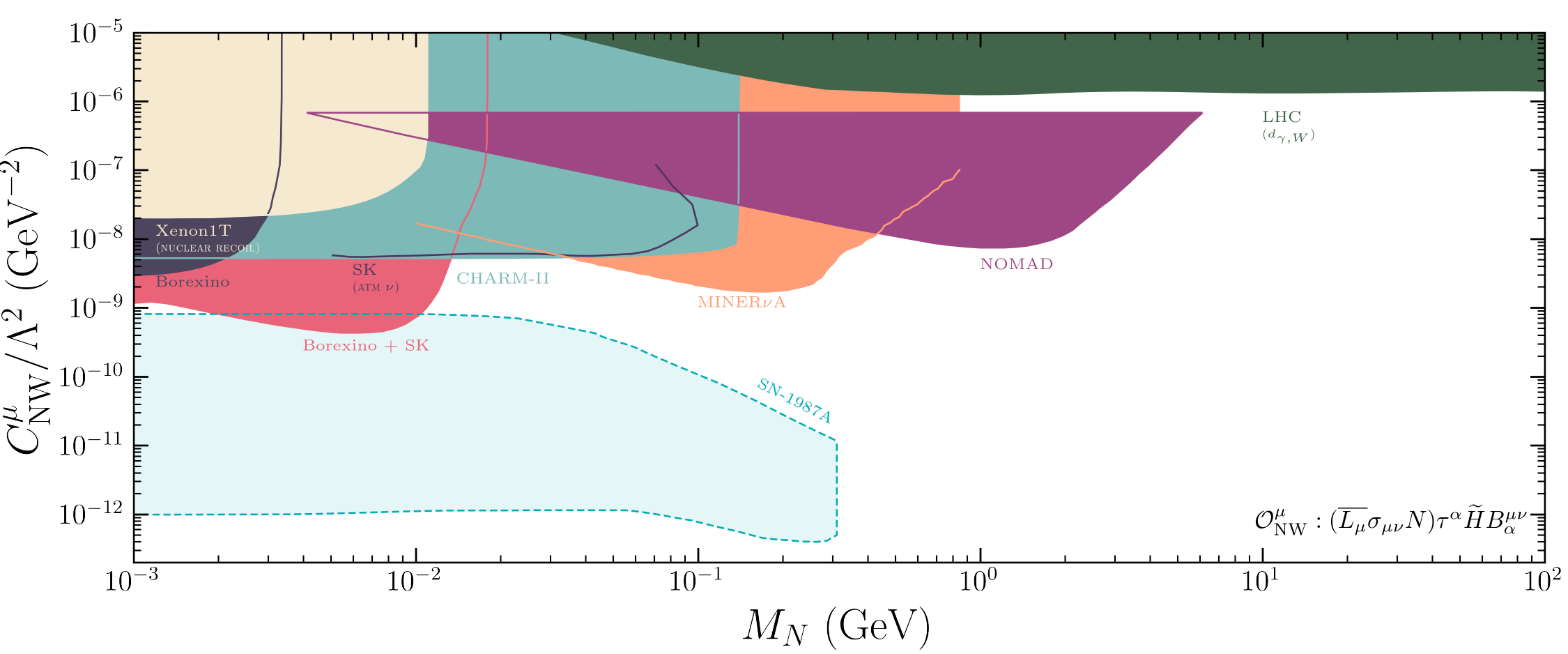}
\includegraphics[width=\columnwidth]{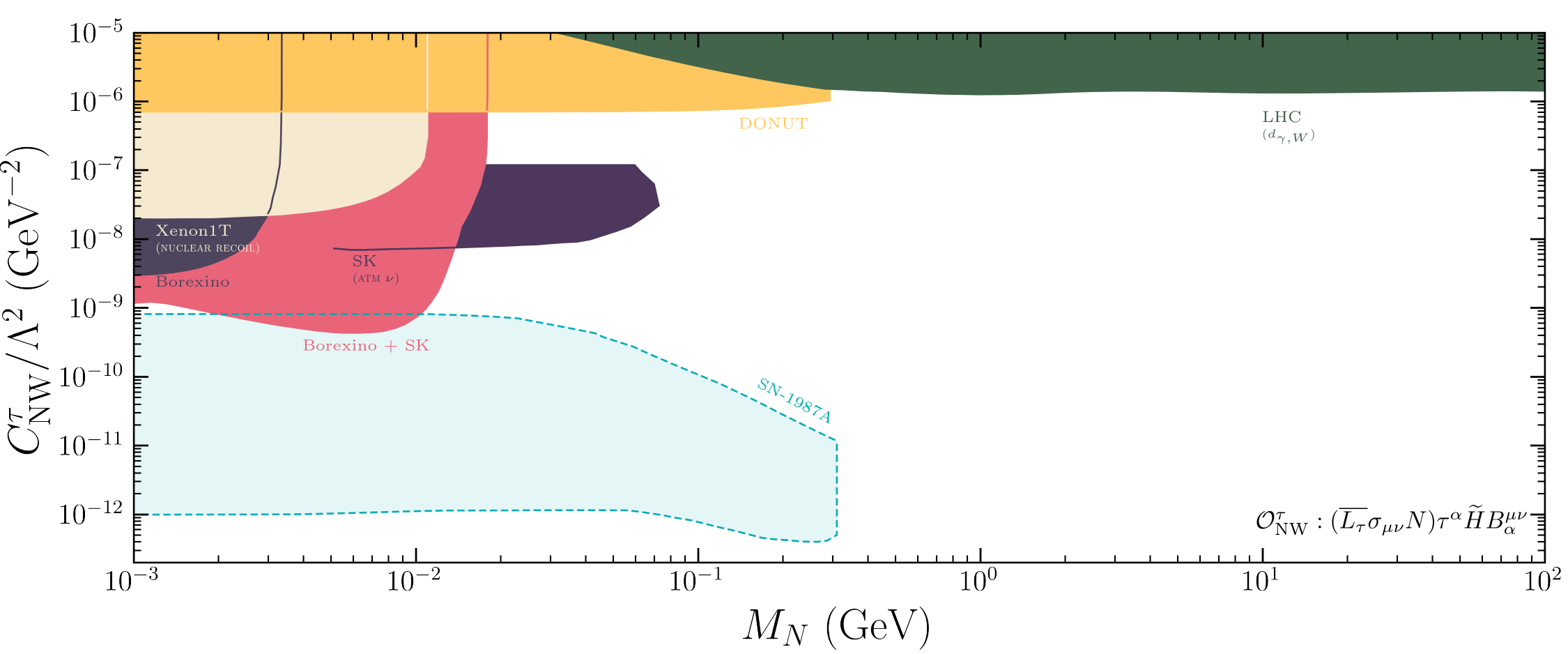}
\caption{The 90\% C.L. limits on the Wilson coefficient of the tensor current operator in \cref{eq:char_tensor}, as a function of the heavy neutrino mass. We display in separate panels the constraints relevant to each lepton flavour, detailing the most dominant sources of limits. See text for details. 
\label{fig:ONW}}
\end{figure}

Bounds on these dipoles can be translated into limits on the Wilson coefficients by applying the relations above. 
Most limits on the Wilson coefficients $C_{\rm NB}$ and $C_{\rm NW}$ are obtained from processes mediated by the magnetic dipole moment $d_\gamma$.
The weak dipoles $d_Z$ and $d_W$ are less important for low-energy processes but can contribute to HNL production at colliders, for example.
We recast the LHC and LEP limits obtained in Ref.~\cite{Magill:2018jla} to match our one-operator-at-a-time approach~\footnote{For the LHC limits on the $d_W^\alpha$ parameter, we take the $a=0.2$ case of Fig.~9 of Ref.~\cite{Magill:2018jla}.}.
We note that a clear flat direction is present for  $C_{\rm NB}/C_{\rm NW} = - \tan{\theta_{\rm W}}$ so that the dipole moment with the photon would vanish even with large Wilson coefficients.
In this case, none of the limits shown in \cref{fig:ONB,fig:ONW} apply, as they rely either on $N$ production or decay via $d_\gamma$.
We leave a study of this particular case to future literature.

Probes of magnetic dipole moments include searches for exotic electromagnetic interactions of light neutrinos~\cite{Brdar:2020quo,Coloma:2017ppo,Gninenko:1998nn,DONUT:2001zvi} or HNL production and decay~\cite{Coloma:2017ppo,Plestid:2020ssy}. 
The LEP monophoton searches and supernova cooling bounds, mentioned in the previous sections, also constrain the dipole moment. 
Finally, this quantity must also be small enough to respect primordial abundances after BBN~\cite{Brdar:2020quo}. 
See~\cite{Schwetz:2020xra} for a recent review on all these bounds. 
However, as we have already discussed in \cref{sec:HNL_bounds}, these relevant bounds ultimately depend on the cosmological model, and we regard them as complementary to the laboratory constraints studied here.

In the scenario in which the couplings lie along the flat direction, couplings to the $W$ and/or $Z$ must be considered instead. In principle, the operator $\mathcal{O}_{\rm NW}$ would also mediate CC interactions, such as the ones involved in the standard bounds on the mixing $U_{\alpha N}$, but with a different Lorentz structure. Similarly, $\mathcal{O}_{\rm NB}$ would induce couplings between HNLs and neutral mesons. Nevertheless, the couplings with mesons vanish due to their particular Lorentz structure. Thus, all bounds arising from HNL production or decay via meson interactions do not apply to these operators either. 

Finally, the operator $\mathcal{O}_{\rm NB}$ would mediate a new invisible $Z$ decay channel, into an HNL and a light neutrino. Imposing its rate to be consistent with current measurements on invisible $Z$ decay would yield yet another limit on $C_{\rm NB}/\Lambda^2$. However, we find this bound to be an order of magnitude worse than the one arising from monophoton searches, and do not include it in our plots.

\Cref{fig:ONB,fig:ONW} show the present bounds at 90\%~C.L. for $C_{\rm NB}/\Lambda^2$ and $C_{\rm NW}/\Lambda^2$, respectively. Both of them contain three panels, one for each lepton flavour, following the single flavour dominance assumption. Most bounds on the dipole moments depend on the flavour of the involved neutrino, so the different panels will, in general, contain limits provided by different experiments. Some sources of constraints are flavour independent, such as those provided by LEP or by supernova cooling arguments.

\section{Four-fermion interactions}
\label{sec:four-ferm}

Several different four-fermion operators involving HNLs can be written at $d=6$. We will separate their study into two main sections, depending on whether the operators mediate CC-like or NC-like processes.

Analyzing their potentially complex flavour structure leads to a large number of operators of each type, a priori independent. As in the previous sections, we will consider separately the operators that affect electrons, muons, and taus, treating the corresponding Wilson coefficients as independent parameters, to be bounded by different observables.

To derive the constraints on the effective operators involving quark fields, their flavour structure becomes very relevant. Indeed, one of the most advantageous ways of producing HNLs is to exploit decays with a relatively low rate in the SM to compete against, such as $K$ or $B$ meson decays, which must proceed through CKM-suppressed weak processes. Thus, these searches only apply if quark flavour violation is allowed in the effective operators. 

However, some operators contain both charged and neutral couplings and, if allowed to be flavour-violating, could lead to much more strongly constrained flavour-changing neutral processes, such as neutral meson oscillations at the loop level. 
In the spirit of deriving conservative limits on the Wilson coefficients, we will assume that the effective operators are flavour universal and diagonal in the SM flavour basis. Thus, the only source of flavour violation in the quark sector will be the SM CKM matrix, which will appear in all CC-like operators, in an analogous way to the SM CC interactions when going to the mass basis. Indeed, we will also assume that the corresponding rotation for the right-handed fields, which will be physical for some operators, also corresponds to that of the CKM matrix~\footnote{Alternatively, more elaborate prescriptions, such as Minimal Flavour Violation~\cite{Chivukula:1987py,DAmbrosio:2002vsn}, could be considered. However, the additional Yukawa suppression present in these scenarios would be too strong for some operators, so that the mass scales probed would be at odds with EFT validity. 
For the type of operators analyzed here, such a level of protection against quark flavour violation is unnecessary.}.

For comparison, for the flavour-changing CC-like operators, we will show our results both under the assumption of a CKM suppression and for the flavour-blind case, where all flavours couple with the same intensity (all flavour copies share a common Wilson coefficient). In this way, it will be apparent which parts of the bounds derive purely from the experimental results, and which rather stem from the flavour alignment prescription. However, notice that, in the flavour-blind case, stronger constraints from FCNC processes could generally apply; we only display this case for the sake of comparison.

\subsection{Neutral currents}
Three sets of four-fermion operators mediate NC-like processes. The first one reads
\begin{equation}
    \mathcal{O}_{\rm ff} =\frac{C_{\rm ff}}{\Lambda^2} (\overline{f} \gamma^\mu f) (\overline{N} \gamma_\mu N)\,,
    \label{eq:Off}
\end{equation}
where $f$ can stand for any right-handed fermion in the SM, either leptons or quarks. 
In the latter case, we will consider couplings to the up- and down-type quarks independently, through the coefficients $C_{\rm uu}$ and $C_{\rm dd}$.
For simplicity, we assume that this coefficient is generation independent.

Analogously, very similar operators can be written out of left-handed SM fields,
\begin{align}
    \label{eq:OLL}
    \mathcal{O}_{\rm LN}^\alpha &= \frac{C_{\rm LN}^\alpha}{\Lambda^2}(\overline{L_\alpha} \gamma^\mu L_\alpha) (\overline{N} \gamma_\mu N)\,,
    \\
    \label{eq:OQN}
    \mathcal{O}_{\rm QN} &= \frac{C_{\rm QN}}{\Lambda^2} \sum_i (\overline{Q_i} \gamma^\mu Q_i) (\overline{N} \gamma_\mu N)\,,
\end{align} 
where $Q_i$ denotes a quark doublet of flavour $i$.
In the case of leptons, we will once again treat the three flavour copies of $\mathcal{O}_{\rm LN}$ as operators with independent coefficients.
In the case of the quarks, we assume that the coefficients are generation independent.

Finally, it is also possible to construct an independent operator with a scalar structure:
\begin{equation}
        \mathcal{O}_{\rm LNL\ell}^{\alpha\beta} = \frac{C_{\rm LNL\ell}^{\alpha\beta}}{\Lambda^2} (\overline{L_\alpha} N)\epsilon (\overline{L_\alpha} \ell_\beta)\,.
        \label{eq:op_4lep}
\end{equation}
The fact that none of these operators mediates CC-like processes makes them difficult to probe. 
For instance, most bounds on HNL mixing cannot be reinterpreted for this purpose, as they rely on the production of HNLs in the decays of charged mesons.
Even the constraints from DELPHI, based on HNL production in $Z$ decays, cannot readily be used. 

Nevertheless, $\mathcal{O}_{\rm uu}$, ${O}_{\rm dd}$ and $\mathcal{O}_{\rm QN}$ contribute to invisible decay channels for neutral mesons, with the $\pi^0$ and the $\Upsilon$(1S) being the most stringent. 
These operators can also induce quark scatterings in which the only visible signal (aside from the HNLs) is a single jet, produced mainly by a gluon emitted by any of the quarks. 
Such monojet searches have been performed in the LHC. Ref.~\cite{Alcaide:2019pnf} recasted the experimental limits into bounds on the corresponding Wilson coefficients. 
As assumed in~\cite{Alcaide:2019pnf}, the HNL mass can be neglected in this process for the range of masses we explore, corresponding to a straight line in our plots.

We find no observables able to constrain the operators involving two muons or two taus. In the case of two electrons, LEP monophoton searches and supernova cooling arguments apply analogously to previous sections, as electrons are present in the possible production of HNLs. 
As discussed before, a rescaling is required to account for the different cross section each operator would lead to (for details, see \cref{app:supernova}). 
The operators $\mathcal{O}_{\rm ee}$ and $\mathcal{O}_{\rm LN}$ have different chiral structures; however, the bounds on their corresponding Wilson coefficients turn out to be the same. 
The same observables also constrain $C_{\rm LNL\ell}^{\rm \alpha \beta}$, provided $\alpha=\beta=e$. 
If $\alpha\neq\beta$, $\ell_\alpha\to\ell_\beta \nu N$ decays are mediated by this operator. 
We perform a $\chi^2$ fit of the corresponding rates to the measured $\mu$ and $\tau$ leptonic decays, obtaining bounds for $C_{\rm LNL\ell}^{\rm e\mu}$, $C_{\rm LNL\ell}^{\rm e\tau}$ and $C_{\rm LNL\ell}^{\rm \mu\tau}$.

\begin{figure}[t]
\includegraphics[width=\columnwidth]{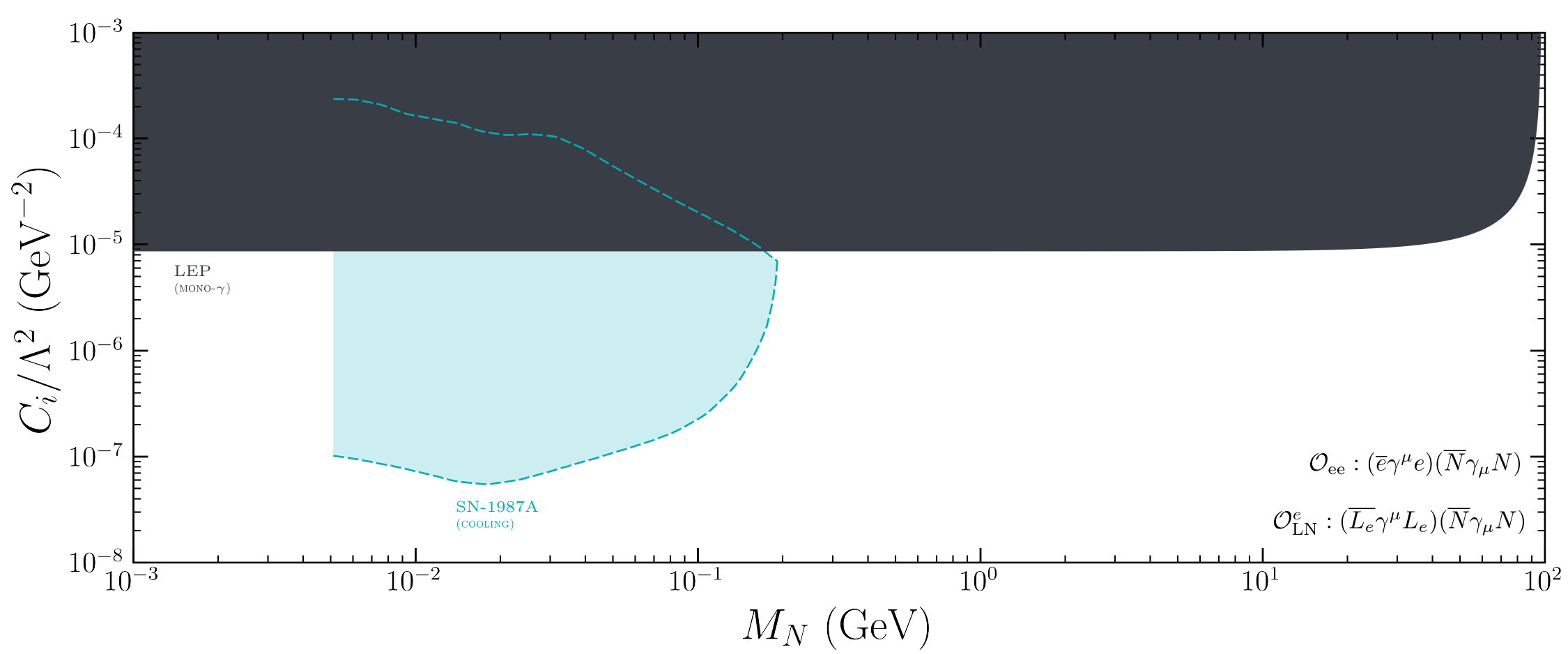}
\includegraphics[width=\columnwidth]{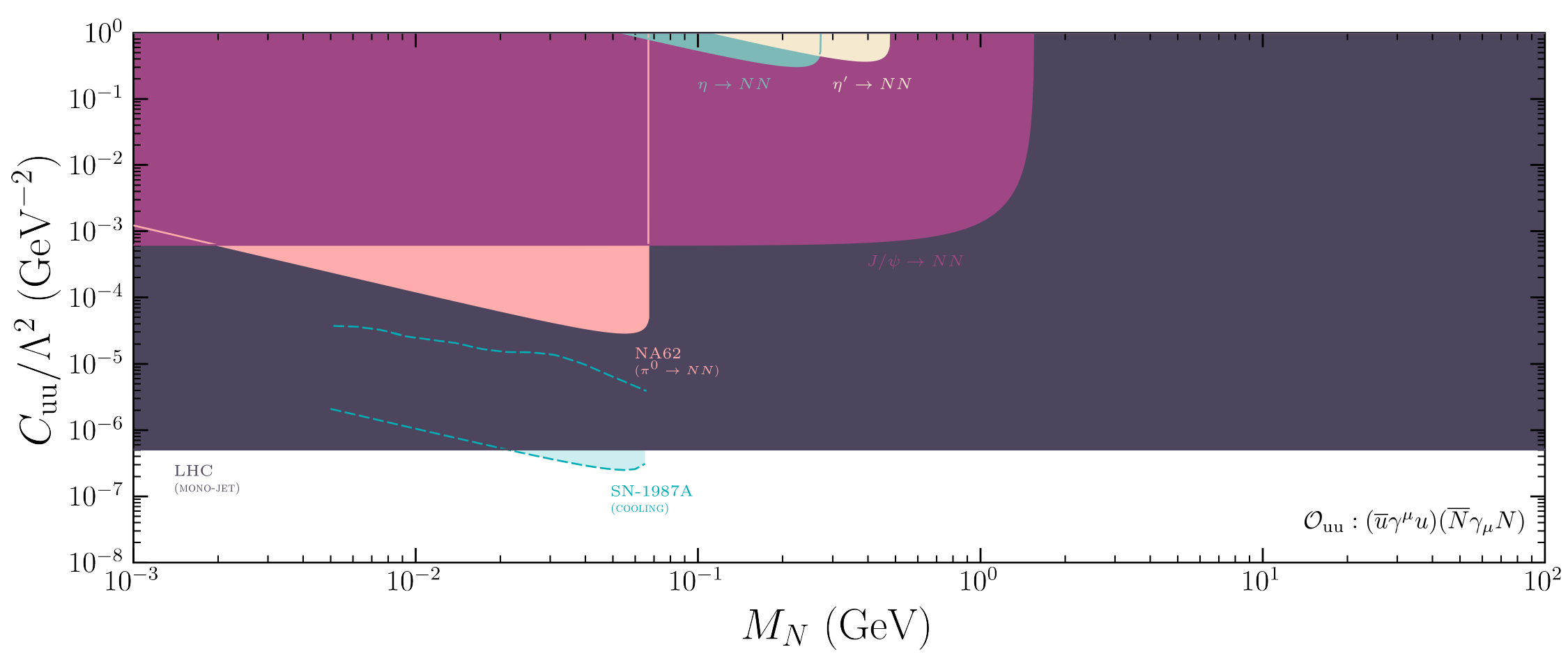}
\includegraphics[width=\columnwidth]{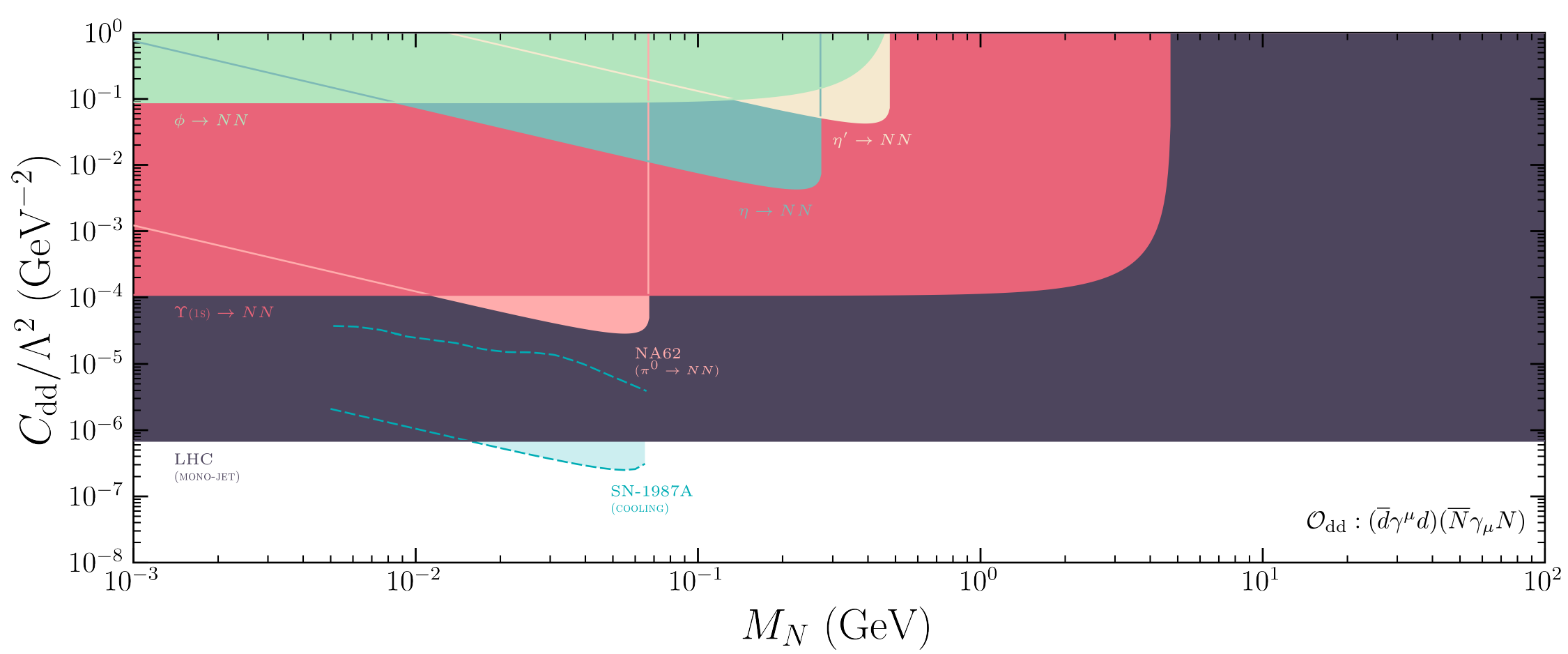}
\caption{The 90\% C.L. limits on some of the Wilson coefficients of 4-fermion NC operators $\mathcal{O}_{\rm ff}$ and $\mathcal{O}_{\rm LN}$, given by \cref{eq:Off,eq:OLL} respectively.
The limits in the top panel apply to both operators, $O_{\rm ee}$ and $O_{\rm LN}^e$.}
\label{fig:4ferm_neut_ff}
\end{figure}

\begin{figure}[t]
\includegraphics[width=\columnwidth]{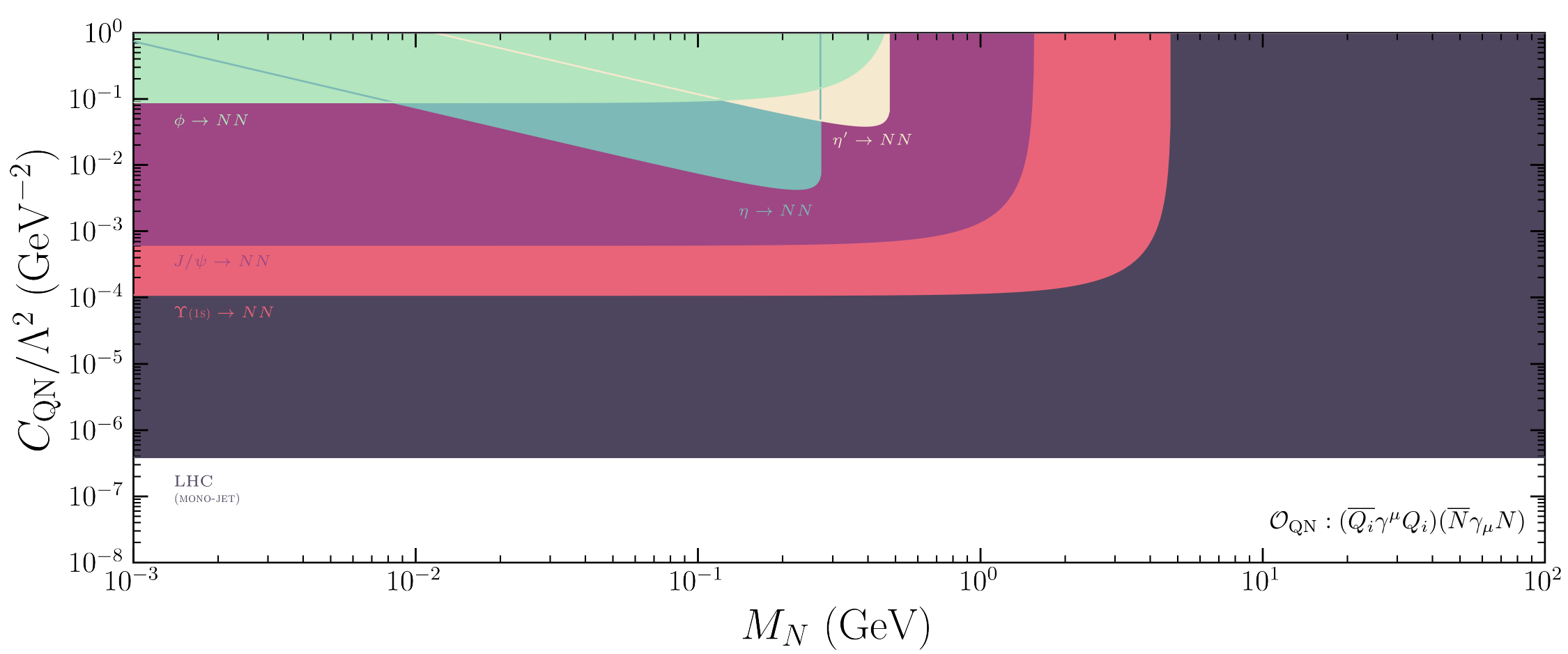}
\caption{The 90\% C.L. limits on the Wilson coefficient of the 4-fermion NC operator $\mathcal{O}_{\rm QN}$, namely \cref{eq:OQN}.}
\label{fig:4ferm_neut_left}
\end{figure}

\begin{figure}[t]
\includegraphics[width=\columnwidth]{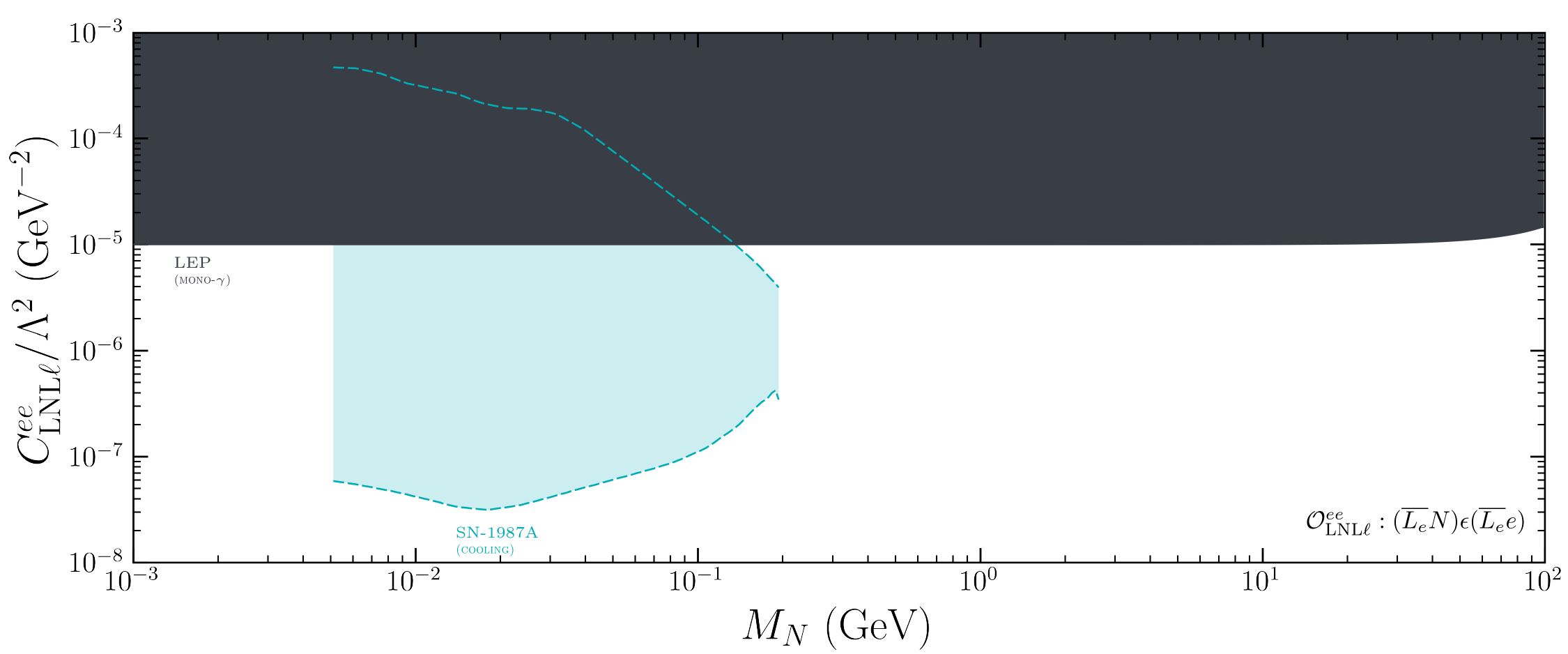}
\includegraphics[width=\columnwidth]{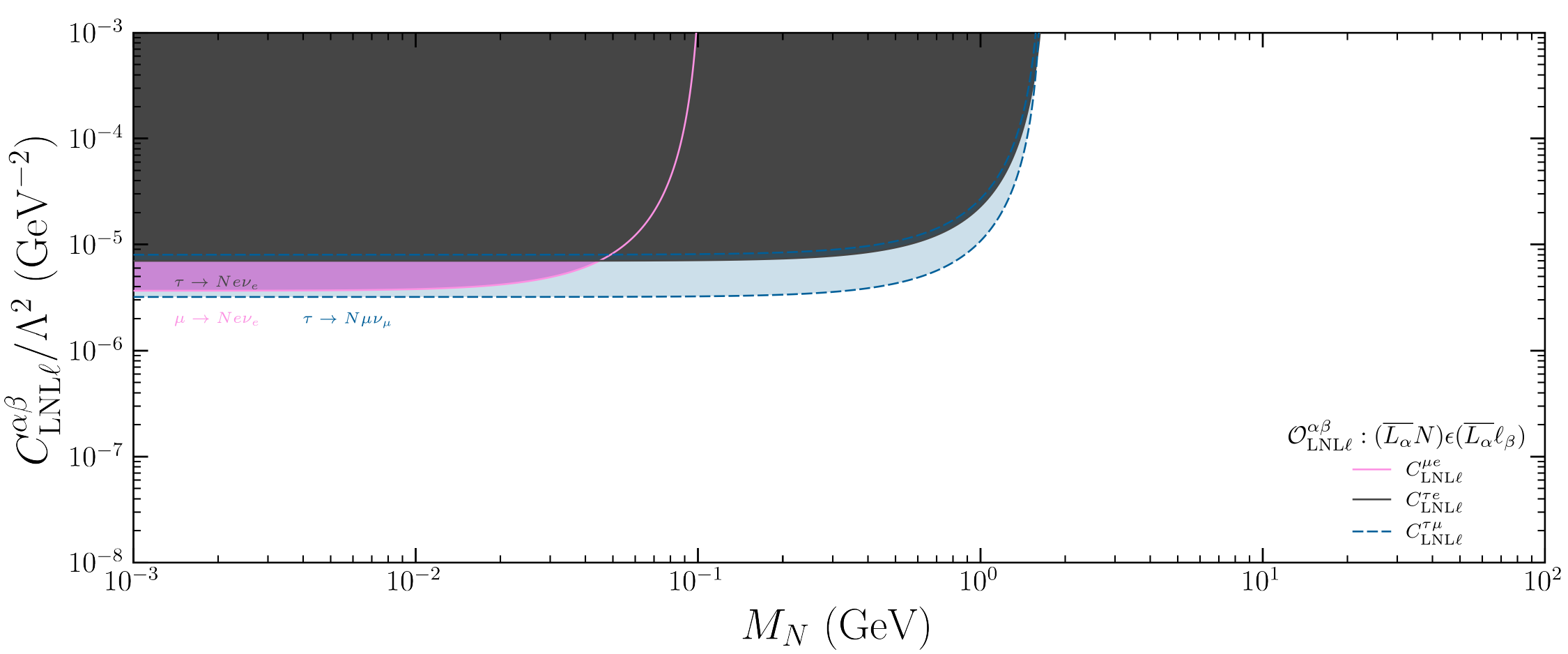}
\caption{The 90\% C.L. limits on some of the Wilson coefficients of the 4-fermion NC operator $\mathcal{O}_{\rm LNL\ell}$ in \cref{eq:op_4lep}.}
\label{fig:4ferm_neut_scalar}
\end{figure}

\Cref{fig:4ferm_neut_ff,fig:4ferm_neut_left,fig:4ferm_neut_scalar} show bounds, at 90\% C.L., on the mentioned Wilson coefficients. The main limits arise from invisible meson decay and monojet searches in the case of quarks, monophoton searches and supernova cooling for electrons and muon and tau decays for the other lepton flavours.. 

\begin{figure}[t]
\includegraphics[width=\columnwidth]{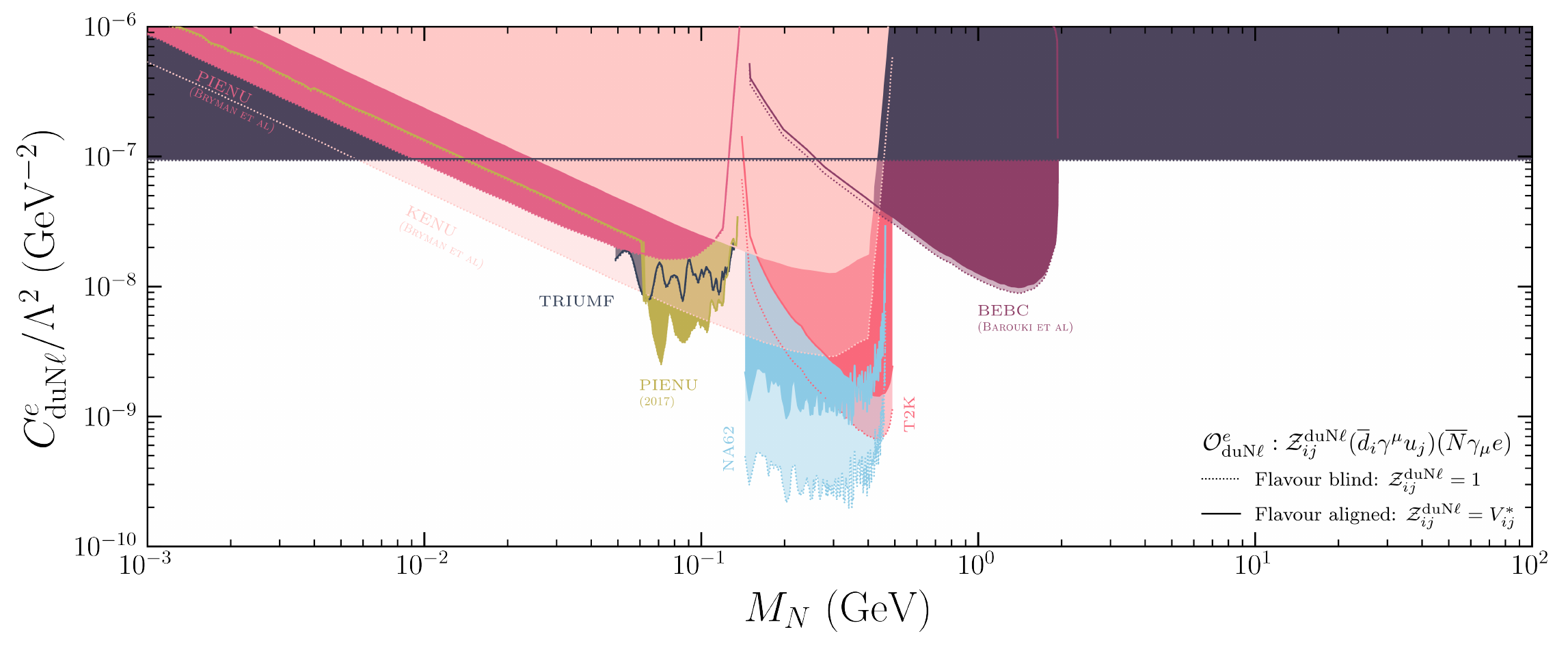}
\includegraphics[width=\columnwidth]{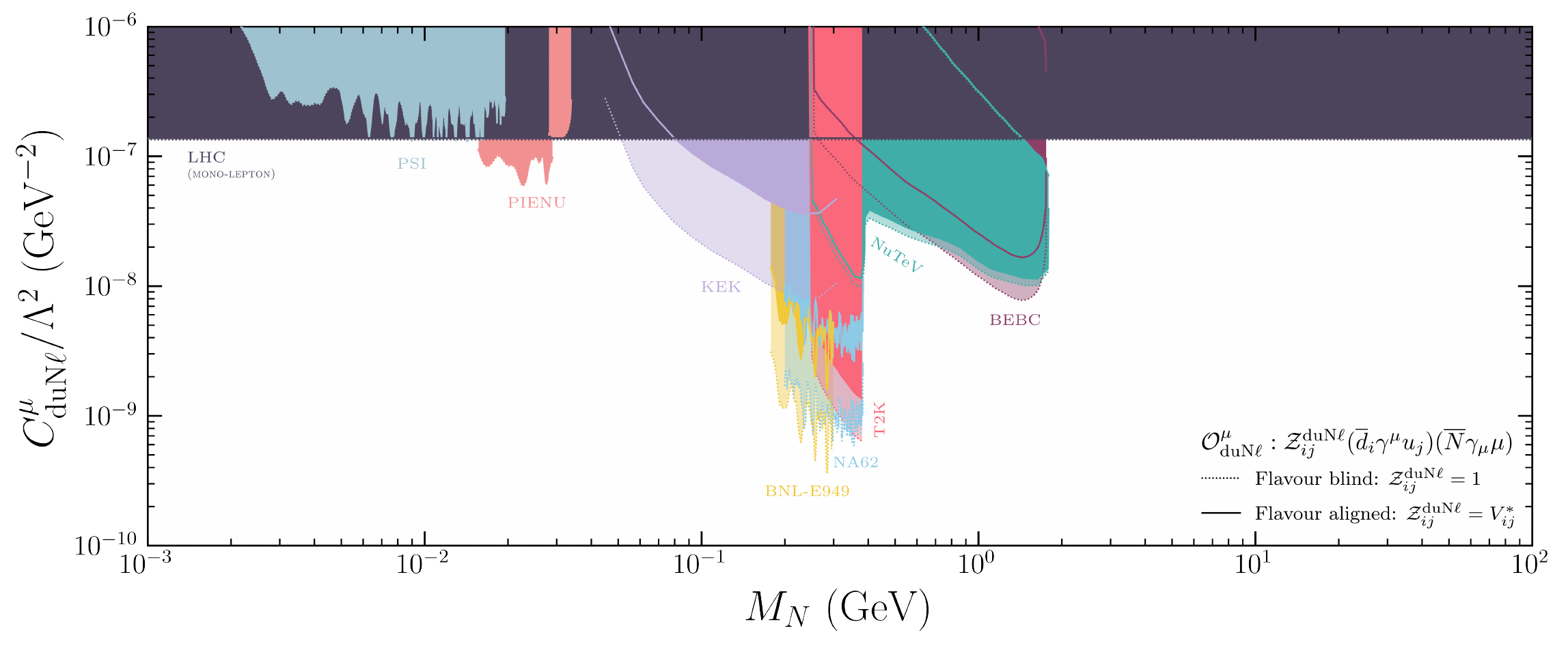}
\includegraphics[width=\columnwidth]{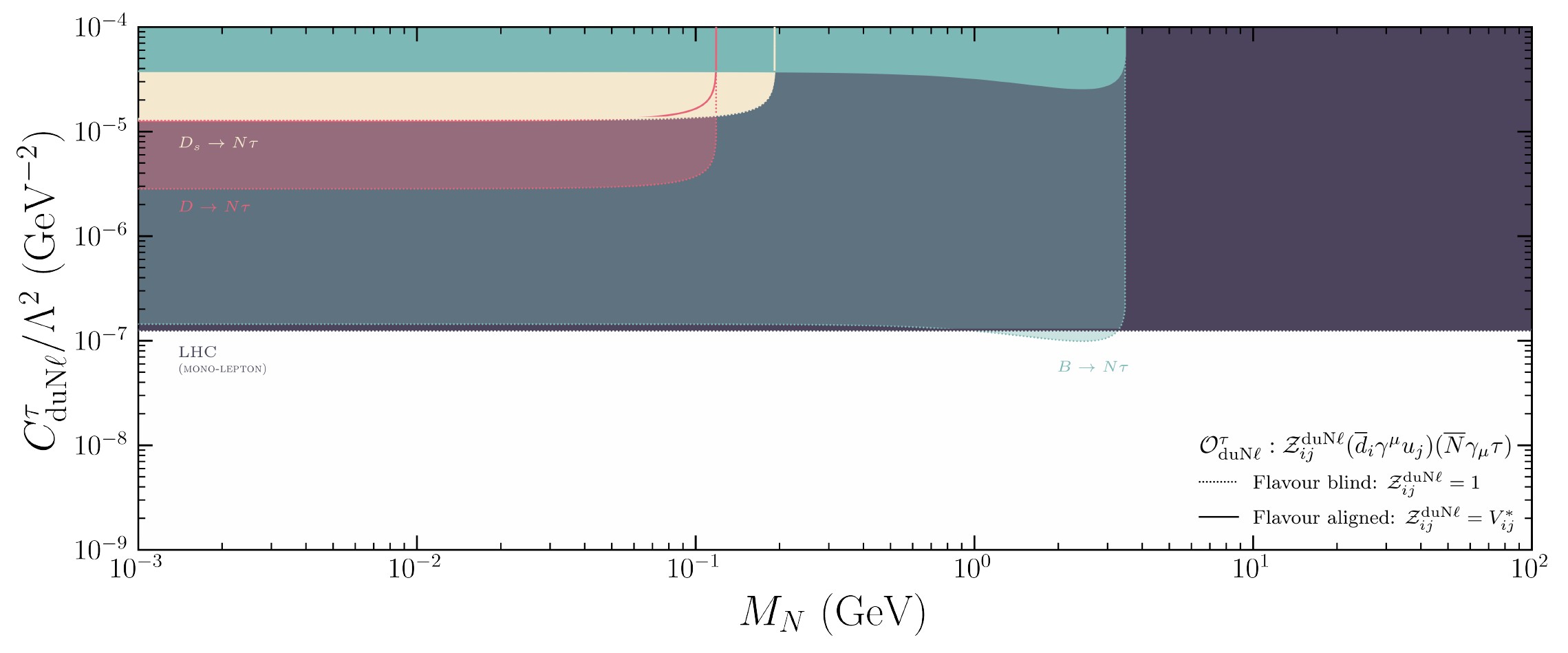}
\caption{The 90\% C.L. limits on the Wilson coefficient of the 4-fermion operator $\mathcal{O}_{\rm duN\ell}^\alpha$ in \cref{eq:CCFFvectorRR}, as a function of the HNL mass. Dotted light (solid dark) regions represent the flavour blind (aligned) scenario. See text for details.}
\label{fig:4ferm_dune}
\end{figure}

\begin{figure}[t]
\includegraphics[width=\columnwidth]{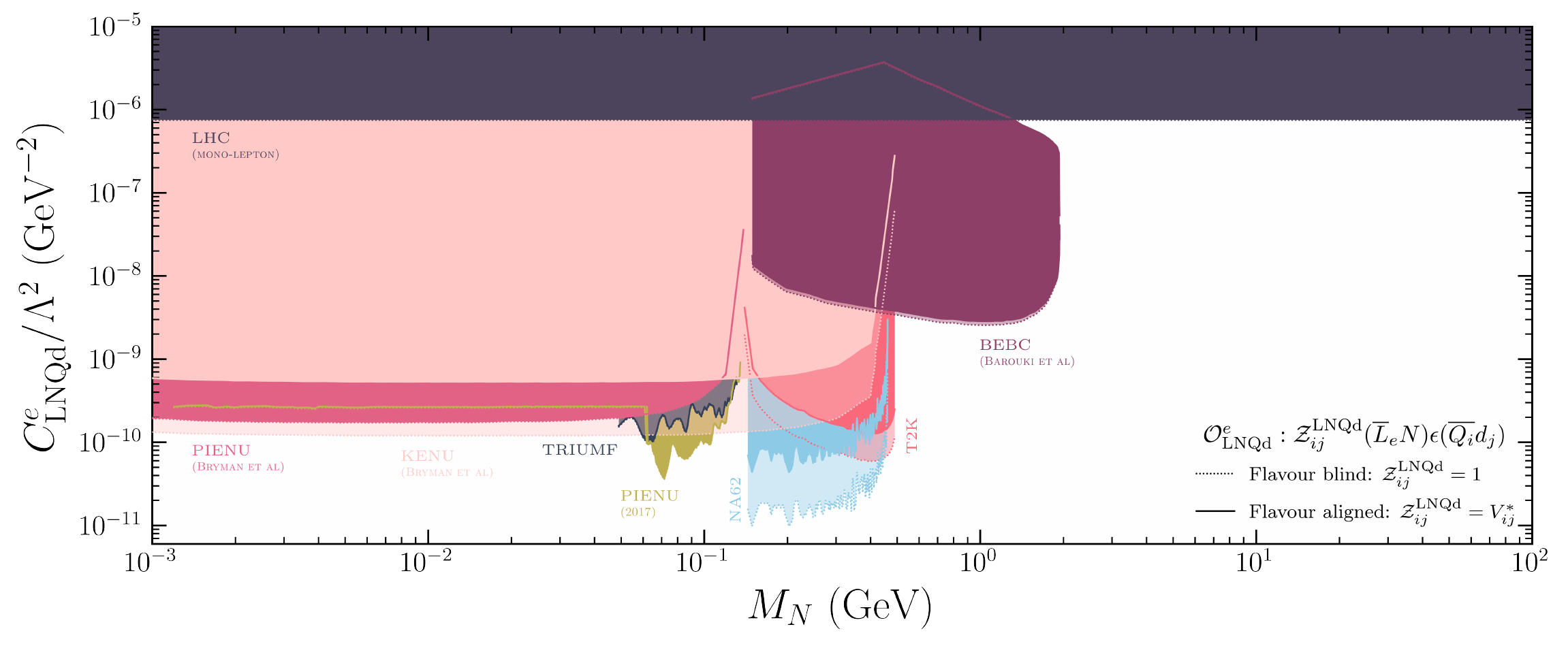}
\includegraphics[width=\columnwidth]{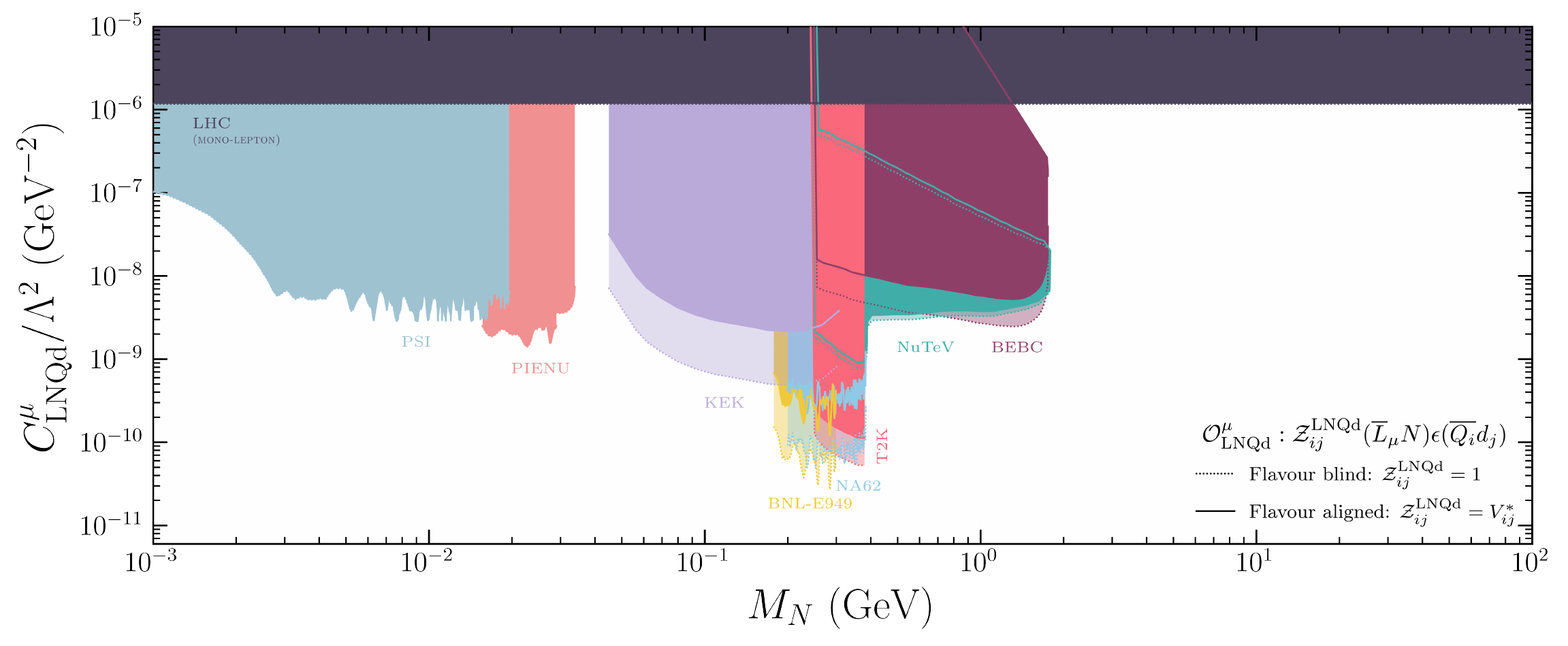}
\includegraphics[width=\columnwidth]{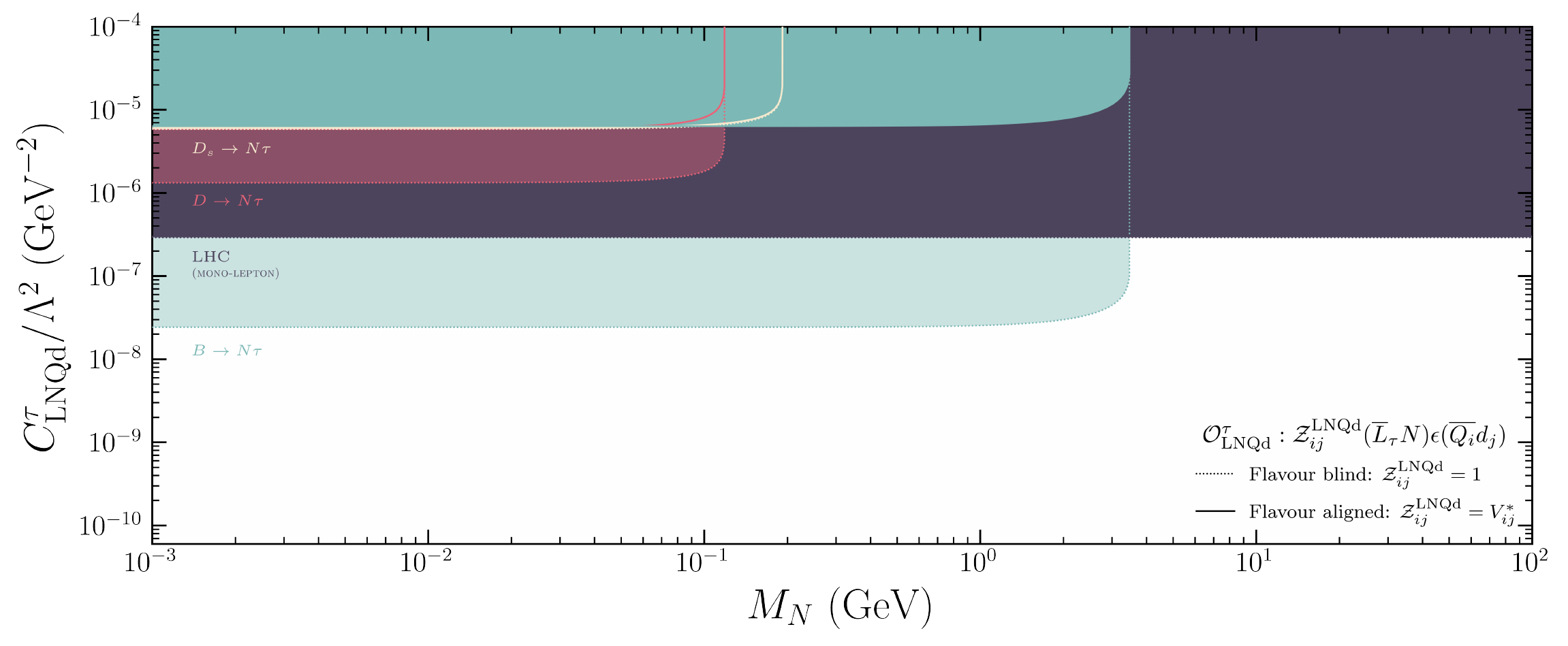}
\caption{The 90\% C.L. limits on the Wilson coefficient of the 4-fermion operator $\mathcal{O}_{\rm LNQd}$ in \cref{eq:CCFFscalarLR_1}, as a function of the HNL mass. Dotted light (solid dark) regions represent the flavour blind (aligned) scenario. See text for details.}
\label{fig:4ferm_lnqd}
\end{figure}

\begin{figure}[t]
\includegraphics[width=\columnwidth]{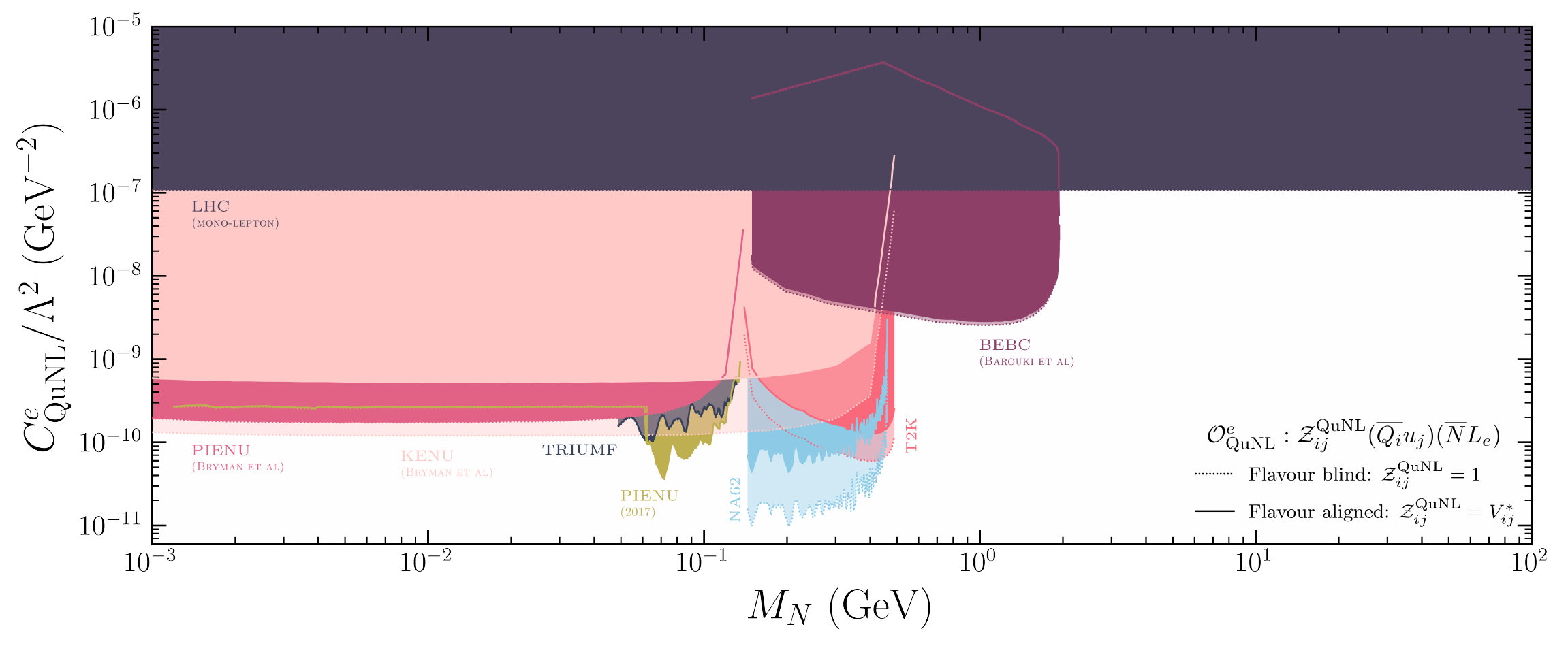}
\includegraphics[width=\columnwidth]{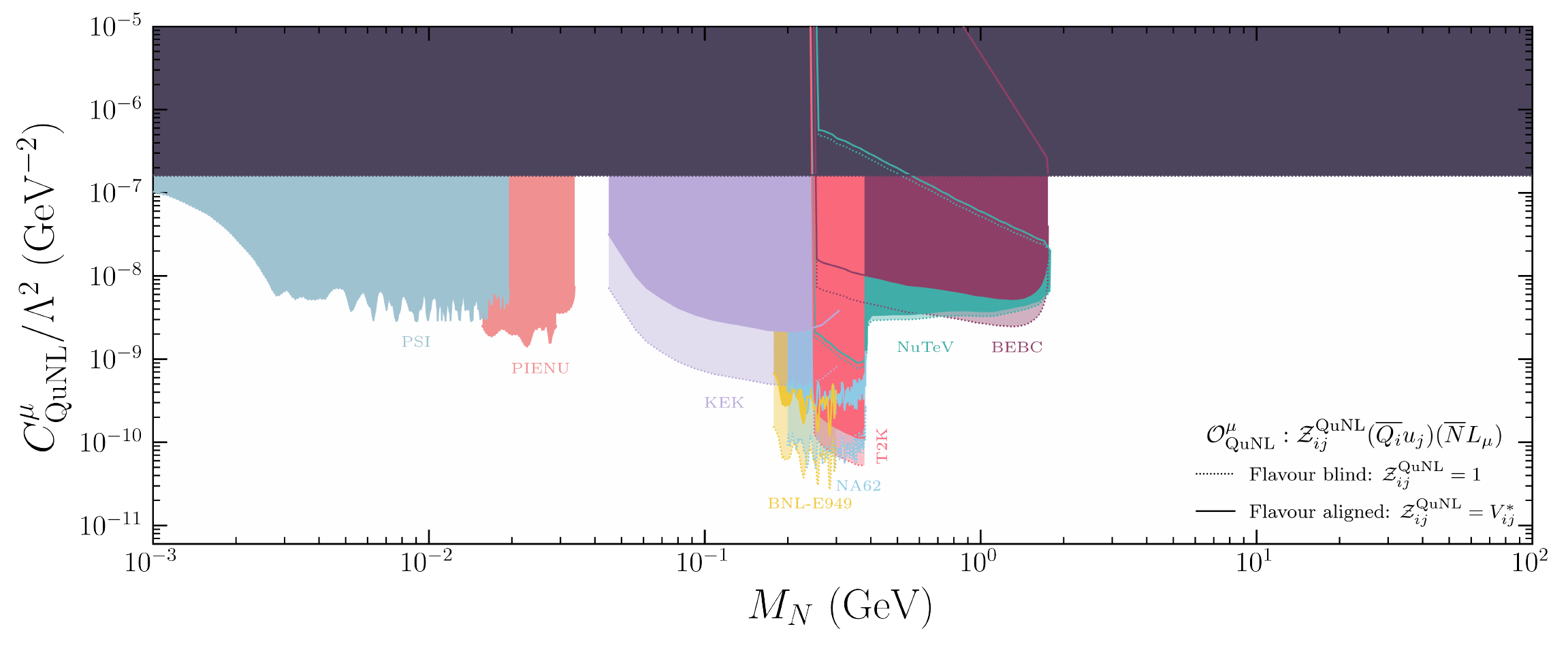}
\includegraphics[width=\columnwidth]{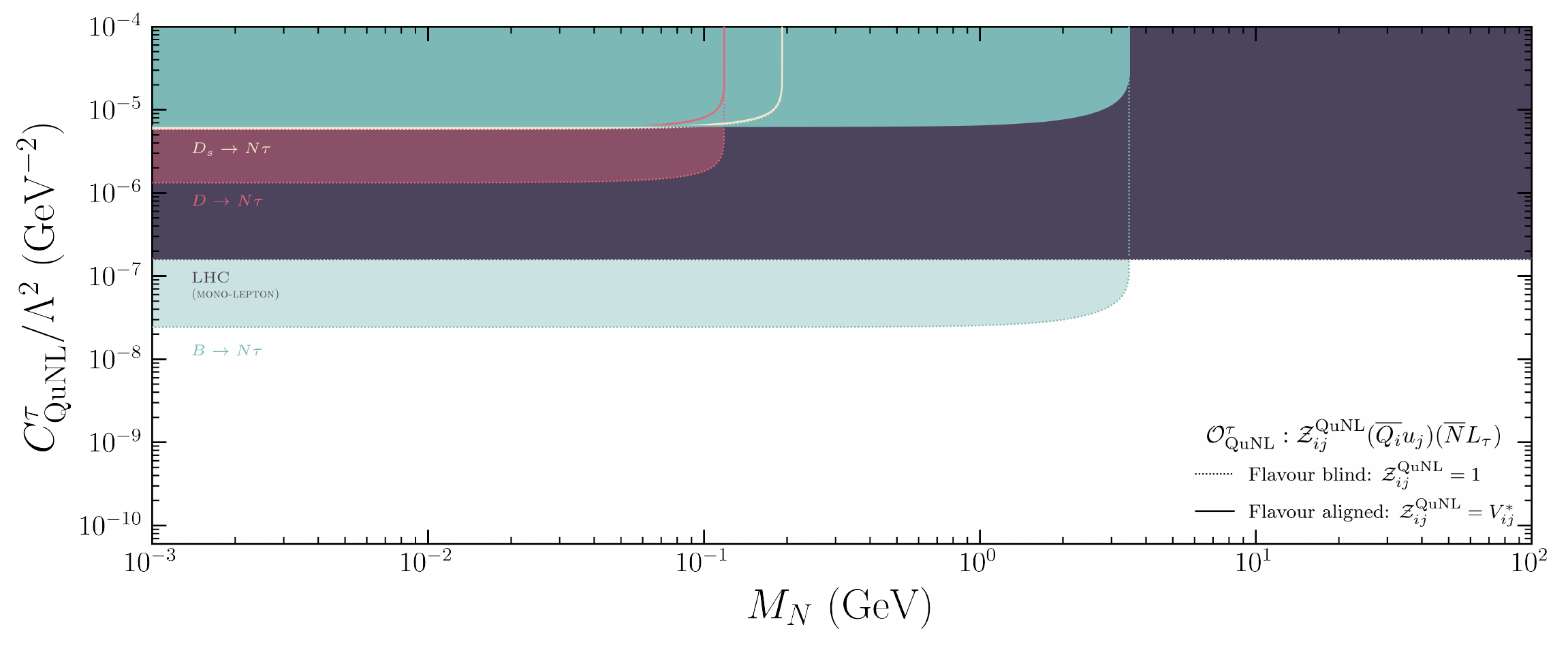}
\caption{The 90\% C.L. limits on the Wilson coefficient of the 4-fermion operators $\mathcal{O}_{\rm QuNL}$ in \cref{eq:opldqn}, as a function of the HNL mass. Dotted light (solid dark) regions represent the flavour blind(aligned) scenario. See text for details.}
\label{fig:4ferm_qunl}
\end{figure}

\subsection{Charged currents}

Three types of four-fermion operators appear at $d=6$, mediating CC-like interactions. 
The first one involves a right-handed charged lepton, a right-handed down-type quark of flavour $i$, and a right-handed up-type quark of flavour $j$, namely
\begin{equation}
    \mathcal{O}_{\rm duN\ell}^{\alpha} = \frac{C_{\rm duN\ell}^{\alpha}}{\Lambda^2}\sum_{i,j}\mathcal{Z}_{ij}^{\rm duN\ell}(\overline{d_i} \gamma^\mu u_j) (\overline{N} \gamma_\mu \ell_\alpha)\,,
        \label{eq:CCFFvectorRR}
\end{equation}
where $u_i$ and $d_i$ are the right-handed quark fields and $i$ is the generation index.
In the flavour-blind case, $\mathcal{Z}_{ij}^{\rm duN\ell}=1$, while, under the flavour alignment hypothesis, an insertion of the corresponding CKM matrix element $V_{ij}$ would be needed:
\begin{equation}
    \mathcal{Z}_{ij}^{\rm duN\ell}=V_{ij}^*\,.
\end{equation}
It is also possible to write a scalar coupling of the form 
\begin{equation}\label{eq:CCFFscalarLR_1}
    \mathcal{O}_{\rm LNQd}^\alpha = \frac{C_{\rm LNQd}^\alpha}{\Lambda^2}\sum_{ij}\mathcal{Z}^{\rm LNQd}_{ij} (\overline{L_\alpha} N)\epsilon (\overline{Q_i} d_j)\,.
\end{equation}
As before, under the flavour alignment hypothesis, we assume this operator is flavour universal and diagonal in the flavour basis, and, upon rotating to the mass basis, the CKM mixing matrix will control the degree of flavour violation.
\begin{equation}
\mathcal{Z}^{\rm LNQd}_{ij}=V_{ij}\,,
\end{equation}
while in the flavour-blind scenario $\mathcal{Z}^{\rm LNQd}_{ij}=1$. 
Note that exchanging the down-type quark and the HNL fields leads to a different operator,
\begin{equation}
        \mathcal{O}_{\rm LdQN}^\alpha = \frac{C_{\rm LdQN}^\alpha}{\Lambda^2}\sum_{ij}\mathcal{Z}^{\rm LdQN}_{ij} (\overline{L_\alpha} d_i)\epsilon (\overline{Q_j} N)\,,
\end{equation}
which shares the flavour coefficients of the previous operator. 
However, through a Fierz transformation, it is possible to rewrite this operator as 
\begin{equation}
\label{eq:opldqn}
       \mathcal{O}_{\rm LdQN}^\alpha = \frac{C_{\rm LdQN}^\alpha}{\Lambda^2}\sum_{ij}\mathcal{Z}^{\rm LdQN}_{ij}\left[ \frac{1}{2}(\overline{Q_j} d_i)\epsilon (\overline{L_\alpha} N) + \frac{1}{8} (\overline{Q_j}\sigma_{\mu\nu} d_i)\epsilon (\overline{L_\alpha}\sigma^{\mu\nu} N)\right]\,.
\end{equation}
so that the bounds on the previous operator will also apply to this one.

Finally, the last independent operator takes the form 
\begin{equation}\label{eq:CCFFscalarLR_2}
    \mathcal{O}_{\rm QuNL}^\alpha = \frac{C_{\rm QuNL}^\alpha}{\Lambda^2}\sum_{ij}\mathcal{Z}^{\rm QuNL}_{ij}(\overline{Q_i} u_j)(\overline{N} L_\alpha)\,,
\end{equation}
with
\begin{equation}
\mathcal{Z}^{\rm QuNL}_{ij}=V_{ij}^*\,,
\end{equation}
in the case of flavour alignment, and $\mathcal{Z}^{\rm QuNL}_{ij}=1$ in the flavour-blind scenario.

The set of operators in \cref{eq:CCFFvectorRR,eq:CCFFscalarLR_1,eq:CCFFscalarLR_2} mediate charged currents involving quarks and HNLs, producing interactions between charged mesons and HNLs. 
These processes are the main source of bounds on the standard HNL mixing, so the corresponding limits can be translated into constraints on the Wilson coefficients. 
The interactions mediated by the effective operators exhibit different Lorentz structures than the SM, so they yield HNL production and decay rates different from those mediated by standard mixing. 
Thus, as for the previous analyses, a rescaling of the existing bounds is necessary, to account both for the absence of neutral currents and for the different production and decay rates of the HNL. This procedure was recently advocated and applied to this type of operator in Ref.~\cite{Beltran:2023nli} for some example observables.

All these operators could also induce monolepton processes in colliders, in which the final state consists of a single observed lepton and missing energy (carried by the HNL). Searches for these signatures have been performed at the LHC and were recasted to constrain the Wilson coefficients of these operators in Ref.~\cite{Alcaide:2019pnf}. We directly employ their bounds, which are independent on the mass of the HNL (as the typical energies at the LHC are much larger than the range of masses under discussion). 

We will not discuss the phenomenology associated with $\mathcal{O}_{\rm QuNL}$, as it is identical to that of $\mathcal{O}_{\rm LNQd}$. This can be inferred from \cref{eq:opldqn}. 
The term with a tensor structure does not contribute to interactions with mesons. 
On the other hand, the term with a scalar Lorentz structure is identical to $\mathcal{O}_{\rm LNQd}$ (modulo a factor of 2), thus yielding no different phenomenology. The bounds on $C_{\rm LdQN}$ can be directly read from those on $\mathcal{O}_{\rm LNQd}$ by rescaling them by a factor of two. 

The operators $\mathcal{O}_{\rm LNQd}$, $\mathcal{O}_{\rm LdQN}$ and $\mathcal{O}_{\rm QuNL}$ also mediate NC-like processes involving an HNL, a light neutrino, and a $q\bar{q}$ pair. 
This would induce invisible decays of neutral mesons. Imposing the experimental limits on those processes could also constrain the corresponding Wilson coefficients. However, these bounds are always considerably looser than those arising from charged currents. As the limits would apply to the same Wilson coefficients, we will focus on those derived from CC-like processes.

\Cref{fig:4ferm_dune,fig:4ferm_lnqd,fig:4ferm_qunl} show the bounds on the Wilson coefficients of these operators, as a function of $M_N$. 
Each of these figures displays two different regions: in the ones defined by the dotted lines with lighter colours, flavour blindness is assumed ($\mathcal{Z}_{ij}= 1$), whereas, in the regions delimited by the solid lines with darker colours, we adopt the flavour alignment prescription.

\section{Conclusions}
\label{sec:conclus}

The addition of right-handed neutrinos, or HNLs, to the SM particle content, is arguably the simplest way to explain the evidence for neutrino masses and mixing. 
The mass scale of these new particles is thus a new dimensionful parameter of the model, to be determined empirically. 
If this mass scale is below EWSB, the HNLs should be considered, together with the other SM fields, as fundamental building blocks of any EFT aiming to probe new physics at a higher energy scale. 

In this work, we have systematically discussed the operators that would extend the basis of the successful SMEFT paradigm with the extra HNL building blocks.
This new EFT is referred to as $\nu$SMEFT.
For each new operator in the $\nu$SMEFT basis, we have discussed the laboratory and astrophysical bounds that would apply in absence of any other new physics contribution.
That is, we consider one operator at a time and neglect also any contribution from the mixing of the HNLs with the SM neutrinos. 
This is the first step towards a more thorough and complete exploration of the operator basis.
Furthermore, this procedure is generally conservative, as it prevents the possibility of different operators reinforcing each other's contributions or having different new physics at the production and detection of a given process.
Nevertheless, we have discussed a few instances where flat directions in the operator basis could avoid certain bounds. 

It should also be noted that these constraints are complementary to the ones that can be derived from the impact of the HNLs in the cosmological evolution of the Universe. 
Indeed, interactions between the HNLs and the SM fields would lead to their production in the early Universe. 
Depending on their lifetime, this could lead to too large contributions to the matter abundance, or their decay products could violate the present bounds on the radiation energy density present during CMB. 
They could also affect the CMB and BBN predictions. 
Since, unlike the laboratory constraints discussed in this work, these constraints rely on assumptions on the cosmological model, they should be explored together as a consistency check of those assumptions.

We find that the same experiments that constrain the mixing of HNLs with the SM neutrinos, such as peak searches and beam dump experiments, as well as collider searches at higher HNL masses, can be easily reinterpreted, as suggested in Ref.~\cite{Beltran:2023nli}, to provide the most stringent direct bounds on the Wilson coefficients of CC-like operators with only one HNL field. 
Conversely, operators with two HNL fields of NC type are significantly more difficult to constrain in absence of mixing with the SM neutrinos. 
We find that these operators can instead be bounded from constraints on invisible decay of the $Z$ boson or neutral mesons, supernova cooling limits, and monophoton and monojet searches at colliders. 

We have presented plots summarizing the most relevant bounds derived on each Wilson coefficient as a function of the HNL mass over a very wide range of values, spanning from the MeV to the electroweak scale,\footnote{For higher masses, it would be more appropriate also to integrate out the HNLs in the EFT approach, recovering the SMEFT paradigm.} and made them available on \gitlink. 
\Cref{tab:main_table} summarizes the operators and the corresponding limits.
These results provide an overview of the present constraints applying to each operator in the EFT basis, enlarged upon the addition of HNLs, for future searches and analyses to improve upon. 
They also provide a first step towards more ambitious and realistic analyses allowing the presence of several operators, as well as mixing between the HNLs and the SM neutrinos simultaneously.

\section*{Acknowledgments}
The authors warmly thank José Manuel Cano, Pilar Coloma, María José Herrero, Xabier Marcano, Luca Merlo, José Miguel No, and Tomás Rodríguez for very useful discussions. 
We are also grateful to Daniel Naredo-Tuero for providing the latest results on PMNS non-unitarity.
The authors would like to thank the Instituto de Física Teórica (IFT UAM-CSIC) in Madrid for support via the Centro de Excelencia Severo Ochoa Program under Grant CEX2020- 001007-S, during the Extended Workshop “Neutrino Theories”, where this work developed.
The research of MH was supported by Perimeter Institute for Theoretical Physics. 
Research at Perimeter Institute is supported by the Government of Canada through the Department of Innovation, Science and Economic Development and by the Province of Ontario through the Ministry of Research, Innovation and Science. JLP acknowledges support from Generalitat Valenciana through the plan GenT program (CIDEGENT/2018/019) and from the Spanish Ministerio de Ciencia e Innovacion through the project PID2020-113644GB-I00. EFM and MGL were also supported by the Spanish Research Agency (Agencia Estatal de Investigaci\'on) through the grant PID2019-108892RB-I00 funded by MCIN/AEI/ 10.13039/501100011033. EFM, MGL, and JLP also acknowledge support from European Union's Horizon 2020 research and innovation programme under the Marie Sklodowska-Curie grant agreements No 860881-HIDDeN and No 101086085-ASYMMETRY. JHG warmly thanks the hospitality of Albert de Roeck and the EP Neutrino group  during his stay at CERN; where this project has been completed. 

\appendix

\section{Rescaling procedure}
\label{app:recasting}

\subsection{Meson decays and decay-in-flight searches}
\label{app:rescaling}

Some effective operators considered here induce interactions akin to those generated by mixing between heavy and active neutrinos. 
While the standard mixing allows all the processes that could be possible in the SM, by simply replacing a light neutrino with a heavy one, the new operators may only generate a subset of those. 
We discuss how to reinterpret constraints on the mixing as constraints on the Wilson coefficients of the $d=6$ operators. A similar rescaling procedure was also advocated in Ref.~\cite{Beltran:2023nli}.

\paragraph{Charged bosonic currents.}
The operator in \cref{eq:OHNe} induces an effective CC interaction.
In analogy to the standard mixing, in \cref{eq:effectiveCCmixing} we defined the variable $U^{\rm CC}_{\alpha N} \equiv \frac{C_{\rm HN\ell}^\alpha v^2}{\sqrt{2}\Lambda^2}$,
which can be related to the standard mixing, $U_{\alpha N}$, in an experiment-dependent way.

The bounds set by peak search experiments apply equally to $U_{\alpha N}$ and $U^{\rm CC}_{\alpha N}$, as the HNLs are always produced via charged currents, and their decay is not observed. 
In this case, the rescaling is trivial.
Constraints obtained from HNL production in $Z$ boson decay, such as the ones by DELPHI, would not apply. 

For decay-in-flight searches, HNL production is identical to the standard mixing case, as it takes place primarily through charged meson decays.\footnote{In principle, the leptonic $\ell_\alpha \to \ell_\beta \nu_\beta N$ decays also produce HNLs via $|U_{\alpha N}|^2$. 
However, these channels are subdominant to the meson decays, $M\to \ell_\alpha N$, except in the mass region $m_{M} - m_{\ell_\alpha} < M_N < m_{\ell_\alpha} - m_{\ell_\beta}$.
Nevertheless, as we assume single-flavour dominance, the HNL CC decays are unobservable in this case, since the HNL cannot decay into its parent lepton, $N \, \slashed{\to} \, \ell_\alpha \pi$.}
The decays, however, are modified, and we take into account that NC channels are no longer available.
By equating the number of events, the limits on the effective mixing of \cref{eq:effectiveCCmixing}, can be obtained by,
\begin{equation}\label{eq:rescaling_beamdump}
     |U^{\rm CC}_{\alpha N}|^4 = |U_{\alpha N}|^4 \frac{ \sum_X\hat{\Gamma}^{\rm mixing}(N\to X)}{\sum_X \hat{\Gamma}^{\rm CC}(N\to X)}\,,
\end{equation}
where $\Gamma^{\rm mixing}(N \to X)$ is the decay width of the HNL into the signal final state $X$ (e.g., $X = \nu e^+e^-, e^+ \pi^-, \dots$) induced by the standard mixing scenario, and $\Gamma^{\rm CC}(N \to X)$ is the one induced by the charged bosonic current.
The hat indicates that the decay width is deprived of its $|U_{\alpha N}|^2$ mixing factor or EFT coefficient $C^2/\Lambda^4$, such that $\hat{\Gamma}^{\rm mixing} = \Gamma^{\rm mixing}/|U_{\alpha N}|^2$ and $\hat{\Gamma}^{\rm CC} = \Gamma^{\rm CC}/|U_{\alpha N}^{\rm CC}|^2$.
An analogous rescaling applies to collider constraints from ATLAS and CMS, where the HNL is produced in $W^\pm$ decays, and the decay channels vary depending on whether NC was considered in the original search. 
In both cases, the Wilson coefficient will then be obtained by applying consecutively this ratio and the rescaling in \cref{eq:rescaling_beamdump}.

Note that in the case of $\tau$ flavour dominance, decay-in-flight searches are insensitive to this operator, as the only possible decays of the HNL involve a $\tau$ lepton in the final state, which is either kinematically forbidden or not directly searched for.

\paragraph{Charged current four-fermion interactions.}
In a similar fashion to the charged bosonic current, the operators in \cref{eq:CCFFvectorRR,eq:CCFFscalarLR_1,eq:CCFFscalarLR_2} induce a subset of the interactions contained in the standard mixing case. 
As these operators involve quarks, the only processes they can mediate are mesons decay into neutrinos and vice versa. 
In that case, searches at CHARM, DELPHI, ATLAS, and CMS do not apply, as they tag HNL decays to fully leptonic final states.
Furthermore, the rates of HNL production and decay can be modified due to the different Lorentz structures of the operators. 

To illustrate the rescaling procedure, let us take the operator in \cref{eq:CCFFscalarLR_1}, namely $\mathcal{O}_{\rm LNQd}=\frac{C_{\rm LNQd}}{\Lambda^2}\mathcal{Z}^{\rm LNQd}_{ij}(\bar{L}N)\epsilon(\bar{Q_i}d_j)$.
Here, $\mathcal{Z}_{ij}^{\rm LNQd}$ stands for the flavour coefficient. 
If flavour alignment is assumed, $\mathcal{Z}_{ij}^{\rm LNQd}=V_{ij}$.
As usual, the necessary rescaling is achieved by equating the hypothetical number of events between the standard mixing case and the $\nu$SMEFT operator case. 
In the case of a peak search in meson decays, only the production decay rates are relevant. If the decay is, say, a pion into an electron and a neutrino, we can obtain a bound on the Wilson coefficient by means of
\begin{equation}
    \frac{|C_{\rm LNQd}^e|^2}{\Lambda^4}=\frac{|U_{e N}|^2}{\left|\mathcal{Z}_{du}^{\rm LNQd}\right|^2}\frac{\hat{\Gamma}^{\rm mixing}(\pi\to eN)}{\hat{\Gamma}^{\rm LNQd}(\pi\to e N)}\,,
\end{equation}
where the decay width in the numerator is the one induced by mixing and, in the denominator, the one induced by $\mathcal{O}_{\rm LNQd}$.
Here, $\hat{\Gamma}^{\rm LNQd} = \Gamma^{\rm LNQd} \times \Lambda^4/|C_{\rm LNQd}^e|^2$.
This relation can be easily generalized for other meson decays and operators.

In the case of decay-in-flight searches, the decay rates of the HNL need to be taken into account. 
For instance, let us consider the same operator as above, $\mathcal{O}_{\rm LNQd}$, in an electron flavour dominance scenario.
In a search for both $N\to \pi \ell$ and $N\to \ell \ell\nu$ decays in flight from HNLs produced by kaon decays at the target, only the former channel is sensitive to $C_{\rm LNQd}^e$.
The limit on the Wilson coefficient can then be extracted as
\begin{equation}
    \frac{|C_{\rm LNQd}^e|^4}{\Lambda^8}=\frac{|U_{e N}|^4}{\left|\mathcal{Z}^e_{su}\mathcal{Z}^e_{du}\right|^2}
    \frac{\hat{\Gamma}^{\rm mixing }(K\to e N)}{\hat{\Gamma}^{\rm LNQd}(K\to eN)}
    \frac{\hat{\Gamma}^{\rm mixing}(N\to\pi e)+\hat{\Gamma}^{\rm mixing}(N\to \ell\ell\nu)}{\hat{\Gamma}^{\rm LNQd}(N\to\pi e)}\,.
\end{equation}
The generalization to other lepton flavours and operators is straightforward and takes into account the different mixing assumptions and decay widths.

\subsection{Supernova cooling}
\label{app:supernova}

Supernova cooling arguments provide upper and lower bounds in the coupling of light particles. 
If the new states couple too feebly to the SM, their production is not enough to cool the supernova efficiently. 
On the other hand, if their interactions with the SM are stronger, they would be trapped inside the supernova, keeping the energy inside the system. 
In the intermediate regime between these two cases, the new particles can escape from the supernova, effectively cooling it. 

Ref.~\cite{DeRocco:2019jti} constrains the parameter space of a dark matter particle with a four-fermion vector coupling to the SM electromagnetic current,
\begin{equation}\label{eq:FFvectorDM}
    \mathcal{O}_\mathrm{vec}=\frac{C_\mathrm{vec}}{\Lambda^2}\bar{\chi}\gamma^\mu\chi J^\mathrm{em}_\mu\,.
\end{equation}
As HNLs play an identical role to DM fermions in the cooling of the supernova, their bounds can be readily translated as bounds on our operators, upon suitable rescaling,
\begin{equation}
    \frac{|C_i|^2}{\Lambda^4} = \frac{|C_j|^2}{\Lambda^4}\frac{\sigma_j}{\sigma_i}\,,
\end{equation}
where $C_i$ and $\sigma_i$ are, respectively, the Wilson coefficient of a particular operator $\mathcal{O}_i$ and the cross section of a process mediated by this operator. The rescaling factor is then obtained for $\mathcal{O}_i \in \{\mathcal{O}_{\rm HN},\mathcal{O}_{\rm HN\ell}, \mathcal{O}_{\rm ff},\mathcal{O}_{\rm LN},\mathcal{O}_{\rm LNL\ell}\}$, and $\mathcal{O}_j=\mathcal{O}_{\rm vec}$. 

To compute the lower bounds, we employ the cross section for the scattering of the new fermion off the SM particles present in the supernova, whereas to compute the upper bound we employ the cross section for the production of the fermion. The particular processes will differ depending on the considered operator. 

HNLs are mainly produced in $e^-e^+$ annihilations. Thus, $\mathcal{O}_{\rm QN}$ cannot mediate this channel, and the operators involving charged leptons can only do so for electron flavour dominance (their muon and tau flavour copies cannot be constrained by supernova cooling arguments). The upper bound on the Wilson coefficient of the corresponding operator is obtained by simply rescaling the HNL production cross section:
\begin{equation}
    \frac{|C_{i}|^2}{\Lambda^4}=\frac{|C_{\rm vec}|^2}{\Lambda^4} \frac{\overline{\sigma}_{\rm vec}(e^+e^-\to NN)}{\overline{\sigma}_{i}(e^+e^-\to NN)}\,,
\end{equation}
Note that an alternative efficient channel for HNL production is $\gamma\gamma\to\pi^0\to NN$. In fact, supernova cooling arguments provide a strong bound on the invisible $\pi^0$ decay. Imposing this limit yields upper limits for the Wilson coefficients of $\mathcal{O}_{\rm uu}$, $\mathcal{O}_{\rm dd}$ and $\mathcal{O}_{\rm HN}$. 
In the latter case, the mentioned constraints will compete with those given by $e^+e^-$ annihilation, whereas they are the only source of upper bounds for $\mathcal{O}_{\rm uu}$ and $\mathcal{O}_{\rm dd}$. 
Note that the quark content of the operator $\mathcal{O}_{\rm QN}$ is orthogonal to that of the $\pi^0$, so this operator is unable to mediate its decay into HNLs, and thus cannot be constrained by supernova cooling arguments.

The scattering processes that may affect the HNL will depend on the effective operator under consideration. The vectorial operator considered in Ref.~\cite{DeRocco:2019jti} mediates scatterings off electrons and protons; $\mathcal{O}_{\rm HN\ell}$, $\mathcal{O}_{\rm ee}$, $\mathcal{O}_{\rm LN}^{\rm e}$ and $\mathcal{O}_{\rm LNL\ell}^{\rm e}$ only induce $Ne\to Ne$ processes, while $\mathcal{O}_{\rm HN}$ yields HNL scatterings off electrons, protons and neutrons. The upper limits on the corresponding Wilson coefficient can be found through
\begin{equation}
    \frac{|C_{i}|^2}{\Lambda^4}=\frac{|C_\mathrm{vec}|^2}{\Lambda^4} \frac{\overline{\sigma}_\mathrm{vec}(e^-N\to e^-N)+\overline{\sigma}_{\rm vec}(p N \to p N)}{\sum_X\overline{\sigma}_{i}(NX\to NX)}\,,
\end{equation}
where the sum in $X$ comprises all the possible HNL scatterings mediated by the particular operator $\mathcal{O}_{i}$.

Of course, all these cross sections are energy dependent. The temperatures inside a supernova, and thus the energies at which these processes take place, vary considerably as a function of the distance with respect to the centre of the supernova. The abundances of electrons, positrons, protons and neutrons also exhibit radial dependences, which determine how likely their interactions are. In order to capture these effects, we compute averaged cross sections (denoted by $\overline{\sigma}$ in the equations above), integrated over the temperature profile of the supernova and convoluted with the abundance of the involved SM particles:
\begin{equation}
\overline{\sigma}=\frac{\int_0^{100 \text{ km}}n(r)T(r)\sigma(T(r))\dd r }{\int_0^{100\text{ km}} T(r)\dd r}\,,   
\end{equation}
where $T(r)$ is the temperature and $n(r)$ is the number density of a given particle at a distance $r$ from the centre of the supernova. $\sigma$ is the energy-dependent cross section of each process (mediated by the operator under study), in which each particle is assigned the energy corresponding to the temperature. We extract the supernova temperature and abundance profiles from Ref.~\cite{DeRocco:2019jti}. 

\subsection{Monophoton searches}
\label{app:monophoton}

At LEP, searches for $e^+e^-$ collisions with a single photon recoiling against invisible particles were used to derive limits on dark matter particles and HNLs. 
Ref.~\cite{Fox:2011fx} places limits on DM production through effective four-fermion operators using these searches. From a collider perspective, HNLs and DM behave identically, as missing energy, so this information also allows constraining some of our effective operators, which are able to mediate $e^+e^-\to NN\gamma$ processes. Such is the case of the operators $\mathcal{O}_{\rm ee}$, $\mathcal{O}_{\rm LN}^{\rm e}$, $\mathcal{O}_{\rm HN}$, $\mathcal{O}_{\rm HN\ell}^{\rm e}$ and $\mathcal{O}_{\rm LNL\ell}^{\rm ee}$. 

To recast their limits on our Wilson coefficients, we account for the different Lorentz structures of the operators. In particular, one of the operators considered in Ref.~\cite{Fox:2011fx} is a vectorial four-fermion coupling, similar to the one in \cref{eq:FFvectorDM}. The bounds on our operators are then related to those on the vectorial one by
\begin{equation}
    \frac{|C_{i}|^2}{\Lambda^4}=\frac{|C_\mathrm{vec}|^2}{\Lambda^4}\frac{\sigma_\mathrm{vec}}{\sigma_{i}}\,,
\end{equation}
where $i$ runs through the operators mentioned above. To compute this ratio, we calculate the $e^+e^- \to \gamma N N$ cross section, as described below.

The $2 \to 3$ phase space is parameterized by the energy of one of the HNLs, $E_N$, the energy of the photon, $E_\gamma$, the relative azimuthal angle between them, $\phi_{N\gamma}$, and the polar angle of the photon, $\theta_\gamma$. 
The total cross section is given by
\begin{equation}
    \sigma=\frac{1}{256\pi^4E_\mathrm{CM}^2}\int_0^{2\pi}d\phi_{\gamma N}\int_{-1}^1 d\cos{\theta_\gamma}\int_0^{E_\gamma^\mathrm{max}}dE_\gamma\int_{E_N^\mathrm{min}}^{E_N^\mathrm{max}}dE_N\vert\mathcal{M_\mathrm}\vert^2\,,
\end{equation}
where $\vert\mathcal{M}\vert^2$ is the corresponding matrix element and $E_\mathrm{CM} = 200$~GeV is the center-of-mass energy~\cite{Fox:2011fx}. 
The amplitude is determined by each particular operator and can be expressed in terms of the products of the four-momenta of the particles involved.
It depends on the relative polar angle between the photon and the neutrino, $\theta_{\gamma N}$, and the polar angle of the neutrino, $\theta_N$. 
The former is fixed once the energies of both particles are determined,  
\begin{equation}
    \cos{\theta_{\gamma N}}=\frac{E_\mathrm{CM}^2-2E_\mathrm{CM}(E_N+E_\gamma)+2E_NE_\gamma}{2E_\gamma\sqrt{E_N^2-M_N^2}}\,,
\end{equation}
while the latter is geometrically determined in terms of the angles in the CM,
\begin{equation}
    \cos{\theta_N}=\cos{\theta_\gamma}\cos{\theta_{\gamma N}}-\sin{\theta_\gamma}\sin{\theta_{\gamma N}}\cos{\phi_{\gamma N}}\,.
\end{equation}
The energy of the photon can vary from $0$ to $E_\gamma^\mathrm{max} = (E_\mathrm{CM}^2-4M_N^2)/2 E_\mathrm{CM}$, where the latter is obtained when the HNLs travel in the same direction, opposite to that of the photon.
Finally, the maximum (minimum) energy the HNL can carry depends on the energy of the photon and is obtained by solving $\cos{\theta_{\gamma N}}= -1$ $(\cos{\theta_{\gamma N}}=1)$, which gives
\begin{equation}
    E_N^\mathrm{max(min)}=\frac{1}{2}\left(E_\mathrm{CM}-E_\gamma\pm \frac{E_\gamma \sqrt{E_\mathrm{CM}^2-4M_N^2-2E_\gamma E_\mathrm{CM}}}{\sqrt{E_\mathrm{CM}^2-2E_\gamma E_\mathrm{CM}}}\right)\,.
\end{equation}
We integrate over the physical kinematical range considering the cuts employed in the analysis by LEP, requiring the photon energy to be at least a $6\%$ of that of the electron and its polar angle to be at least $45^\circ$ and no larger than $135^\circ$.

Note that the case of the operator $\mathcal{O}_{\rm LNL\ell}^{\rm ee}$ is slightly different, as it gives rise to a final state with a light neutrino and an HNL, instead of two HNLs. However, the recasting procedure is very similar, with some differences in the kinematics described above. The expressions can be easily generalized by assuming that one HNL is massless.

\section{Decay widths}
\label{app:decay_widths}

\renewcommand{\arraystretch}{1.5}
\begin{table}[t!]
\centering
\begin{tabular}{|c|c|c|c|}
\cline{2-4}
\multicolumn{1}{c|}{}& $\mathcal{O}_{\rm uu}$& $\mathcal{O}_{\rm dd}$& $\mathcal{O}_{\rm QN}$
\\
\hline
$\pi^0$&$\frac{f_\pi}{2\sqrt{2}}$&$\frac{-f_\pi}{2\sqrt{2}}$&0\\
\hline
$\eta$&$\frac{1}{2\sqrt{6}}\left(c_8f_8-\sqrt{2}s_0f_0\right)$&$\frac{-1}{2\sqrt{6}}\left(c_8f_8+2\sqrt{2}s_0f_0\right)$&$\frac{-\sqrt{3}}{2}s_0f_0$\\
\hline
$\eta^\prime$&$\frac{1}{2\sqrt{6}}\left(s_8f_8+\sqrt{2}c_0f_0\right)$&$\frac{1}{2\sqrt{6}}\left(-s_8f_8+2\sqrt{2}c_0f_0\right)$&$\frac{\sqrt{3}}{2}c_0f_0$\\
\hline
$\Phi$&0&$\frac{f_\Phi}{2}$&$\frac{f_\Phi}{2}$\\
\hline
$J/\psi$&$\frac{f_{J/\psi}}{2}$&0&$\frac{f_{J/\psi}}{2}$\\
\hline
$\Upsilon$(1S)&0&$\frac{f_\Upsilon}{2}$&$\frac{f_\Upsilon}{2}$\\
\hline
\end{tabular}

\caption{Couplings controlling the interactions between neutral mesons and HNLs mediated by NC-like four-fermion operators. $s_{0(8)}$ and $c_{0(8)}$ stand for the sine and cosine of the mixing angles $\theta_{0(8)}$.}
\label{tab:neut_mesons}
\end{table}

In this appendix, we provide explicit expressions for the relevant decay widths of SM particles decaying into HNLs and vice versa,
for each of the operators considered.

\paragraph{Higgs-dressed $d=5$ operator.} As discussed in \cref{sec:SMEFT}, this operator yields an invisible decay of the Higgs into two HNLs, given by the rate
\begin{equation}
    \Gamma_{\rm Higgs}^{d=5}=\frac{\vert C_{\rm Higgs}^{d=5}\vert ^2}{\Lambda^2}\frac{v^2 M_H}{8\pi}(1-2x_N^2)\sqrt{1-4x_N^2}\,,
\end{equation}
where $M_H$ is the Higgs mass, $v$ its vev and $x_N\equiv M_N/M_H$.

\paragraph{Higgs-dressed mixing.} This operator (see \cref{sec:higgs_mix}) opens an invisible decay channel for the Higgs boson into an HNL and a light neutrino, with a rate given by
\begin{equation}
    \Gamma_{\rm LNH}(H\to \nu N)=\frac{\vert C_{\rm LNH}\vert^2}{\Lambda^4}\frac{9v^4M_H}{64\pi}(1-x_N^2)^2\,,
\end{equation}
where we have neglected the light neutrino mass. Note that no flavour index has been specified; this decay affects all three flavour copies of the operator, so the squared Wilson coefficient in the equation above stands for the sum of all three squared coefficients: $|C_{\rm LNH}|^2=\sum_\alpha|C_{\rm LNH}^\alpha|^2$.

\paragraph{Neutral bosonic current.} A vertex between a $Z$ boson and two HNLs is introduced (\cref{sec:neut_bos_curr}), thus mediating invisible decays for the $Z$ and for neutral mesons, both pseudoscalar, $P^0$, and vector, $V^0$. The corresponding rates are
\begin{align}
 \Gamma_{\rm HN}(Z\to NN) &= \frac{\vert C_{\rm HN}\vert^2}{\Lambda^4}\frac{G_FM_Z^3v^4}{12\sqrt{2}\pi}\left(1-4x_N^2\right)^{3/2}\,,
    \\
    \Gamma_{\rm HN}(P^0\to NN) &= \frac{\vert C_{\rm HN}\vert^2}{\Lambda^4}\frac{G_F^2M_N^2M_{P}v^4f_{P}^2}{4\pi}\sqrt{1-4x_N^2}\,,
    \\
    \Gamma_{\rm HN}(V^0\to NN) &= \frac{\vert C_{\rm HN}\vert^2}{\Lambda^4}\frac{G_F^2M_{V}v^4f_{V}^2g_V^2}{24\pi}\sqrt{1-4x_N^2}\,.
\end{align} 
Here, $M_Z$ stands for the $Z$ mass, $G_F$ for Fermi's constant, and $x_N$ is the ratio of the HNL mass with respect to its parent particle mass. $M_{P}$ and $f_{P}$ are the neutral pseudoscalar meson mass and decay constant,\footnote{Note that the $\eta$ and $\eta^\prime$ mesons are not interaction eigenstates, so their \say{effective} decay constants are given in terms of those of the interaction eigenstates and the corresponding mixing angles: $f_\eta=f_8\cos{\theta_8}/\sqrt{3}+f_0\sin{\theta_0}/\sqrt{6}$, $f_{\eta^\prime}=f_8\sin{\theta_8}/\sqrt{3}-f_0\cos{\theta_0}/\sqrt{6}$~\cite{Coloma:2020lgy}.} and similarly for $M_{V}$ and $f_{V}$ for the neutral vector mesons. $g_V$ accounts for the coupling of the $Z$ boson to the quark content of the meson under consideration. In particular, for the $\Upsilon$(1S), $g_\Upsilon=\sqrt{2}\left(\frac{1}{2}-\frac{2}{3}s_W^2\right)$, where $s_W$ is the sine of the weak mixing angle (see Ref.~\cite{Coloma:2020lgy} for more details).

\paragraph{Charged bosonic current.}  This operator mediates the most relevant processes for HNL production, mainly in charged meson decays. 
Furthermore, there is a direct relation between the Wilson coefficient and the effective mixing they induce (\cref{eq:effectiveCCmixing}). 
Thus, the usual expressions for decay rates involving HNLs apply for this effective operator. We refer the reader to Ref.~\cite{Coloma:2020lgy} for a list of the relevant expressions. 

However, the absence of neutral currents affects some HNL decay channels, which, in the case of standard mixing, involve NC- and CC-like diagrams. This is the case for the $N\to\nu ee$ and $N\to\nu\mu\mu$ processes. The corresponding rate mediated by this operator is analogous to that of the decay $N\to\nu qq\prime$ in the case of standard mixing~\cite{Bondarenko:2018ptm}, and reads
\begin{equation}
    \Gamma_{\rm HN\ell}(N\to\nu\ell_\alpha^\pm\ell_\alpha^\mp)=\frac{\vert C_{\rm HN\ell}^\alpha\vert^2}{\Lambda^4}\frac{G_F^2v^4M_N^5}{384\pi^3}I(x_\alpha)\,,
\end{equation}
where
\begin{equation}
    I(x_\alpha)\equiv 12\int_{x_\alpha^2}^{(1-x_\alpha)^2}\frac{ds}{s}(s-x_\alpha^2)(1+x_\alpha^2-s)\sqrt{\lambda(s,0,x_\alpha^2)\lambda\left(1,s,x_\alpha^2\right)}\,,
\end{equation}
with $x_\alpha\equiv m_{\ell_\alpha}/M_N$ and $\lambda(x,y,z)$ is the K\"allen function. 

\paragraph{Neutral four-fermion interactions.} The operators $\mathcal{O}_{\rm uu}$, $\mathcal{O}_{\rm dd}$ (\cref{eq:Off}) and $\mathcal{O}_{\rm QN}$ (\cref{eq:OQN}) mediate invisible decays of neutral mesons into two HNLs. Although the chiralities of the quark fields are different, the expressions for the widths will be the same, since the vector contribution vanishes. They are given by
\begin{align}
    \Gamma_{\rm NC4F}(P^0\to NN) &= \frac{\vert C\vert^2}{\Lambda^4}\frac{g_P^2M_N^2M_P}{4\pi}\sqrt{1-4x_N^2}\,,
    \\
    \Gamma_{\rm NC4F}(V^0\to NN) &= \frac{\vert C\vert^2}{\Lambda^4}\frac{g_V^2M_V}{24\pi}\sqrt{1-4x_N^2}\,.
\end{align} 
The couplings $g_P$ and $g_V$ account for the overlap of the quark content of the involved meson and the quark structure of the operator under consideration. They are summarized in \cref{tab:neut_mesons} for the most relevant mesons. We refer the reader to Ref.~\cite{Coloma:2020lgy} for the values of most of the decay constants, as well as for the angles $\theta_0$ and $\theta_8$, that parametrize the mixing in the $\eta$ sector. For the heavier mesons we employ $f_{J/\psi}=1.8$ GeV$^2$~\cite{Hwang:1997ie} and $f_{\Upsilon}=9.2$ GeV$^2$~\cite{Merlo:2019anv}.

A four-lepton interaction is induced by the operator in \cref{eq:op_4lep}, inducing $\mu$ and $\tau$ decay into lighter leptons. The corresponding width is given by:
\begin{equation}
    \Gamma_{\rm LNL\ell}(\ell_\alpha\to\ell_\beta\nu N)=\frac{\vert C_{\rm LNL\ell}^{\alpha\beta}\vert^2}{\Lambda^4}\frac{1}{256\pi^3m_\alpha^3}\int_{(m_\beta + M_N)^2}^{m_\alpha^2}dm_{12}^2\int_{m_{\rm 23,min}^2}^{m_{\rm 23,max}^2}dm_{23}^2\,\vert\mathcal{M}\vert^2\,,
\end{equation}
where
\begin{equation}
    \vert\mathcal{M}\vert^2=\frac{1}{2}\left(m_\alpha^2+m_\beta^2-m_{12}^2-m_{23}^2\right)\left(m_{12}^2+m_{23}^2-M_N^2\right)\,,
\end{equation}
and 
\begin{align}
    m_{\rm 23,max(min)}^2 &= \frac{1}{2m_{12}^2}\left[m_\alpha^2\left(m_\beta^2+m_{12}^2-M_N^2\right)+m_{12}^2\left(m_\beta^2+M_N^2-m_{12}^2\right)\right.
    \nonumber
    \\
    &\left.\pm\left(m_\alpha^2-m_{12}^2\right)\sqrt{m_{12}^4-2m_{12}^2\left(m_\beta^2+M_N^2\right)+\left(m_\beta^2-M_N^2\right)^2}\right]\,.
\end{align}

\paragraph{Charged four-fermion interactions.} The operators in \cref{eq:CCFFvectorRR,eq:CCFFscalarLR_1,eq:CCFFscalarLR_2} induce charged meson decays into HNLs and vice versa (we only display here the pseudoscalar case, as it is the main HNL production source). 
The operator $\mathcal{O}_{\rm duN\ell}$  provides the same rates as in the standard mixing case (see Ref.~\cite{Coloma:2020lgy}) but controlled by the ``effective mixing''
\begin{equation}
    U_{\rm duN\ell} =\frac{1}{4G_FV_{ij}}\mathcal{Z}^{\rm duN\ell}_{ij}\frac{C_{\rm duN\ell}}{\Lambda^2}\,,
\end{equation}
where $i,j$ are the flavours of the quarks composing the meson, and $V_{ij}$ is the corresponding CKM element. 

The operators $\mathcal{O}_{\rm LNQd}$ and $\mathcal{O}_{\rm QuNL}$ exhibit the same decay widths for mesons into HNLs and vice versa:
\begin{align}
    \Gamma (N\to P^\pm\ell^\mp) &= \frac{\vert C\mathcal{Z}_{ij}\vert^2}{\Lambda^4}\frac{f_P^2m_P^4}{256M_N(m_i+m_j)^2}\left(M_N^2+m_\ell^2-m_P^2\right)\sqrt{\lambda\left(1,x_\ell^2,x_P^2\right)}\,,
    \\
     \Gamma (P^\pm\to N\ell^\pm) &= \frac{\vert C\mathcal{Z}_{ij}\vert^2}{\Lambda^4}\frac{f_P^2m_P^3}{128(m_i+m_j)^2}\left(m_P^2-m_\ell^2-M_N^2\right)\sqrt{\lambda\left(1,x_\ell^2,x_N^2\right)}\,,
\end{align}
where $m_i$ and $m_j$ are the masses of the quarks that compose the meson under consideration.

\bibliographystyle{apsrev4-1}

\begin{thebibliography}{143}%
\makeatletter
\providecommand \@ifxundefined [1]{%
 \@ifx{#1\undefined}
}%
\providecommand \@ifnum [1]{%
 \ifnum #1\expandafter \@firstoftwo
 \else \expandafter \@secondoftwo
 \fi
}%
\providecommand \@ifx [1]{%
 \ifx #1\expandafter \@firstoftwo
 \else \expandafter \@secondoftwo
 \fi
}%
\providecommand \natexlab [1]{#1}%
\providecommand \enquote  [1]{``#1''}%
\providecommand \bibnamefont  [1]{#1}%
\providecommand \bibfnamefont [1]{#1}%
\providecommand \citenamefont [1]{#1}%
\providecommand \href@noop [0]{\@secondoftwo}%
\providecommand \href [0]{\begingroup \@sanitize@url \@href}%
\providecommand \@href[1]{\@@startlink{#1}\@@href}%
\providecommand \@@href[1]{\endgroup#1\@@endlink}%
\providecommand \@sanitize@url [0]{\catcode `\\12\catcode `\$12\catcode
  `\&12\catcode `\#12\catcode `\^12\catcode `\_12\catcode `\%12\relax}%
\providecommand \@@startlink[1]{}%
\providecommand \@@endlink[0]{}%
\providecommand \url  [0]{\begingroup\@sanitize@url \@url }%
\providecommand \@url [1]{\endgroup\@href {#1}{\urlprefix }}%
\providecommand \urlprefix  [0]{URL }%
\providecommand \Eprint [0]{\href }%
\providecommand \doibase [0]{http://dx.doi.org/}%
\providecommand \selectlanguage [0]{\@gobble}%
\providecommand \bibinfo  [0]{\@secondoftwo}%
\providecommand \bibfield  [0]{\@secondoftwo}%
\providecommand \translation [1]{[#1]}%
\providecommand \BibitemOpen [0]{}%
\providecommand \bibitemStop [0]{}%
\providecommand \bibitemNoStop [0]{.\EOS\space}%
\providecommand \EOS [0]{\spacefactor3000\relax}%
\providecommand \BibitemShut  [1]{\csname bibitem#1\endcsname}%
\let\auto@bib@innerbib\@empty
\bibitem [{\citenamefont {Weinberg}(1979)}]{Weinberg:1979sa}%
  \BibitemOpen
  \bibfield  {author} {\bibinfo {author} {\bibfnamefont {S.}~\bibnamefont
  {Weinberg}},\ }\href {\doibase 10.1103/PhysRevLett.43.1566} {\bibfield
  {journal} {\bibinfo  {journal} {Phys. Rev. Lett.}\ }\textbf {\bibinfo
  {volume} {43}},\ \bibinfo {pages} {1566} (\bibinfo {year}
  {1979})}\BibitemShut {NoStop}%
\bibitem [{\citenamefont {Minkowski}(1977)}]{Minkowski:1977sc}%
  \BibitemOpen
  \bibfield  {author} {\bibinfo {author} {\bibfnamefont {P.}~\bibnamefont
  {Minkowski}},\ }\href {\doibase 10.1016/0370-2693(77)90435-X} {\bibfield
  {journal} {\bibinfo  {journal} {Phys. Lett. B}\ }\textbf {\bibinfo {volume}
  {67}},\ \bibinfo {pages} {421} (\bibinfo {year} {1977})}\BibitemShut
  {NoStop}%
\bibitem [{\citenamefont {Mohapatra}\ and\ \citenamefont
  {Senjanovic}(1980)}]{Mohapatra:1979ia}%
  \BibitemOpen
  \bibfield  {author} {\bibinfo {author} {\bibfnamefont {R.~N.}\ \bibnamefont
  {Mohapatra}}\ and\ \bibinfo {author} {\bibfnamefont {G.}~\bibnamefont
  {Senjanovic}},\ }\href {\doibase 10.1103/PhysRevLett.44.912} {\bibfield
  {journal} {\bibinfo  {journal} {Phys. Rev. Lett.}\ }\textbf {\bibinfo
  {volume} {44}},\ \bibinfo {pages} {912} (\bibinfo {year} {1980})}\BibitemShut
  {NoStop}%
\bibitem [{\citenamefont {Yanagida}(1979)}]{Yanagida:1979as}%
  \BibitemOpen
  \bibfield  {author} {\bibinfo {author} {\bibfnamefont {T.}~\bibnamefont
  {Yanagida}},\ }\href@noop {} {\bibfield  {journal} {\bibinfo  {journal}
  {Conf. Proc. C}\ }\textbf {\bibinfo {volume} {7902131}},\ \bibinfo {pages}
  {95} (\bibinfo {year} {1979})}\BibitemShut {NoStop}%
\bibitem [{\citenamefont {Gell-Mann}\ \emph {et~al.}(1979)\citenamefont
  {Gell-Mann}, \citenamefont {Ramond},\ and\ \citenamefont
  {Slansky}}]{Gell-Mann:1979vob}%
  \BibitemOpen
  \bibfield  {author} {\bibinfo {author} {\bibfnamefont {M.}~\bibnamefont
  {Gell-Mann}}, \bibinfo {author} {\bibfnamefont {P.}~\bibnamefont {Ramond}}, \
  and\ \bibinfo {author} {\bibfnamefont {R.}~\bibnamefont {Slansky}},\
  }\href@noop {} {\bibfield  {journal} {\bibinfo  {journal} {Conf. Proc. C}\
  }\textbf {\bibinfo {volume} {790927}},\ \bibinfo {pages} {315} (\bibinfo
  {year} {1979})},\ \Eprint {http://arxiv.org/abs/1306.4669} {arXiv:1306.4669
  [hep-th]} \BibitemShut {NoStop}%
\bibitem [{\citenamefont {Fukugita}\ and\ \citenamefont
  {Yanagida}(1986)}]{Fukugita:1986hr}%
  \BibitemOpen
  \bibfield  {author} {\bibinfo {author} {\bibfnamefont {M.}~\bibnamefont
  {Fukugita}}\ and\ \bibinfo {author} {\bibfnamefont {T.}~\bibnamefont
  {Yanagida}},\ }\href {\doibase 10.1016/0370-2693(86)91126-3} {\bibfield
  {journal} {\bibinfo  {journal} {Phys. Lett. B}\ }\textbf {\bibinfo {volume}
  {174}},\ \bibinfo {pages} {45} (\bibinfo {year} {1986})}\BibitemShut
  {NoStop}%
\bibitem [{\citenamefont {Vissani}(1998)}]{Vissani:1997ys}%
  \BibitemOpen
  \bibfield  {author} {\bibinfo {author} {\bibfnamefont {F.}~\bibnamefont
  {Vissani}},\ }\href {\doibase 10.1103/PhysRevD.57.7027} {\bibfield  {journal}
  {\bibinfo  {journal} {Phys. Rev. D}\ }\textbf {\bibinfo {volume} {57}},\
  \bibinfo {pages} {7027} (\bibinfo {year} {1998})},\ \Eprint
  {http://arxiv.org/abs/hep-ph/9709409} {arXiv:hep-ph/9709409} \BibitemShut
  {NoStop}%
\bibitem [{\citenamefont {Casas}\ \emph {et~al.}(2004)\citenamefont {Casas},
  \citenamefont {Espinosa},\ and\ \citenamefont {Hidalgo}}]{Casas:2004gh}%
  \BibitemOpen
  \bibfield  {author} {\bibinfo {author} {\bibfnamefont {J.}~\bibnamefont
  {Casas}}, \bibinfo {author} {\bibfnamefont {J.}~\bibnamefont {Espinosa}}, \
  and\ \bibinfo {author} {\bibfnamefont {I.}~\bibnamefont {Hidalgo}},\ }\href
  {\doibase 10.1088/1126-6708/2004/11/057} {\bibfield  {journal} {\bibinfo
  {journal} {JHEP}\ }\textbf {\bibinfo {volume} {11}},\ \bibinfo {pages} {057}
  (\bibinfo {year} {2004})},\ \Eprint {http://arxiv.org/abs/hep-ph/0410298}
  {arXiv:hep-ph/0410298} \BibitemShut {NoStop}%
\bibitem [{\citenamefont {Branco}\ \emph {et~al.}(1989)\citenamefont {Branco},
  \citenamefont {Grimus},\ and\ \citenamefont {Lavoura}}]{Branco:1988ex}%
  \BibitemOpen
  \bibfield  {author} {\bibinfo {author} {\bibfnamefont {G.~C.}\ \bibnamefont
  {Branco}}, \bibinfo {author} {\bibfnamefont {W.}~\bibnamefont {Grimus}}, \
  and\ \bibinfo {author} {\bibfnamefont {L.}~\bibnamefont {Lavoura}},\ }\href
  {\doibase 10.1016/0550-3213(89)90304-0} {\bibfield  {journal} {\bibinfo
  {journal} {Nucl. Phys.}\ }\textbf {\bibinfo {volume} {B312}},\ \bibinfo
  {pages} {492} (\bibinfo {year} {1989})}\BibitemShut {NoStop}%
\bibitem [{\citenamefont {Kersten}\ and\ \citenamefont
  {Smirnov}(2007)}]{Kersten:2007vk}%
  \BibitemOpen
  \bibfield  {author} {\bibinfo {author} {\bibfnamefont {J.}~\bibnamefont
  {Kersten}}\ and\ \bibinfo {author} {\bibfnamefont {A.~{\relax Yu}.}\
  \bibnamefont {Smirnov}},\ }\href {\doibase 10.1103/PhysRevD.76.073005}
  {\bibfield  {journal} {\bibinfo  {journal} {Phys. Rev.}\ }\textbf {\bibinfo
  {volume} {D76}},\ \bibinfo {pages} {073005} (\bibinfo {year} {2007})},\
  \Eprint {http://arxiv.org/abs/0705.3221} {arXiv:0705.3221 [hep-ph]}
  \BibitemShut {NoStop}%
\bibitem [{\citenamefont {Abada}\ \emph {et~al.}(2007)\citenamefont {Abada},
  \citenamefont {Biggio}, \citenamefont {Bonnet}, \citenamefont {Gavela},\ and\
  \citenamefont {Hambye}}]{Abada:2007ux}%
  \BibitemOpen
  \bibfield  {author} {\bibinfo {author} {\bibfnamefont {A.}~\bibnamefont
  {Abada}}, \bibinfo {author} {\bibfnamefont {C.}~\bibnamefont {Biggio}},
  \bibinfo {author} {\bibfnamefont {F.}~\bibnamefont {Bonnet}}, \bibinfo
  {author} {\bibfnamefont {M.~B.}\ \bibnamefont {Gavela}}, \ and\ \bibinfo
  {author} {\bibfnamefont {T.}~\bibnamefont {Hambye}},\ }\href {\doibase
  10.1088/1126-6708/2007/12/061} {\bibfield  {journal} {\bibinfo  {journal}
  {JHEP}\ }\textbf {\bibinfo {volume} {12}},\ \bibinfo {pages} {061} (\bibinfo
  {year} {2007})},\ \Eprint {http://arxiv.org/abs/0707.4058} {arXiv:0707.4058
  [hep-ph]} \BibitemShut {NoStop}%
\bibitem [{\citenamefont {Moffat}\ \emph {et~al.}(2017)\citenamefont {Moffat},
  \citenamefont {Pascoli},\ and\ \citenamefont {Weiland}}]{Moffat:2017feq}%
  \BibitemOpen
  \bibfield  {author} {\bibinfo {author} {\bibfnamefont {K.}~\bibnamefont
  {Moffat}}, \bibinfo {author} {\bibfnamefont {S.}~\bibnamefont {Pascoli}}, \
  and\ \bibinfo {author} {\bibfnamefont {C.}~\bibnamefont {Weiland}},\
  }\href@noop {} {\  (\bibinfo {year} {2017})},\ \Eprint
  {http://arxiv.org/abs/1712.07611} {arXiv:1712.07611 [hep-ph]} \BibitemShut
  {NoStop}%
\bibitem [{\citenamefont {Mohapatra}(1986)}]{Mohapatra:1986aw}%
  \BibitemOpen
  \bibfield  {author} {\bibinfo {author} {\bibfnamefont {R.~N.}\ \bibnamefont
  {Mohapatra}},\ }\href {\doibase 10.1103/PhysRevLett.56.561} {\bibfield
  {journal} {\bibinfo  {journal} {Phys. Rev. Lett.}\ }\textbf {\bibinfo
  {volume} {56}},\ \bibinfo {pages} {561} (\bibinfo {year} {1986})}\BibitemShut
  {NoStop}%
\bibitem [{\citenamefont {Mohapatra}\ and\ \citenamefont
  {Valle}(1986)}]{Mohapatra:1986bd}%
  \BibitemOpen
  \bibfield  {author} {\bibinfo {author} {\bibfnamefont {R.~N.}\ \bibnamefont
  {Mohapatra}}\ and\ \bibinfo {author} {\bibfnamefont {J.~W.~F.}\ \bibnamefont
  {Valle}},\ }\href {\doibase 10.1103/PhysRevD.34.1642} {\bibfield  {journal}
  {\bibinfo  {journal} {Phys. Rev.}\ }\textbf {\bibinfo {volume} {D34}},\
  \bibinfo {pages} {1642} (\bibinfo {year} {1986})}\BibitemShut {NoStop}%
\bibitem [{\citenamefont {Akhmedov}\ \emph {et~al.}(1996)\citenamefont
  {Akhmedov}, \citenamefont {Lindner}, \citenamefont {Schnapka},\ and\
  \citenamefont {Valle}}]{Akhmedov:1995ip}%
  \BibitemOpen
  \bibfield  {author} {\bibinfo {author} {\bibfnamefont {E.~K.}\ \bibnamefont
  {Akhmedov}}, \bibinfo {author} {\bibfnamefont {M.}~\bibnamefont {Lindner}},
  \bibinfo {author} {\bibfnamefont {E.}~\bibnamefont {Schnapka}}, \ and\
  \bibinfo {author} {\bibfnamefont {J.~W.~F.}\ \bibnamefont {Valle}},\ }\href
  {\doibase 10.1016/0370-2693(95)01504-3} {\bibfield  {journal} {\bibinfo
  {journal} {Phys. Lett. B}\ }\textbf {\bibinfo {volume} {368}},\ \bibinfo
  {pages} {270} (\bibinfo {year} {1996})},\ \Eprint
  {http://arxiv.org/abs/hep-ph/9507275} {arXiv:hep-ph/9507275} \BibitemShut
  {NoStop}%
\bibitem [{\citenamefont {Malinsky}\ \emph {et~al.}(2005)\citenamefont
  {Malinsky}, \citenamefont {Romao},\ and\ \citenamefont
  {Valle}}]{Malinsky:2005bi}%
  \BibitemOpen
  \bibfield  {author} {\bibinfo {author} {\bibfnamefont {M.}~\bibnamefont
  {Malinsky}}, \bibinfo {author} {\bibfnamefont {J.~C.}\ \bibnamefont {Romao}},
  \ and\ \bibinfo {author} {\bibfnamefont {J.~W.~F.}\ \bibnamefont {Valle}},\
  }\href {\doibase 10.1103/PhysRevLett.95.161801} {\bibfield  {journal}
  {\bibinfo  {journal} {Phys. Rev. Lett.}\ }\textbf {\bibinfo {volume} {95}},\
  \bibinfo {pages} {161801} (\bibinfo {year} {2005})},\ \Eprint
  {http://arxiv.org/abs/hep-ph/0506296} {arXiv:hep-ph/0506296 [hep-ph]}
  \BibitemShut {NoStop}%
\bibitem [{\citenamefont {Brivio}\ and\ \citenamefont
  {Trott}(2019)}]{Brivio:2017vri}%
  \BibitemOpen
  \bibfield  {author} {\bibinfo {author} {\bibfnamefont {I.}~\bibnamefont
  {Brivio}}\ and\ \bibinfo {author} {\bibfnamefont {M.}~\bibnamefont {Trott}},\
  }\href {\doibase 10.1016/j.physrep.2018.11.002} {\bibfield  {journal}
  {\bibinfo  {journal} {Phys. Rept.}\ }\textbf {\bibinfo {volume} {793}},\
  \bibinfo {pages} {1} (\bibinfo {year} {2019})},\ \Eprint
  {http://arxiv.org/abs/1706.08945} {arXiv:1706.08945 [hep-ph]} \BibitemShut
  {NoStop}%
\bibitem [{\citenamefont {Graesser}(2007)}]{Graesser:2007yj}%
  \BibitemOpen
  \bibfield  {author} {\bibinfo {author} {\bibfnamefont {M.~L.}\ \bibnamefont
  {Graesser}},\ }\href {\doibase 10.1103/PhysRevD.76.075006} {\bibfield
  {journal} {\bibinfo  {journal} {Phys. Rev. D}\ }\textbf {\bibinfo {volume}
  {76}},\ \bibinfo {pages} {075006} (\bibinfo {year} {2007})},\ \Eprint
  {http://arxiv.org/abs/0704.0438} {arXiv:0704.0438 [hep-ph]} \BibitemShut
  {NoStop}%
\bibitem [{\citenamefont {Graesser}()}]{Graesser:2007pc}%
  \BibitemOpen
  \bibfield  {author} {\bibinfo {author} {\bibfnamefont {M.~L.}\ \bibnamefont
  {Graesser}},\ }\href@noop {} {\ }\Eprint {http://arxiv.org/abs/0705.2190}
  {arXiv:0705.2190 [hep-ph]} \BibitemShut {NoStop}%
\bibitem [{\citenamefont {del Aguila}\ \emph {et~al.}(2009)\citenamefont {del
  Aguila}, \citenamefont {Bar-Shalom}, \citenamefont {Soni},\ and\
  \citenamefont {Wudka}}]{delAguila:2008ir}%
  \BibitemOpen
  \bibfield  {author} {\bibinfo {author} {\bibfnamefont {F.}~\bibnamefont {del
  Aguila}}, \bibinfo {author} {\bibfnamefont {S.}~\bibnamefont {Bar-Shalom}},
  \bibinfo {author} {\bibfnamefont {A.}~\bibnamefont {Soni}}, \ and\ \bibinfo
  {author} {\bibfnamefont {J.}~\bibnamefont {Wudka}},\ }\href {\doibase
  10.1016/j.physletb.2008.11.031} {\bibfield  {journal} {\bibinfo  {journal}
  {Phys. Lett. B}\ }\textbf {\bibinfo {volume} {670}},\ \bibinfo {pages} {399}
  (\bibinfo {year} {2009})},\ \Eprint {http://arxiv.org/abs/0806.0876}
  {arXiv:0806.0876 [hep-ph]} \BibitemShut {NoStop}%
\bibitem [{\citenamefont {Aparici}\ \emph {et~al.}(2009)\citenamefont
  {Aparici}, \citenamefont {Kim}, \citenamefont {Santamaria},\ and\
  \citenamefont {Wudka}}]{Aparici:2009fh}%
  \BibitemOpen
  \bibfield  {author} {\bibinfo {author} {\bibfnamefont {A.}~\bibnamefont
  {Aparici}}, \bibinfo {author} {\bibfnamefont {K.}~\bibnamefont {Kim}},
  \bibinfo {author} {\bibfnamefont {A.}~\bibnamefont {Santamaria}}, \ and\
  \bibinfo {author} {\bibfnamefont {J.}~\bibnamefont {Wudka}},\ }\href
  {\doibase 10.1103/PhysRevD.80.013010} {\bibfield  {journal} {\bibinfo
  {journal} {Phys. Rev. D}\ }\textbf {\bibinfo {volume} {80}},\ \bibinfo
  {pages} {013010} (\bibinfo {year} {2009})},\ \Eprint
  {http://arxiv.org/abs/0904.3244} {arXiv:0904.3244 [hep-ph]} \BibitemShut
  {NoStop}%
\bibitem [{\citenamefont {Peressutti}\ and\ \citenamefont
  {Sampayo}(2014)}]{Peressutti:2014lka}%
  \BibitemOpen
  \bibfield  {author} {\bibinfo {author} {\bibfnamefont {J.}~\bibnamefont
  {Peressutti}}\ and\ \bibinfo {author} {\bibfnamefont {O.~A.}\ \bibnamefont
  {Sampayo}},\ }\href {\doibase 10.1103/PhysRevD.90.013003} {\bibfield
  {journal} {\bibinfo  {journal} {Phys. Rev. D}\ }\textbf {\bibinfo {volume}
  {90}},\ \bibinfo {pages} {013003} (\bibinfo {year} {2014})}\BibitemShut
  {NoStop}%
\bibitem [{\citenamefont {Duarte}\ \emph
  {et~al.}(2015{\natexlab{a}})\citenamefont {Duarte}, \citenamefont
  {Gonz\'alez-Sprinberg},\ and\ \citenamefont {Sampayo}}]{Duarte:2014zea}%
  \BibitemOpen
  \bibfield  {author} {\bibinfo {author} {\bibfnamefont {L.}~\bibnamefont
  {Duarte}}, \bibinfo {author} {\bibfnamefont {G.~A.}\ \bibnamefont
  {Gonz\'alez-Sprinberg}}, \ and\ \bibinfo {author} {\bibfnamefont {O.~A.}\
  \bibnamefont {Sampayo}},\ }\href {\doibase 10.1103/PhysRevD.91.053007}
  {\bibfield  {journal} {\bibinfo  {journal} {Phys. Rev. D}\ }\textbf {\bibinfo
  {volume} {91}},\ \bibinfo {pages} {053007} (\bibinfo {year}
  {2015}{\natexlab{a}})},\ \Eprint {http://arxiv.org/abs/1412.1433}
  {arXiv:1412.1433 [hep-ph]} \BibitemShut {NoStop}%
\bibitem [{\citenamefont {Bhattacharya}\ and\ \citenamefont
  {Wudka}(2016)}]{Bhattacharya:2015vja}%
  \BibitemOpen
  \bibfield  {author} {\bibinfo {author} {\bibfnamefont {S.}~\bibnamefont
  {Bhattacharya}}\ and\ \bibinfo {author} {\bibfnamefont {J.}~\bibnamefont
  {Wudka}},\ }\href {\doibase 10.1103/PhysRevD.94.055022} {\bibfield  {journal}
  {\bibinfo  {journal} {Phys. Rev. D}\ }\textbf {\bibinfo {volume} {94}},\
  \bibinfo {pages} {055022} (\bibinfo {year} {2016})},\ \bibinfo {note}
  {[Erratum: Phys.Rev.D 95, 039904 (2017)]},\ \Eprint
  {http://arxiv.org/abs/1505.05264} {arXiv:1505.05264 [hep-ph]} \BibitemShut
  {NoStop}%
\bibitem [{\citenamefont {Duarte}\ \emph
  {et~al.}(2015{\natexlab{b}})\citenamefont {Duarte}, \citenamefont
  {Peressutti},\ and\ \citenamefont {Sampayo}}]{Duarte:2015iba}%
  \BibitemOpen
  \bibfield  {author} {\bibinfo {author} {\bibfnamefont {L.}~\bibnamefont
  {Duarte}}, \bibinfo {author} {\bibfnamefont {J.}~\bibnamefont {Peressutti}},
  \ and\ \bibinfo {author} {\bibfnamefont {O.~A.}\ \bibnamefont {Sampayo}},\
  }\href {\doibase 10.1103/PhysRevD.92.093002} {\bibfield  {journal} {\bibinfo
  {journal} {Phys. Rev. D}\ }\textbf {\bibinfo {volume} {92}},\ \bibinfo
  {pages} {093002} (\bibinfo {year} {2015}{\natexlab{b}})},\ \Eprint
  {http://arxiv.org/abs/1508.01588} {arXiv:1508.01588 [hep-ph]} \BibitemShut
  {NoStop}%
\bibitem [{\citenamefont {Duarte}\ \emph
  {et~al.}(2018{\natexlab{a}})\citenamefont {Duarte}, \citenamefont
  {Peressutti},\ and\ \citenamefont {Sampayo}}]{Duarte:2016caz}%
  \BibitemOpen
  \bibfield  {author} {\bibinfo {author} {\bibfnamefont {L.}~\bibnamefont
  {Duarte}}, \bibinfo {author} {\bibfnamefont {J.}~\bibnamefont {Peressutti}},
  \ and\ \bibinfo {author} {\bibfnamefont {O.~A.}\ \bibnamefont {Sampayo}},\
  }\href {\doibase 10.1088/1361-6471/aa99f5} {\bibfield  {journal} {\bibinfo
  {journal} {J. Phys. G}\ }\textbf {\bibinfo {volume} {45}},\ \bibinfo {pages}
  {025001} (\bibinfo {year} {2018}{\natexlab{a}})},\ \Eprint
  {http://arxiv.org/abs/1610.03894} {arXiv:1610.03894 [hep-ph]} \BibitemShut
  {NoStop}%
\bibitem [{\citenamefont {Liao}\ and\ \citenamefont {Ma}(2017)}]{Liao:2016qyd}%
  \BibitemOpen
  \bibfield  {author} {\bibinfo {author} {\bibfnamefont {Y.}~\bibnamefont
  {Liao}}\ and\ \bibinfo {author} {\bibfnamefont {X.-D.}\ \bibnamefont {Ma}},\
  }\href {\doibase 10.1103/PhysRevD.96.015012} {\bibfield  {journal} {\bibinfo
  {journal} {Phys. Rev. D}\ }\textbf {\bibinfo {volume} {96}},\ \bibinfo
  {pages} {015012} (\bibinfo {year} {2017})},\ \Eprint
  {http://arxiv.org/abs/1612.04527} {arXiv:1612.04527 [hep-ph]} \BibitemShut
  {NoStop}%
\bibitem [{\citenamefont {Duarte}\ \emph {et~al.}(2016)\citenamefont {Duarte},
  \citenamefont {Romero}, \citenamefont {Peressutti},\ and\ \citenamefont
  {Sampayo}}]{Duarte:2016miz}%
  \BibitemOpen
  \bibfield  {author} {\bibinfo {author} {\bibfnamefont {L.}~\bibnamefont
  {Duarte}}, \bibinfo {author} {\bibfnamefont {I.}~\bibnamefont {Romero}},
  \bibinfo {author} {\bibfnamefont {J.}~\bibnamefont {Peressutti}}, \ and\
  \bibinfo {author} {\bibfnamefont {O.~A.}\ \bibnamefont {Sampayo}},\ }\href
  {\doibase 10.1140/epjc/s10052-016-4301-8} {\bibfield  {journal} {\bibinfo
  {journal} {Eur. Phys. J. C}\ }\textbf {\bibinfo {volume} {76}},\ \bibinfo
  {pages} {453} (\bibinfo {year} {2016})},\ \Eprint
  {http://arxiv.org/abs/1603.08052} {arXiv:1603.08052 [hep-ph]} \BibitemShut
  {NoStop}%
\bibitem [{\citenamefont {Caputo}\ \emph {et~al.}(2017)\citenamefont {Caputo},
  \citenamefont {Hernandez}, \citenamefont {Lopez-Pavon},\ and\ \citenamefont
  {Salvado}}]{Caputo:2017pit}%
  \BibitemOpen
  \bibfield  {author} {\bibinfo {author} {\bibfnamefont {A.}~\bibnamefont
  {Caputo}}, \bibinfo {author} {\bibfnamefont {P.}~\bibnamefont {Hernandez}},
  \bibinfo {author} {\bibfnamefont {J.}~\bibnamefont {Lopez-Pavon}}, \ and\
  \bibinfo {author} {\bibfnamefont {J.}~\bibnamefont {Salvado}},\ }\href
  {\doibase 10.1007/JHEP06(2017)112} {\bibfield  {journal} {\bibinfo  {journal}
  {JHEP}\ }\textbf {\bibinfo {volume} {06}},\ \bibinfo {pages} {112} (\bibinfo
  {year} {2017})},\ \Eprint {http://arxiv.org/abs/1704.08721} {arXiv:1704.08721
  [hep-ph]} \BibitemShut {NoStop}%
\bibitem [{\citenamefont {Duarte}\ \emph
  {et~al.}(2018{\natexlab{b}})\citenamefont {Duarte}, \citenamefont {Zapata},\
  and\ \citenamefont {Sampayo}}]{Duarte:2018xst}%
  \BibitemOpen
  \bibfield  {author} {\bibinfo {author} {\bibfnamefont {L.}~\bibnamefont
  {Duarte}}, \bibinfo {author} {\bibfnamefont {G.}~\bibnamefont {Zapata}}, \
  and\ \bibinfo {author} {\bibfnamefont {O.~A.}\ \bibnamefont {Sampayo}},\
  }\href {\doibase 10.1140/epjc/s10052-018-5833-x} {\bibfield  {journal}
  {\bibinfo  {journal} {Eur. Phys. J. C}\ }\textbf {\bibinfo {volume} {78}},\
  \bibinfo {pages} {352} (\bibinfo {year} {2018}{\natexlab{b}})},\ \Eprint
  {http://arxiv.org/abs/1802.07620} {arXiv:1802.07620 [hep-ph]} \BibitemShut
  {NoStop}%
\bibitem [{\citenamefont {Alcaide}\ \emph {et~al.}(2019)\citenamefont
  {Alcaide}, \citenamefont {Banerjee}, \citenamefont {Chala},\ and\
  \citenamefont {Titov}}]{Alcaide:2019pnf}%
  \BibitemOpen
  \bibfield  {author} {\bibinfo {author} {\bibfnamefont {J.}~\bibnamefont
  {Alcaide}}, \bibinfo {author} {\bibfnamefont {S.}~\bibnamefont {Banerjee}},
  \bibinfo {author} {\bibfnamefont {M.}~\bibnamefont {Chala}}, \ and\ \bibinfo
  {author} {\bibfnamefont {A.}~\bibnamefont {Titov}},\ }\href {\doibase
  10.1007/JHEP08(2019)031} {\bibfield  {journal} {\bibinfo  {journal} {JHEP}\
  }\textbf {\bibinfo {volume} {08}},\ \bibinfo {pages} {031} (\bibinfo {year}
  {2019})},\ \Eprint {http://arxiv.org/abs/1905.11375} {arXiv:1905.11375
  [hep-ph]} \BibitemShut {NoStop}%
\bibitem [{\citenamefont {Chala}\ and\ \citenamefont
  {Titov}(2020)}]{Chala:2020vqp}%
  \BibitemOpen
  \bibfield  {author} {\bibinfo {author} {\bibfnamefont {M.}~\bibnamefont
  {Chala}}\ and\ \bibinfo {author} {\bibfnamefont {A.}~\bibnamefont {Titov}},\
  }\href {\doibase 10.1007/JHEP05(2020)139} {\bibfield  {journal} {\bibinfo
  {journal} {JHEP}\ }\textbf {\bibinfo {volume} {05}},\ \bibinfo {pages} {139}
  (\bibinfo {year} {2020})},\ \Eprint {http://arxiv.org/abs/2001.07732}
  {arXiv:2001.07732 [hep-ph]} \BibitemShut {NoStop}%
\bibitem [{\citenamefont {Barducci}\ \emph {et~al.}(2021)\citenamefont
  {Barducci}, \citenamefont {Bertuzzo}, \citenamefont {Caputo}, \citenamefont
  {Hernandez},\ and\ \citenamefont {Mele}}]{Barducci:2020icf}%
  \BibitemOpen
  \bibfield  {author} {\bibinfo {author} {\bibfnamefont {D.}~\bibnamefont
  {Barducci}}, \bibinfo {author} {\bibfnamefont {E.}~\bibnamefont {Bertuzzo}},
  \bibinfo {author} {\bibfnamefont {A.}~\bibnamefont {Caputo}}, \bibinfo
  {author} {\bibfnamefont {P.}~\bibnamefont {Hernandez}}, \ and\ \bibinfo
  {author} {\bibfnamefont {B.}~\bibnamefont {Mele}},\ }\href {\doibase
  10.1007/JHEP03(2021)117} {\bibfield  {journal} {\bibinfo  {journal} {JHEP}\
  }\textbf {\bibinfo {volume} {03}},\ \bibinfo {pages} {117} (\bibinfo {year}
  {2021})},\ \Eprint {http://arxiv.org/abs/2011.04725} {arXiv:2011.04725
  [hep-ph]} \BibitemShut {NoStop}%
\bibitem [{\citenamefont {Dekens}\ \emph {et~al.}(2020)\citenamefont {Dekens},
  \citenamefont {de~Vries}, \citenamefont {Fuyuto}, \citenamefont
  {Mereghetti},\ and\ \citenamefont {Zhou}}]{Dekens:2020ttz}%
  \BibitemOpen
  \bibfield  {author} {\bibinfo {author} {\bibfnamefont {W.}~\bibnamefont
  {Dekens}}, \bibinfo {author} {\bibfnamefont {J.}~\bibnamefont {de~Vries}},
  \bibinfo {author} {\bibfnamefont {K.}~\bibnamefont {Fuyuto}}, \bibinfo
  {author} {\bibfnamefont {E.}~\bibnamefont {Mereghetti}}, \ and\ \bibinfo
  {author} {\bibfnamefont {G.}~\bibnamefont {Zhou}},\ }\href {\doibase
  10.1007/JHEP06(2020)097} {\bibfield  {journal} {\bibinfo  {journal} {JHEP}\
  }\textbf {\bibinfo {volume} {06}},\ \bibinfo {pages} {097} (\bibinfo {year}
  {2020})},\ \Eprint {http://arxiv.org/abs/2002.07182} {arXiv:2002.07182
  [hep-ph]} \BibitemShut {NoStop}%
\bibitem [{\citenamefont {Biek\"otter}\ \emph {et~al.}(2020)\citenamefont
  {Biek\"otter}, \citenamefont {Chala},\ and\ \citenamefont
  {Spannowsky}}]{Biekotter:2020tbd}%
  \BibitemOpen
  \bibfield  {author} {\bibinfo {author} {\bibfnamefont {A.}~\bibnamefont
  {Biek\"otter}}, \bibinfo {author} {\bibfnamefont {M.}~\bibnamefont {Chala}},
  \ and\ \bibinfo {author} {\bibfnamefont {M.}~\bibnamefont {Spannowsky}},\
  }\href {\doibase 10.1140/s10052-020-8339-2} {\bibfield  {journal} {\bibinfo
  {journal} {Eur. Phys. J. C}\ }\textbf {\bibinfo {volume} {80}},\ \bibinfo
  {pages} {743} (\bibinfo {year} {2020})},\ \Eprint
  {http://arxiv.org/abs/2007.00673} {arXiv:2007.00673 [hep-ph]} \BibitemShut
  {NoStop}%
\bibitem [{\citenamefont {Duarte}\ \emph {et~al.}(2020)\citenamefont {Duarte},
  \citenamefont {Zapata},\ and\ \citenamefont {Sampayo}}]{Duarte:2020vgj}%
  \BibitemOpen
  \bibfield  {author} {\bibinfo {author} {\bibfnamefont {L.}~\bibnamefont
  {Duarte}}, \bibinfo {author} {\bibfnamefont {G.}~\bibnamefont {Zapata}}, \
  and\ \bibinfo {author} {\bibfnamefont {O.~A.}\ \bibnamefont {Sampayo}},\
  }\href {\doibase 10.1140/epjc/s10052-020-08471-0} {\bibfield  {journal}
  {\bibinfo  {journal} {Eur. Phys. J. C}\ }\textbf {\bibinfo {volume} {80}},\
  \bibinfo {pages} {896} (\bibinfo {year} {2020})},\ \Eprint
  {http://arxiv.org/abs/2006.11216} {arXiv:2006.11216 [hep-ph]} \BibitemShut
  {NoStop}%
\bibitem [{\citenamefont {Barducci}\ \emph {et~al.}(2020)\citenamefont
  {Barducci}, \citenamefont {Bertuzzo}, \citenamefont {Caputo},\ and\
  \citenamefont {Hernandez}}]{Barducci:2020ncz}%
  \BibitemOpen
  \bibfield  {author} {\bibinfo {author} {\bibfnamefont {D.}~\bibnamefont
  {Barducci}}, \bibinfo {author} {\bibfnamefont {E.}~\bibnamefont {Bertuzzo}},
  \bibinfo {author} {\bibfnamefont {A.}~\bibnamefont {Caputo}}, \ and\ \bibinfo
  {author} {\bibfnamefont {P.}~\bibnamefont {Hernandez}},\ }\href {\doibase
  10.1007/JHEP06(2020)185} {\bibfield  {journal} {\bibinfo  {journal} {JHEP}\
  }\textbf {\bibinfo {volume} {06}},\ \bibinfo {pages} {185} (\bibinfo {year}
  {2020})},\ \Eprint {http://arxiv.org/abs/2003.08391} {arXiv:2003.08391
  [hep-ph]} \BibitemShut {NoStop}%
\bibitem [{\citenamefont {Dekens}\ \emph {et~al.}(2021)\citenamefont {Dekens},
  \citenamefont {de~Vries},\ and\ \citenamefont {Tong}}]{Dekens:2021qch}%
  \BibitemOpen
  \bibfield  {author} {\bibinfo {author} {\bibfnamefont {W.}~\bibnamefont
  {Dekens}}, \bibinfo {author} {\bibfnamefont {J.}~\bibnamefont {de~Vries}}, \
  and\ \bibinfo {author} {\bibfnamefont {T.}~\bibnamefont {Tong}},\ }\href
  {\doibase 10.1007/JHEP08(2021)128} {\bibfield  {journal} {\bibinfo  {journal}
  {JHEP}\ }\textbf {\bibinfo {volume} {08}},\ \bibinfo {pages} {128} (\bibinfo
  {year} {2021})},\ \Eprint {http://arxiv.org/abs/2104.00140} {arXiv:2104.00140
  [hep-ph]} \BibitemShut {NoStop}%
\bibitem [{\citenamefont {Cottin}\ \emph {et~al.}(2021)\citenamefont {Cottin},
  \citenamefont {Helo}, \citenamefont {Hirsch}, \citenamefont {Titov},\ and\
  \citenamefont {Wang}}]{Cottin:2021lzz}%
  \BibitemOpen
  \bibfield  {author} {\bibinfo {author} {\bibfnamefont {G.}~\bibnamefont
  {Cottin}}, \bibinfo {author} {\bibfnamefont {J.~C.}\ \bibnamefont {Helo}},
  \bibinfo {author} {\bibfnamefont {M.}~\bibnamefont {Hirsch}}, \bibinfo
  {author} {\bibfnamefont {A.}~\bibnamefont {Titov}}, \ and\ \bibinfo {author}
  {\bibfnamefont {Z.~S.}\ \bibnamefont {Wang}},\ }\href {\doibase
  10.1007/JHEP09(2021)039} {\bibfield  {journal} {\bibinfo  {journal} {JHEP}\
  }\textbf {\bibinfo {volume} {09}},\ \bibinfo {pages} {039} (\bibinfo {year}
  {2021})},\ \Eprint {http://arxiv.org/abs/2105.13851} {arXiv:2105.13851
  [hep-ph]} \BibitemShut {NoStop}%
\bibitem [{\citenamefont {Li}\ \emph {et~al.}(2021)\citenamefont {Li},
  \citenamefont {Ren}, \citenamefont {Xiao}, \citenamefont {Yu},\ and\
  \citenamefont {Zheng}}]{Li:2021tsq}%
  \BibitemOpen
  \bibfield  {author} {\bibinfo {author} {\bibfnamefont {H.-L.}\ \bibnamefont
  {Li}}, \bibinfo {author} {\bibfnamefont {Z.}~\bibnamefont {Ren}}, \bibinfo
  {author} {\bibfnamefont {M.-L.}\ \bibnamefont {Xiao}}, \bibinfo {author}
  {\bibfnamefont {J.-H.}\ \bibnamefont {Yu}}, \ and\ \bibinfo {author}
  {\bibfnamefont {Y.-H.}\ \bibnamefont {Zheng}},\ }\href {\doibase
  10.1007/JHEP11(2021)003} {\bibfield  {journal} {\bibinfo  {journal} {JHEP}\
  }\textbf {\bibinfo {volume} {11}},\ \bibinfo {pages} {003} (\bibinfo {year}
  {2021})},\ \Eprint {http://arxiv.org/abs/2105.09329} {arXiv:2105.09329
  [hep-ph]} \BibitemShut {NoStop}%
\bibitem [{\citenamefont {Cirigliano}\ \emph {et~al.}(2021)\citenamefont
  {Cirigliano}, \citenamefont {Dekens}, \citenamefont {de~Vries}, \citenamefont
  {Fuyuto}, \citenamefont {Mereghetti},\ and\ \citenamefont
  {Ruiz}}]{Cirigliano:2021peb}%
  \BibitemOpen
  \bibfield  {author} {\bibinfo {author} {\bibfnamefont {V.}~\bibnamefont
  {Cirigliano}}, \bibinfo {author} {\bibfnamefont {W.}~\bibnamefont {Dekens}},
  \bibinfo {author} {\bibfnamefont {J.}~\bibnamefont {de~Vries}}, \bibinfo
  {author} {\bibfnamefont {K.}~\bibnamefont {Fuyuto}}, \bibinfo {author}
  {\bibfnamefont {E.}~\bibnamefont {Mereghetti}}, \ and\ \bibinfo {author}
  {\bibfnamefont {R.}~\bibnamefont {Ruiz}},\ }\href {\doibase
  10.1007/JHEP08(2021)103} {\bibfield  {journal} {\bibinfo  {journal} {JHEP}\
  }\textbf {\bibinfo {volume} {08}},\ \bibinfo {pages} {103} (\bibinfo {year}
  {2021})},\ \Eprint {http://arxiv.org/abs/2105.11462} {arXiv:2105.11462
  [hep-ph]} \BibitemShut {NoStop}%
\bibitem [{\citenamefont {De~Vries}\ \emph {et~al.}(2021)\citenamefont
  {De~Vries}, \citenamefont {Dreiner}, \citenamefont {G\"unther}, \citenamefont
  {Wang},\ and\ \citenamefont {Zhou}}]{DeVries:2020jbs}%
  \BibitemOpen
  \bibfield  {author} {\bibinfo {author} {\bibfnamefont {J.}~\bibnamefont
  {De~Vries}}, \bibinfo {author} {\bibfnamefont {H.~K.}\ \bibnamefont
  {Dreiner}}, \bibinfo {author} {\bibfnamefont {J.~Y.}\ \bibnamefont
  {G\"unther}}, \bibinfo {author} {\bibfnamefont {Z.~S.}\ \bibnamefont {Wang}},
  \ and\ \bibinfo {author} {\bibfnamefont {G.}~\bibnamefont {Zhou}},\ }\href
  {\doibase 10.1007/JHEP03(2021)148} {\bibfield  {journal} {\bibinfo  {journal}
  {JHEP}\ }\textbf {\bibinfo {volume} {03}},\ \bibinfo {pages} {148} (\bibinfo
  {year} {2021})},\ \Eprint {http://arxiv.org/abs/2010.07305} {arXiv:2010.07305
  [hep-ph]} \BibitemShut {NoStop}%
\bibitem [{\citenamefont {Zhou}\ \emph {et~al.}(2022)\citenamefont {Zhou},
  \citenamefont {G\"unther}, \citenamefont {Wang}, \citenamefont {de~Vries},\
  and\ \citenamefont {Dreiner}}]{Zhou:2021ylt}%
  \BibitemOpen
  \bibfield  {author} {\bibinfo {author} {\bibfnamefont {G.}~\bibnamefont
  {Zhou}}, \bibinfo {author} {\bibfnamefont {J.~Y.}\ \bibnamefont {G\"unther}},
  \bibinfo {author} {\bibfnamefont {Z.~S.}\ \bibnamefont {Wang}}, \bibinfo
  {author} {\bibfnamefont {J.}~\bibnamefont {de~Vries}}, \ and\ \bibinfo
  {author} {\bibfnamefont {H.~K.}\ \bibnamefont {Dreiner}},\ }\href {\doibase
  10.1007/JHEP04(2022)057} {\bibfield  {journal} {\bibinfo  {journal} {JHEP}\
  }\textbf {\bibinfo {volume} {04}},\ \bibinfo {pages} {057} (\bibinfo {year}
  {2022})},\ \Eprint {http://arxiv.org/abs/2111.04403} {arXiv:2111.04403
  [hep-ph]} \BibitemShut {NoStop}%
\bibitem [{\citenamefont {Zhou}(2022)}]{Zhou:2021lnl}%
  \BibitemOpen
  \bibfield  {author} {\bibinfo {author} {\bibfnamefont {G.}~\bibnamefont
  {Zhou}},\ }\href {\doibase 10.1007/JHEP06(2022)127} {\bibfield  {journal}
  {\bibinfo  {journal} {JHEP}\ }\textbf {\bibinfo {volume} {06}},\ \bibinfo
  {pages} {127} (\bibinfo {year} {2022})},\ \Eprint
  {http://arxiv.org/abs/2112.00767} {arXiv:2112.00767 [hep-ph]} \BibitemShut
  {NoStop}%
\bibitem [{\citenamefont {Beltr\'an}\ \emph
  {et~al.}(2023{\natexlab{a}})\citenamefont {Beltr\'an}, \citenamefont
  {Cottin}, \citenamefont {Helo}, \citenamefont {Hirsch}, \citenamefont
  {Titov},\ and\ \citenamefont {Wang}}]{Beltran:2022ast}%
  \BibitemOpen
  \bibfield  {author} {\bibinfo {author} {\bibfnamefont {R.}~\bibnamefont
  {Beltr\'an}}, \bibinfo {author} {\bibfnamefont {G.}~\bibnamefont {Cottin}},
  \bibinfo {author} {\bibfnamefont {J.~C.}\ \bibnamefont {Helo}}, \bibinfo
  {author} {\bibfnamefont {M.}~\bibnamefont {Hirsch}}, \bibinfo {author}
  {\bibfnamefont {A.}~\bibnamefont {Titov}}, \ and\ \bibinfo {author}
  {\bibfnamefont {Z.~S.}\ \bibnamefont {Wang}},\ }\href {\doibase
  10.1007/JHEP01(2023)015} {\bibfield  {journal} {\bibinfo  {journal} {JHEP}\
  }\textbf {\bibinfo {volume} {01}},\ \bibinfo {pages} {015} (\bibinfo {year}
  {2023}{\natexlab{a}})},\ \Eprint {http://arxiv.org/abs/2210.02461}
  {arXiv:2210.02461 [hep-ph]} \BibitemShut {NoStop}%
\bibitem [{\citenamefont {Delgado}\ \emph {et~al.}(2022)\citenamefont
  {Delgado}, \citenamefont {Duarte}, \citenamefont {Jones-Perez}, \citenamefont
  {Manrique-Chavil},\ and\ \citenamefont {Pe\~na}}]{Delgado:2022fea}%
  \BibitemOpen
  \bibfield  {author} {\bibinfo {author} {\bibfnamefont {F.}~\bibnamefont
  {Delgado}}, \bibinfo {author} {\bibfnamefont {L.}~\bibnamefont {Duarte}},
  \bibinfo {author} {\bibfnamefont {J.}~\bibnamefont {Jones-Perez}}, \bibinfo
  {author} {\bibfnamefont {C.}~\bibnamefont {Manrique-Chavil}}, \ and\ \bibinfo
  {author} {\bibfnamefont {S.}~\bibnamefont {Pe\~na}},\ }\href {\doibase
  10.1007/JHEP09(2022)079} {\bibfield  {journal} {\bibinfo  {journal} {JHEP}\
  }\textbf {\bibinfo {volume} {09}},\ \bibinfo {pages} {079} (\bibinfo {year}
  {2022})},\ \Eprint {http://arxiv.org/abs/2205.13550} {arXiv:2205.13550
  [hep-ph]} \BibitemShut {NoStop}%
\bibitem [{\citenamefont {Barducci}\ \emph {et~al.}(2022)\citenamefont
  {Barducci}, \citenamefont {Bertuzzo}, \citenamefont {Taoso},\ and\
  \citenamefont {Toni}}]{Barducci:2022gdv}%
  \BibitemOpen
  \bibfield  {author} {\bibinfo {author} {\bibfnamefont {D.}~\bibnamefont
  {Barducci}}, \bibinfo {author} {\bibfnamefont {E.}~\bibnamefont {Bertuzzo}},
  \bibinfo {author} {\bibfnamefont {M.}~\bibnamefont {Taoso}}, \ and\ \bibinfo
  {author} {\bibfnamefont {C.}~\bibnamefont {Toni}},\ }\href {\doibase
  10.1007/JHEP03(2023)239} {\bibfield  {journal} {\bibinfo  {journal} {JHEP}\
  }\textbf {\bibinfo {volume} {03}},\ \bibinfo {pages} {239} (\bibinfo {year}
  {2022})},\ \Eprint {http://arxiv.org/abs/2209.13469} {arXiv:2209.13469
  [hep-ph]} \BibitemShut {NoStop}%
\bibitem [{\citenamefont {Talbert}(2023)}]{Talbert:2022unj}%
  \BibitemOpen
  \bibfield  {author} {\bibinfo {author} {\bibfnamefont {J.}~\bibnamefont
  {Talbert}},\ }\href {\doibase 10.1007/JHEP01(2023)069} {\bibfield  {journal}
  {\bibinfo  {journal} {JHEP}\ }\textbf {\bibinfo {volume} {01}},\ \bibinfo
  {pages} {069} (\bibinfo {year} {2023})},\ \Eprint
  {http://arxiv.org/abs/2208.11139} {arXiv:2208.11139 [hep-ph]} \BibitemShut
  {NoStop}%
\bibitem [{\citenamefont {Barducci}\ and\ \citenamefont
  {Bertuzzo}(2022)}]{Barducci:2022hll}%
  \BibitemOpen
  \bibfield  {author} {\bibinfo {author} {\bibfnamefont {D.}~\bibnamefont
  {Barducci}}\ and\ \bibinfo {author} {\bibfnamefont {E.}~\bibnamefont
  {Bertuzzo}},\ }\href {\doibase 10.1007/JHEP06(2022)077} {\bibfield  {journal}
  {\bibinfo  {journal} {JHEP}\ }\textbf {\bibinfo {volume} {06}},\ \bibinfo
  {pages} {077} (\bibinfo {year} {2022})},\ \Eprint
  {http://arxiv.org/abs/2201.11754} {arXiv:2201.11754 [hep-ph]} \BibitemShut
  {NoStop}%
\bibitem [{\citenamefont {Zapata}\ \emph {et~al.}(2022)\citenamefont {Zapata},
  \citenamefont {Urruzola}, \citenamefont {Sampayo},\ and\ \citenamefont
  {Duarte}}]{Zapata:2022qwo}%
  \BibitemOpen
  \bibfield  {author} {\bibinfo {author} {\bibfnamefont {G.}~\bibnamefont
  {Zapata}}, \bibinfo {author} {\bibfnamefont {T.}~\bibnamefont {Urruzola}},
  \bibinfo {author} {\bibfnamefont {O.~A.}\ \bibnamefont {Sampayo}}, \ and\
  \bibinfo {author} {\bibfnamefont {L.}~\bibnamefont {Duarte}},\ }\href
  {\doibase 10.1140/epjc/s10052-022-10448-0} {\bibfield  {journal} {\bibinfo
  {journal} {Eur. Phys. J. C}\ }\textbf {\bibinfo {volume} {82}},\ \bibinfo
  {pages} {544} (\bibinfo {year} {2022})},\ \Eprint
  {http://arxiv.org/abs/2201.02480} {arXiv:2201.02480 [hep-ph]} \BibitemShut
  {NoStop}%
\bibitem [{\citenamefont {Mitra}\ \emph {et~al.}(2022)\citenamefont {Mitra},
  \citenamefont {Mandal}, \citenamefont {Padhan}, \citenamefont {Sarkar},\ and\
  \citenamefont {Spannowsky}}]{Mitra:2022nri}%
  \BibitemOpen
  \bibfield  {author} {\bibinfo {author} {\bibfnamefont {M.}~\bibnamefont
  {Mitra}}, \bibinfo {author} {\bibfnamefont {S.}~\bibnamefont {Mandal}},
  \bibinfo {author} {\bibfnamefont {R.}~\bibnamefont {Padhan}}, \bibinfo
  {author} {\bibfnamefont {A.}~\bibnamefont {Sarkar}}, \ and\ \bibinfo {author}
  {\bibfnamefont {M.}~\bibnamefont {Spannowsky}},\ }\href {\doibase
  10.1103/PhysRevD.106.113008} {\bibfield  {journal} {\bibinfo  {journal}
  {Phys. Rev. D}\ }\textbf {\bibinfo {volume} {106}},\ \bibinfo {pages}
  {113008} (\bibinfo {year} {2022})},\ \Eprint
  {http://arxiv.org/abs/2210.12404} {arXiv:2210.12404 [hep-ph]} \BibitemShut
  {NoStop}%
\bibitem [{\citenamefont {Beltr\'an}\ \emph
  {et~al.}(2023{\natexlab{b}})\citenamefont {Beltr\'an}, \citenamefont
  {Cottin}, \citenamefont {Hirsch}, \citenamefont {Titov},\ and\ \citenamefont
  {Wang}}]{Beltran:2023nli}%
  \BibitemOpen
  \bibfield  {author} {\bibinfo {author} {\bibfnamefont {R.}~\bibnamefont
  {Beltr\'an}}, \bibinfo {author} {\bibfnamefont {G.}~\bibnamefont {Cottin}},
  \bibinfo {author} {\bibfnamefont {M.}~\bibnamefont {Hirsch}}, \bibinfo
  {author} {\bibfnamefont {A.}~\bibnamefont {Titov}}, \ and\ \bibinfo {author}
  {\bibfnamefont {Z.~S.}\ \bibnamefont {Wang}},\ }\href@noop {} {\  (\bibinfo
  {year} {2023}{\natexlab{b}})},\ \Eprint {http://arxiv.org/abs/2302.03216}
  {arXiv:2302.03216 [hep-ph]} \BibitemShut {NoStop}%
\bibitem [{\citenamefont {Dekens}\ \emph {et~al.}(2023)\citenamefont {Dekens},
  \citenamefont {de~Vries}, \citenamefont {Mereghetti}, \citenamefont
  {Men\'endez}, \citenamefont {Soriano},\ and\ \citenamefont
  {Zhou}}]{Dekens:2023iyc}%
  \BibitemOpen
  \bibfield  {author} {\bibinfo {author} {\bibfnamefont {W.}~\bibnamefont
  {Dekens}}, \bibinfo {author} {\bibfnamefont {J.}~\bibnamefont {de~Vries}},
  \bibinfo {author} {\bibfnamefont {E.}~\bibnamefont {Mereghetti}}, \bibinfo
  {author} {\bibfnamefont {J.}~\bibnamefont {Men\'endez}}, \bibinfo {author}
  {\bibfnamefont {P.}~\bibnamefont {Soriano}}, \ and\ \bibinfo {author}
  {\bibfnamefont {G.}~\bibnamefont {Zhou}},\ }\href@noop {} {\  (\bibinfo
  {year} {2023})},\ \Eprint {http://arxiv.org/abs/2303.04168} {arXiv:2303.04168
  [hep-ph]} \BibitemShut {NoStop}%
\bibitem [{\citenamefont {Bischer}\ and\ \citenamefont
  {Rodejohann}(2019)}]{Bischer:2019ttk}%
  \BibitemOpen
  \bibfield  {author} {\bibinfo {author} {\bibfnamefont {I.}~\bibnamefont
  {Bischer}}\ and\ \bibinfo {author} {\bibfnamefont {W.}~\bibnamefont
  {Rodejohann}},\ }\href {\doibase 10.1016/j.nuclphysb.2019.114746} {\bibfield
  {journal} {\bibinfo  {journal} {Nucl. Phys. B}\ }\textbf {\bibinfo {volume}
  {947}},\ \bibinfo {pages} {114746} (\bibinfo {year} {2019})},\ \Eprint
  {http://arxiv.org/abs/1905.08699} {arXiv:1905.08699 [hep-ph]} \BibitemShut
  {NoStop}%
\bibitem [{\citenamefont {Drewes}\ \emph {et~al.}(2018)\citenamefont {Drewes},
  \citenamefont {Hajer}, \citenamefont {Klaric},\ and\ \citenamefont
  {Lanfranchi}}]{Drewes:2018gkc}%
  \BibitemOpen
  \bibfield  {author} {\bibinfo {author} {\bibfnamefont {M.}~\bibnamefont
  {Drewes}}, \bibinfo {author} {\bibfnamefont {J.}~\bibnamefont {Hajer}},
  \bibinfo {author} {\bibfnamefont {J.}~\bibnamefont {Klaric}}, \ and\ \bibinfo
  {author} {\bibfnamefont {G.}~\bibnamefont {Lanfranchi}},\ }\href {\doibase
  10.1007/JHEP07(2018)105} {\bibfield  {journal} {\bibinfo  {journal} {JHEP}\
  }\textbf {\bibinfo {volume} {07}},\ \bibinfo {pages} {105} (\bibinfo {year}
  {2018})},\ \Eprint {http://arxiv.org/abs/1801.04207} {arXiv:1801.04207
  [hep-ph]} \BibitemShut {NoStop}%
\bibitem [{\citenamefont {Ahdida}\ \emph {et~al.}(2019)\citenamefont {Ahdida}
  \emph {et~al.}}]{SHiP:2018xqw}%
  \BibitemOpen
  \bibfield  {author} {\bibinfo {author} {\bibfnamefont {C.}~\bibnamefont
  {Ahdida}} \emph {et~al.} (\bibinfo {collaboration} {SHiP}),\ }\href {\doibase
  10.1007/JHEP04(2019)077} {\bibfield  {journal} {\bibinfo  {journal} {JHEP}\
  }\textbf {\bibinfo {volume} {04}},\ \bibinfo {pages} {077} (\bibinfo {year}
  {2019})},\ \Eprint {http://arxiv.org/abs/1811.00930} {arXiv:1811.00930
  [hep-ph]} \BibitemShut {NoStop}%
\bibitem [{\citenamefont {Bondarenko}\ \emph {et~al.}(2021)\citenamefont
  {Bondarenko}, \citenamefont {Boyarsky}, \citenamefont {Klaric}, \citenamefont
  {Mikulenko}, \citenamefont {Ruchayskiy}, \citenamefont {Syvolap},\ and\
  \citenamefont {Timiryasov}}]{Bondarenko:2021cpc}%
  \BibitemOpen
  \bibfield  {author} {\bibinfo {author} {\bibfnamefont {K.}~\bibnamefont
  {Bondarenko}}, \bibinfo {author} {\bibfnamefont {A.}~\bibnamefont
  {Boyarsky}}, \bibinfo {author} {\bibfnamefont {J.}~\bibnamefont {Klaric}},
  \bibinfo {author} {\bibfnamefont {O.}~\bibnamefont {Mikulenko}}, \bibinfo
  {author} {\bibfnamefont {O.}~\bibnamefont {Ruchayskiy}}, \bibinfo {author}
  {\bibfnamefont {V.}~\bibnamefont {Syvolap}}, \ and\ \bibinfo {author}
  {\bibfnamefont {I.}~\bibnamefont {Timiryasov}},\ }\href {\doibase
  10.1007/JHEP07(2021)193} {\bibfield  {journal} {\bibinfo  {journal} {JHEP}\
  }\textbf {\bibinfo {volume} {07}},\ \bibinfo {pages} {193} (\bibinfo {year}
  {2021})},\ \Eprint {http://arxiv.org/abs/2101.09255} {arXiv:2101.09255
  [hep-ph]} \BibitemShut {NoStop}%
\bibitem [{\citenamefont {Tastet}\ \emph {et~al.}(2021)\citenamefont {Tastet},
  \citenamefont {Ruchayskiy},\ and\ \citenamefont
  {Timiryasov}}]{Tastet:2021vwp}%
  \BibitemOpen
  \bibfield  {author} {\bibinfo {author} {\bibfnamefont {J.-L.}\ \bibnamefont
  {Tastet}}, \bibinfo {author} {\bibfnamefont {O.}~\bibnamefont {Ruchayskiy}},
  \ and\ \bibinfo {author} {\bibfnamefont {I.}~\bibnamefont {Timiryasov}},\
  }\href {\doibase 10.1007/JHEP12(2021)182} {\bibfield  {journal} {\bibinfo
  {journal} {JHEP}\ }\textbf {\bibinfo {volume} {12}},\ \bibinfo {pages} {182}
  (\bibinfo {year} {2021})},\ \Eprint {http://arxiv.org/abs/2107.12980}
  {arXiv:2107.12980 [hep-ph]} \BibitemShut {NoStop}%
\bibitem [{\citenamefont {Drewes}\ \emph {et~al.}(2022)\citenamefont {Drewes},
  \citenamefont {Klari\'c},\ and\ \citenamefont
  {L\'opez-Pav\'on}}]{Drewes:2022akb}%
  \BibitemOpen
  \bibfield  {author} {\bibinfo {author} {\bibfnamefont {M.}~\bibnamefont
  {Drewes}}, \bibinfo {author} {\bibfnamefont {J.}~\bibnamefont {Klari\'c}}, \
  and\ \bibinfo {author} {\bibfnamefont {J.}~\bibnamefont {L\'opez-Pav\'on}},\
  }\href {\doibase 10.1140/epjc/s10052-022-11100-7} {\bibfield  {journal}
  {\bibinfo  {journal} {Eur. Phys. J. C}\ }\textbf {\bibinfo {volume} {82}},\
  \bibinfo {pages} {1176} (\bibinfo {year} {2022})},\ \Eprint
  {http://arxiv.org/abs/2207.02742} {arXiv:2207.02742 [hep-ph]} \BibitemShut
  {NoStop}%
\bibitem [{\citenamefont {Ilten}\ \emph {et~al.}(2018)\citenamefont {Ilten},
  \citenamefont {Soreq}, \citenamefont {Williams},\ and\ \citenamefont
  {Xue}}]{Ilten:2018crw}%
  \BibitemOpen
  \bibfield  {author} {\bibinfo {author} {\bibfnamefont {P.}~\bibnamefont
  {Ilten}}, \bibinfo {author} {\bibfnamefont {Y.}~\bibnamefont {Soreq}},
  \bibinfo {author} {\bibfnamefont {M.}~\bibnamefont {Williams}}, \ and\
  \bibinfo {author} {\bibfnamefont {W.}~\bibnamefont {Xue}},\ }\href {\doibase
  10.1007/JHEP06(2018)004} {\bibfield  {journal} {\bibinfo  {journal} {JHEP}\
  }\textbf {\bibinfo {volume} {06}},\ \bibinfo {pages} {004} (\bibinfo {year}
  {2018})},\ \Eprint {http://arxiv.org/abs/1801.04847} {arXiv:1801.04847
  [hep-ph]} \BibitemShut {NoStop}%
\bibitem [{\citenamefont {O'Hare}(2020)}]{AxionLimits}%
  \BibitemOpen
  \bibfield  {author} {\bibinfo {author} {\bibfnamefont {C.}~\bibnamefont
  {O'Hare}},\ }\href {\doibase 10.5281/zenodo.3932430} {}\bibinfo
  {howpublished} {\url{https://cajohare.github.io/AxionLimits/}} (\bibinfo
  {year} {2020})\BibitemShut {NoStop}%
\bibitem [{\citenamefont {Bolton}\ \emph {et~al.}(2020)\citenamefont {Bolton},
  \citenamefont {Deppisch},\ and\ \citenamefont {Bhupal~Dev}}]{Bolton:2019pcu}%
  \BibitemOpen
  \bibfield  {author} {\bibinfo {author} {\bibfnamefont {P.~D.}\ \bibnamefont
  {Bolton}}, \bibinfo {author} {\bibfnamefont {F.~F.}\ \bibnamefont
  {Deppisch}}, \ and\ \bibinfo {author} {\bibfnamefont {P.~S.}\ \bibnamefont
  {Bhupal~Dev}},\ }\href {\doibase 10.1007/JHEP03(2020)170} {\bibfield
  {journal} {\bibinfo  {journal} {JHEP}\ }\textbf {\bibinfo {volume} {03}},\
  \bibinfo {pages} {170} (\bibinfo {year} {2020})},\ \Eprint
  {http://arxiv.org/abs/1912.03058} {arXiv:1912.03058 [hep-ph]} \BibitemShut
  {NoStop}%
\bibitem [{\citenamefont {Atre}\ \emph {et~al.}(2009)\citenamefont {Atre},
  \citenamefont {Han}, \citenamefont {Silvia},\ and\ \citenamefont
  {Zhang}}]{Atre:2009rg}%
  \BibitemOpen
  \bibfield  {author} {\bibinfo {author} {\bibfnamefont {A.}~\bibnamefont
  {Atre}}, \bibinfo {author} {\bibfnamefont {T.}~\bibnamefont {Han}}, \bibinfo
  {author} {\bibfnamefont {P.}~\bibnamefont {Silvia}}, \ and\ \bibinfo {author}
  {\bibfnamefont {B.}~\bibnamefont {Zhang}},\ }\href {\doibase
  10.1088/1126-6708/2009/05/030} {\bibfield  {journal} {\bibinfo  {journal}
  {JHEP}\ }\textbf {\bibinfo {volume} {05}},\ \bibinfo {pages} {030} (\bibinfo
  {year} {2009})},\ \Eprint {http://arxiv.org/abs/0901.3589} {arXiv:0901.3589
  [hep-ph]} \BibitemShut {NoStop}%
\bibitem [{\citenamefont {Ruchayskiy}\ and\ \citenamefont
  {Ivashko}(2012{\natexlab{a}})}]{Ruchayskiy:2011aa}%
  \BibitemOpen
  \bibfield  {author} {\bibinfo {author} {\bibfnamefont {O.}~\bibnamefont
  {Ruchayskiy}}\ and\ \bibinfo {author} {\bibfnamefont {A.}~\bibnamefont
  {Ivashko}},\ }\href {\doibase 10.1007/JHEP06(2012)100} {\bibfield  {journal}
  {\bibinfo  {journal} {JHEP}\ }\textbf {\bibinfo {volume} {06}},\ \bibinfo
  {pages} {100} (\bibinfo {year} {2012}{\natexlab{a}})},\ \Eprint
  {http://arxiv.org/abs/1112.3319} {arXiv:1112.3319 [hep-ph]} \BibitemShut
  {NoStop}%
\bibitem [{\citenamefont {de~Gouv\^ea}\ and\ \citenamefont
  {Kobach}(2016)}]{deGouvea:2015euy}%
  \BibitemOpen
  \bibfield  {author} {\bibinfo {author} {\bibfnamefont {A.}~\bibnamefont
  {de~Gouv\^ea}}\ and\ \bibinfo {author} {\bibfnamefont {A.}~\bibnamefont
  {Kobach}},\ }\href {\doibase 10.1103/PhysRevD.93.033005} {\bibfield
  {journal} {\bibinfo  {journal} {Phys. Rev. D}\ }\textbf {\bibinfo {volume}
  {93}},\ \bibinfo {pages} {033005} (\bibinfo {year} {2016})},\ \Eprint
  {http://arxiv.org/abs/1511.00683} {arXiv:1511.00683 [hep-ph]} \BibitemShut
  {NoStop}%
\bibitem [{\citenamefont {Drewes}\ and\ \citenamefont
  {Garbrecht}(2017)}]{Drewes:2015iva}%
  \BibitemOpen
  \bibfield  {author} {\bibinfo {author} {\bibfnamefont {M.}~\bibnamefont
  {Drewes}}\ and\ \bibinfo {author} {\bibfnamefont {B.}~\bibnamefont
  {Garbrecht}},\ }\href {\doibase 10.1016/j.nuclphysb.2017.05.001} {\bibfield
  {journal} {\bibinfo  {journal} {Nucl. Phys. B}\ }\textbf {\bibinfo {volume}
  {921}},\ \bibinfo {pages} {250} (\bibinfo {year} {2017})},\ \Eprint
  {http://arxiv.org/abs/1502.00477} {arXiv:1502.00477 [hep-ph]} \BibitemShut
  {NoStop}%
\bibitem [{\citenamefont {Antusch}\ and\ \citenamefont
  {Fischer}(2015)}]{Antusch:2015mia}%
  \BibitemOpen
  \bibfield  {author} {\bibinfo {author} {\bibfnamefont {S.}~\bibnamefont
  {Antusch}}\ and\ \bibinfo {author} {\bibfnamefont {O.}~\bibnamefont
  {Fischer}},\ }\href {\doibase 10.1007/JHEP05(2015)053} {\bibfield  {journal}
  {\bibinfo  {journal} {JHEP}\ }\textbf {\bibinfo {volume} {05}},\ \bibinfo
  {pages} {053} (\bibinfo {year} {2015})},\ \Eprint
  {http://arxiv.org/abs/1502.05915} {arXiv:1502.05915 [hep-ph]} \BibitemShut
  {NoStop}%
\bibitem [{\citenamefont {Fernandez-Martinez}\ \emph
  {et~al.}(2016)\citenamefont {Fernandez-Martinez}, \citenamefont
  {Hernandez-Garcia},\ and\ \citenamefont
  {Lopez-Pavon}}]{Fernandez-Martinez:2016lgt}%
  \BibitemOpen
  \bibfield  {author} {\bibinfo {author} {\bibfnamefont {E.}~\bibnamefont
  {Fernandez-Martinez}}, \bibinfo {author} {\bibfnamefont {J.}~\bibnamefont
  {Hernandez-Garcia}}, \ and\ \bibinfo {author} {\bibfnamefont
  {J.}~\bibnamefont {Lopez-Pavon}},\ }\href {\doibase 10.1007/JHEP08(2016)033}
  {\bibfield  {journal} {\bibinfo  {journal} {JHEP}\ }\textbf {\bibinfo
  {volume} {08}},\ \bibinfo {pages} {033} (\bibinfo {year} {2016})},\ \Eprint
  {http://arxiv.org/abs/1605.08774} {arXiv:1605.08774 [hep-ph]} \BibitemShut
  {NoStop}%
\bibitem [{\citenamefont {Chrzaszcz}\ \emph {et~al.}(2020)\citenamefont
  {Chrzaszcz}, \citenamefont {Drewes}, \citenamefont {Gonzalo}, \citenamefont
  {Harz}, \citenamefont {Krishnamurthy},\ and\ \citenamefont
  {Weniger}}]{Chrzaszcz:2019inj}%
  \BibitemOpen
  \bibfield  {author} {\bibinfo {author} {\bibfnamefont {M.}~\bibnamefont
  {Chrzaszcz}}, \bibinfo {author} {\bibfnamefont {M.}~\bibnamefont {Drewes}},
  \bibinfo {author} {\bibfnamefont {T.~E.}\ \bibnamefont {Gonzalo}}, \bibinfo
  {author} {\bibfnamefont {J.}~\bibnamefont {Harz}}, \bibinfo {author}
  {\bibfnamefont {S.}~\bibnamefont {Krishnamurthy}}, \ and\ \bibinfo {author}
  {\bibfnamefont {C.}~\bibnamefont {Weniger}},\ }\href {\doibase
  10.1140/epjc/s10052-020-8073-9} {\bibfield  {journal} {\bibinfo  {journal}
  {Eur. Phys. J. C}\ }\textbf {\bibinfo {volume} {80}},\ \bibinfo {pages} {569}
  (\bibinfo {year} {2020})},\ \Eprint {http://arxiv.org/abs/1908.02302}
  {arXiv:1908.02302 [hep-ph]} \BibitemShut {NoStop}%
\bibitem [{\citenamefont {Agrawal}\ \emph {et~al.}(2021)\citenamefont {Agrawal}
  \emph {et~al.}}]{Agrawal:2021dbo}%
  \BibitemOpen
  \bibfield  {author} {\bibinfo {author} {\bibfnamefont {P.}~\bibnamefont
  {Agrawal}} \emph {et~al.},\ }\href {\doibase 10.1140/epjc/s10052-021-09703-7}
  {\bibfield  {journal} {\bibinfo  {journal} {Eur. Phys. J. C}\ }\textbf
  {\bibinfo {volume} {81}},\ \bibinfo {pages} {1015} (\bibinfo {year}
  {2021})},\ \Eprint {http://arxiv.org/abs/2102.12143} {arXiv:2102.12143
  [hep-ph]} \BibitemShut {NoStop}%
\bibitem [{\citenamefont {Abdullahi}\ \emph {et~al.}(2023)\citenamefont
  {Abdullahi} \emph {et~al.}}]{Abdullahi:2022jlv}%
  \BibitemOpen
  \bibfield  {author} {\bibinfo {author} {\bibfnamefont {A.~M.}\ \bibnamefont
  {Abdullahi}} \emph {et~al.},\ }\href {\doibase 10.1088/1361-6471/ac98f9}
  {\bibfield  {journal} {\bibinfo  {journal} {J. Phys. G}\ }\textbf {\bibinfo
  {volume} {50}},\ \bibinfo {pages} {020501} (\bibinfo {year} {2023})},\
  \Eprint {http://arxiv.org/abs/2203.08039} {arXiv:2203.08039 [hep-ph]}
  \BibitemShut {NoStop}%
\bibitem [{\citenamefont {Britton}\ \emph {et~al.}(1992)\citenamefont {Britton}
  \emph {et~al.}}]{Britton:1992xv}%
  \BibitemOpen
  \bibfield  {author} {\bibinfo {author} {\bibfnamefont {D.}~\bibnamefont
  {Britton}} \emph {et~al.},\ }\href {\doibase 10.1103/PhysRevD.46.R885}
  {\bibfield  {journal} {\bibinfo  {journal} {Phys. Rev. D}\ }\textbf {\bibinfo
  {volume} {46}},\ \bibinfo {pages} {885} (\bibinfo {year} {1992})}\BibitemShut
  {NoStop}%
\bibitem [{\citenamefont {Aguilar-Arevalo}\ \emph {et~al.}(2018)\citenamefont
  {Aguilar-Arevalo} \emph {et~al.}}]{Aguilar-Arevalo:2017vlf}%
  \BibitemOpen
  \bibfield  {author} {\bibinfo {author} {\bibfnamefont {A.}~\bibnamefont
  {Aguilar-Arevalo}} \emph {et~al.} (\bibinfo {collaboration} {PIENU}),\ }\href
  {\doibase 10.1103/PhysRevD.97.072012} {\bibfield  {journal} {\bibinfo
  {journal} {Phys. Rev. D}\ }\textbf {\bibinfo {volume} {97}},\ \bibinfo
  {pages} {072012} (\bibinfo {year} {2018})},\ \Eprint
  {http://arxiv.org/abs/1712.03275} {arXiv:1712.03275 [hep-ex]} \BibitemShut
  {NoStop}%
\bibitem [{\citenamefont {Bryman}\ and\ \citenamefont
  {Shrock}(2019)}]{Bryman:2019bjg}%
  \BibitemOpen
  \bibfield  {author} {\bibinfo {author} {\bibfnamefont {D.~A.}\ \bibnamefont
  {Bryman}}\ and\ \bibinfo {author} {\bibfnamefont {R.}~\bibnamefont
  {Shrock}},\ }\href {\doibase 10.1103/PhysRevD.100.073011} {\bibfield
  {journal} {\bibinfo  {journal} {Phys. Rev. D}\ }\textbf {\bibinfo {volume}
  {100}},\ \bibinfo {pages} {073011} (\bibinfo {year} {2019})},\ \Eprint
  {http://arxiv.org/abs/1909.11198} {arXiv:1909.11198 [hep-ph]} \BibitemShut
  {NoStop}%
\bibitem [{\citenamefont {Bellini}\ \emph {et~al.}(2013)\citenamefont {Bellini}
  \emph {et~al.}}]{Borexino:2013bot}%
  \BibitemOpen
  \bibfield  {author} {\bibinfo {author} {\bibfnamefont {G.}~\bibnamefont
  {Bellini}} \emph {et~al.} (\bibinfo {collaboration} {Borexino}),\ }\href
  {\doibase 10.1103/PhysRevD.88.072010} {\bibfield  {journal} {\bibinfo
  {journal} {Phys. Rev. D}\ }\textbf {\bibinfo {volume} {88}},\ \bibinfo
  {pages} {072010} (\bibinfo {year} {2013})},\ \Eprint
  {http://arxiv.org/abs/1311.5347} {arXiv:1311.5347 [hep-ex]} \BibitemShut
  {NoStop}%
\bibitem [{\citenamefont {Cortina~Gil}\ \emph
  {et~al.}(2020{\natexlab{a}})\citenamefont {Cortina~Gil} \emph
  {et~al.}}]{NA62:2020mcv}%
  \BibitemOpen
  \bibfield  {author} {\bibinfo {author} {\bibfnamefont {E.}~\bibnamefont
  {Cortina~Gil}} \emph {et~al.} (\bibinfo {collaboration} {NA62}),\ }\href
  {\doibase 10.1016/j.physletb.2020.135599} {\bibfield  {journal} {\bibinfo
  {journal} {Phys. Lett. B}\ }\textbf {\bibinfo {volume} {807}},\ \bibinfo
  {pages} {135599} (\bibinfo {year} {2020}{\natexlab{a}})},\ \Eprint
  {http://arxiv.org/abs/2005.09575} {arXiv:2005.09575 [hep-ex]} \BibitemShut
  {NoStop}%
\bibitem [{\citenamefont {Abe}\ \emph {et~al.}(2019)\citenamefont {Abe} \emph
  {et~al.}}]{Abe:2019kgx}%
  \BibitemOpen
  \bibfield  {author} {\bibinfo {author} {\bibfnamefont {K.}~\bibnamefont
  {Abe}} \emph {et~al.} (\bibinfo {collaboration} {T2K}),\ }\href {\doibase
  10.1103/PhysRevD.100.052006} {\bibfield  {journal} {\bibinfo  {journal}
  {Phys. Rev.}\ }\textbf {\bibinfo {volume} {D100}},\ \bibinfo {pages} {052006}
  (\bibinfo {year} {2019})},\ \Eprint {http://arxiv.org/abs/1902.07598}
  {arXiv:1902.07598 [hep-ex]} \BibitemShut {NoStop}%
\bibitem [{\citenamefont {Bergsma}\ \emph {et~al.}(1986)\citenamefont {Bergsma}
  \emph {et~al.}}]{Bergsma:1985is}%
  \BibitemOpen
  \bibfield  {author} {\bibinfo {author} {\bibfnamefont {F.}~\bibnamefont
  {Bergsma}} \emph {et~al.} (\bibinfo {collaboration} {CHARM}),\ }\href
  {\doibase 10.1016/0370-2693(86)91601-1} {\bibfield  {journal} {\bibinfo
  {journal} {Phys. Lett. B}\ }\textbf {\bibinfo {volume} {166}},\ \bibinfo
  {pages} {473} (\bibinfo {year} {1986})}\BibitemShut {NoStop}%
\bibitem [{\citenamefont {Cooper-Sarkar}\ \emph
  {et~al.}(1985{\natexlab{a}})\citenamefont {Cooper-Sarkar} \emph
  {et~al.}}]{CooperSarkar:1985nh}%
  \BibitemOpen
  \bibfield  {author} {\bibinfo {author} {\bibfnamefont {A.~M.}\ \bibnamefont
  {Cooper-Sarkar}} \emph {et~al.} (\bibinfo {collaboration} {WA66}),\ }\href
  {\doibase 10.1016/0370-2693(85)91493-5} {\bibfield  {journal} {\bibinfo
  {journal} {Phys. Lett. B}\ }\textbf {\bibinfo {volume} {160}},\ \bibinfo
  {pages} {207} (\bibinfo {year} {1985}{\natexlab{a}})}\BibitemShut {NoStop}%
\bibitem [{\citenamefont {Barouki}\ \emph {et~al.}(2022)\citenamefont
  {Barouki}, \citenamefont {Marocco},\ and\ \citenamefont
  {Sarkar}}]{Barouki:2022bkt}%
  \BibitemOpen
  \bibfield  {author} {\bibinfo {author} {\bibfnamefont {R.}~\bibnamefont
  {Barouki}}, \bibinfo {author} {\bibfnamefont {G.}~\bibnamefont {Marocco}}, \
  and\ \bibinfo {author} {\bibfnamefont {S.}~\bibnamefont {Sarkar}},\ }\href
  {\doibase 10.21468/SciPostPhys.13.5.118} {\bibfield  {journal} {\bibinfo
  {journal} {SciPost Phys.}\ }\textbf {\bibinfo {volume} {13}},\ \bibinfo
  {pages} {118} (\bibinfo {year} {2022})},\ \Eprint
  {http://arxiv.org/abs/2208.00416} {arXiv:2208.00416 [hep-ph]} \BibitemShut
  {NoStop}%
\bibitem [{\citenamefont {Abreu}\ \emph {et~al.}(1997)\citenamefont {Abreu}
  \emph {et~al.}}]{Abreu:1996pa}%
  \BibitemOpen
  \bibfield  {author} {\bibinfo {author} {\bibfnamefont {P.}~\bibnamefont
  {Abreu}} \emph {et~al.} (\bibinfo {collaboration} {DELPHI}),\ }\href
  {\doibase 10.1007/s002880050370} {\bibfield  {journal} {\bibinfo  {journal}
  {Z. Phys. C}\ }\textbf {\bibinfo {volume} {74}},\ \bibinfo {pages} {57}
  (\bibinfo {year} {1997})},\ \bibinfo {note} {[Erratum: Z.Phys.C 75, 580
  (1997)]}\BibitemShut {NoStop}%
\bibitem [{\citenamefont {Aad}\ \emph {et~al.}(2019)\citenamefont {Aad} \emph
  {et~al.}}]{ATLAS:2019kpx}%
  \BibitemOpen
  \bibfield  {author} {\bibinfo {author} {\bibfnamefont {G.}~\bibnamefont
  {Aad}} \emph {et~al.} (\bibinfo {collaboration} {ATLAS}),\ }\href {\doibase
  10.1007/JHEP10(2019)265} {\bibfield  {journal} {\bibinfo  {journal} {JHEP}\
  }\textbf {\bibinfo {volume} {10}},\ \bibinfo {pages} {265} (\bibinfo {year}
  {2019})},\ \Eprint {http://arxiv.org/abs/1905.09787} {arXiv:1905.09787
  [hep-ex]} \BibitemShut {NoStop}%
\bibitem [{\citenamefont {ATLAS}()}]{ATLAS:2022atq}%
  \BibitemOpen
  \bibfield  {author} {\bibinfo {author} {\bibnamefont {ATLAS}},\ }\href@noop
  {} {\ }\Eprint {http://arxiv.org/abs/2204.11988} {arXiv:2204.11988 [hep-ex]}
  \BibitemShut {NoStop}%
\bibitem [{\citenamefont {Tumasyan}\ \emph
  {et~al.}(2022{\natexlab{a}})\citenamefont {Tumasyan} \emph
  {et~al.}}]{CMS:2022fut}%
  \BibitemOpen
  \bibfield  {author} {\bibinfo {author} {\bibfnamefont {A.}~\bibnamefont
  {Tumasyan}} \emph {et~al.} (\bibinfo {collaboration} {CMS}),\ }\href
  {\doibase 10.1007/JHEP07(2022)081} {\bibfield  {journal} {\bibinfo  {journal}
  {JHEP}\ }\textbf {\bibinfo {volume} {07}},\ \bibinfo {pages} {081} (\bibinfo
  {year} {2022}{\natexlab{a}})},\ \Eprint {http://arxiv.org/abs/2201.05578}
  {arXiv:2201.05578 [hep-ex]} \BibitemShut {NoStop}%
\bibitem [{\citenamefont {Naredo-Tuero}()}]{dani_in_progress}%
  \BibitemOpen
  \bibfield  {author} {\bibinfo {author} {\bibfnamefont {D.}~\bibnamefont
  {Naredo-Tuero}},\ }\href@noop {} {}\bibinfo {note} {Private
  communication}\BibitemShut {NoStop}%
\bibitem [{\citenamefont {Daum}\ \emph {et~al.}(1987)\citenamefont {Daum},
  \citenamefont {Jost}, \citenamefont {Marshall}, \citenamefont {Minehart},
  \citenamefont {Stephens},\ and\ \citenamefont {Ziock}}]{Daum:1987bg}%
  \BibitemOpen
  \bibfield  {author} {\bibinfo {author} {\bibfnamefont {M.}~\bibnamefont
  {Daum}}, \bibinfo {author} {\bibfnamefont {B.}~\bibnamefont {Jost}}, \bibinfo
  {author} {\bibfnamefont {R.}~\bibnamefont {Marshall}}, \bibinfo {author}
  {\bibfnamefont {R.}~\bibnamefont {Minehart}}, \bibinfo {author}
  {\bibfnamefont {W.}~\bibnamefont {Stephens}}, \ and\ \bibinfo {author}
  {\bibfnamefont {K.}~\bibnamefont {Ziock}},\ }\href {\doibase
  10.1103/PhysRevD.36.2624} {\bibfield  {journal} {\bibinfo  {journal} {Phys.
  Rev. D}\ }\textbf {\bibinfo {volume} {36}},\ \bibinfo {pages} {2624}
  (\bibinfo {year} {1987})}\BibitemShut {NoStop}%
\bibitem [{\citenamefont {Aguilar-Arevalo}\ \emph {et~al.}(2019)\citenamefont
  {Aguilar-Arevalo} \emph {et~al.}}]{Aguilar-Arevalo:2019owf}%
  \BibitemOpen
  \bibfield  {author} {\bibinfo {author} {\bibfnamefont {A.}~\bibnamefont
  {Aguilar-Arevalo}} \emph {et~al.} (\bibinfo {collaboration} {PIENU}),\ }\href
  {\doibase 10.1016/j.physletb.2019.134980} {\bibfield  {journal} {\bibinfo
  {journal} {Phys. Lett. B}\ }\textbf {\bibinfo {volume} {798}},\ \bibinfo
  {pages} {134980} (\bibinfo {year} {2019})},\ \Eprint
  {http://arxiv.org/abs/1904.03269} {arXiv:1904.03269 [hep-ex]} \BibitemShut
  {NoStop}%
\bibitem [{\citenamefont {Bernardi}\ \emph {et~al.}(1986)\citenamefont
  {Bernardi} \emph {et~al.}}]{Bernardi:1985ny}%
  \BibitemOpen
  \bibfield  {author} {\bibinfo {author} {\bibfnamefont {G.}~\bibnamefont
  {Bernardi}} \emph {et~al.},\ }\href {\doibase 10.1016/0370-2693(86)91602-3}
  {\bibfield  {journal} {\bibinfo  {journal} {Phys. Lett. B}\ }\textbf
  {\bibinfo {volume} {166}},\ \bibinfo {pages} {479} (\bibinfo {year}
  {1986})}\BibitemShut {NoStop}%
\bibitem [{\citenamefont {Bernardi}\ \emph {et~al.}(1988)\citenamefont
  {Bernardi} \emph {et~al.}}]{Bernardi:1987ek}%
  \BibitemOpen
  \bibfield  {author} {\bibinfo {author} {\bibfnamefont {G.}~\bibnamefont
  {Bernardi}} \emph {et~al.},\ }\href {\doibase 10.1016/0370-2693(88)90563-1}
  {\bibfield  {journal} {\bibinfo  {journal} {Phys. Lett. B}\ }\textbf
  {\bibinfo {volume} {203}},\ \bibinfo {pages} {332} (\bibinfo {year}
  {1988})}\BibitemShut {NoStop}%
\bibitem [{\citenamefont {Arg\"uelles}\ \emph {et~al.}(2022)\citenamefont
  {Arg\"uelles}, \citenamefont {Foppiani},\ and\ \citenamefont
  {Hostert}}]{Arguelles:2021dqn}%
  \BibitemOpen
  \bibfield  {author} {\bibinfo {author} {\bibfnamefont {C.~A.}\ \bibnamefont
  {Arg\"uelles}}, \bibinfo {author} {\bibfnamefont {N.}~\bibnamefont
  {Foppiani}}, \ and\ \bibinfo {author} {\bibfnamefont {M.}~\bibnamefont
  {Hostert}},\ }\href {\doibase 10.1103/PhysRevD.105.095006} {\bibfield
  {journal} {\bibinfo  {journal} {Phys. Rev. D}\ }\textbf {\bibinfo {volume}
  {105}},\ \bibinfo {pages} {095006} (\bibinfo {year} {2022})},\ \Eprint
  {http://arxiv.org/abs/2109.03831} {arXiv:2109.03831 [hep-ph]} \BibitemShut
  {NoStop}%
\bibitem [{\citenamefont {Gorbunov}\ \emph {et~al.}(2022)\citenamefont
  {Gorbunov}, \citenamefont {Krasnov},\ and\ \citenamefont
  {Suvorov}}]{Gorbunov:2021wua}%
  \BibitemOpen
  \bibfield  {author} {\bibinfo {author} {\bibfnamefont {D.}~\bibnamefont
  {Gorbunov}}, \bibinfo {author} {\bibfnamefont {I.}~\bibnamefont {Krasnov}}, \
  and\ \bibinfo {author} {\bibfnamefont {S.}~\bibnamefont {Suvorov}},\ }\href
  {\doibase 10.1016/j.physletb.2022.137173} {\bibfield  {journal} {\bibinfo
  {journal} {Phys. Lett. B}\ }\textbf {\bibinfo {volume} {830}},\ \bibinfo
  {pages} {137173} (\bibinfo {year} {2022})},\ \Eprint
  {http://arxiv.org/abs/2112.06800} {arXiv:2112.06800 [hep-ph]} \BibitemShut
  {NoStop}%
\bibitem [{\citenamefont {Kelly}\ and\ \citenamefont
  {Machado}(2021)}]{Kelly:2021xbv}%
  \BibitemOpen
  \bibfield  {author} {\bibinfo {author} {\bibfnamefont {K.~J.}\ \bibnamefont
  {Kelly}}\ and\ \bibinfo {author} {\bibfnamefont {P.~A.~N.}\ \bibnamefont
  {Machado}},\ }\href {\doibase 10.1103/PhysRevD.104.055015} {\bibfield
  {journal} {\bibinfo  {journal} {Phys. Rev. D}\ }\textbf {\bibinfo {volume}
  {104}},\ \bibinfo {pages} {055015} (\bibinfo {year} {2021})},\ \Eprint
  {http://arxiv.org/abs/2106.06548} {arXiv:2106.06548 [hep-ph]} \BibitemShut
  {NoStop}%
\bibitem [{\citenamefont {Hayano}\ \emph {et~al.}(1982)\citenamefont {Hayano}
  \emph {et~al.}}]{Hayano:1982wu}%
  \BibitemOpen
  \bibfield  {author} {\bibinfo {author} {\bibfnamefont {R.}~\bibnamefont
  {Hayano}} \emph {et~al.},\ }\href {\doibase 10.1103/PhysRevLett.49.1305}
  {\bibfield  {journal} {\bibinfo  {journal} {Phys. Rev. Lett.}\ }\textbf
  {\bibinfo {volume} {49}},\ \bibinfo {pages} {1305} (\bibinfo {year}
  {1982})}\BibitemShut {NoStop}%
\bibitem [{\citenamefont {Yamazaki}\ \emph {et~al.}(1984)\citenamefont
  {Yamazaki} \emph {et~al.}}]{Yamazaki:1984sj}%
  \BibitemOpen
  \bibfield  {author} {\bibinfo {author} {\bibfnamefont {T.}~\bibnamefont
  {Yamazaki}} \emph {et~al.},\ }\href@noop {} {\bibfield  {journal} {\bibinfo
  {journal} {Conf. Proc. C}\ }\textbf {\bibinfo {volume} {840719}},\ \bibinfo
  {pages} {262} (\bibinfo {year} {1984})}\BibitemShut {NoStop}%
\bibitem [{\citenamefont {Artamonov}\ \emph {et~al.}(2009)\citenamefont
  {Artamonov} \emph {et~al.}}]{Artamonov:2009sz}%
  \BibitemOpen
  \bibfield  {author} {\bibinfo {author} {\bibfnamefont {A.}~\bibnamefont
  {Artamonov}} \emph {et~al.} (\bibinfo {collaboration} {BNL-E949}),\ }\href
  {\doibase 10.1103/PhysRevD.79.092004} {\bibfield  {journal} {\bibinfo
  {journal} {Phys. Rev. D}\ }\textbf {\bibinfo {volume} {79}},\ \bibinfo
  {pages} {092004} (\bibinfo {year} {2009})},\ \Eprint
  {http://arxiv.org/abs/0903.0030} {arXiv:0903.0030 [hep-ex]} \BibitemShut
  {NoStop}%
\bibitem [{\citenamefont {Cortina~Gil}\ \emph {et~al.}(2021)\citenamefont
  {Cortina~Gil} \emph {et~al.}}]{NA62:2021bji}%
  \BibitemOpen
  \bibfield  {author} {\bibinfo {author} {\bibfnamefont {E.}~\bibnamefont
  {Cortina~Gil}} \emph {et~al.} (\bibinfo {collaboration} {NA62}),\ }\href
  {\doibase 10.1016/j.physletb.2021.136259} {\bibfield  {journal} {\bibinfo
  {journal} {Phys. Lett. B}\ }\textbf {\bibinfo {volume} {816}},\ \bibinfo
  {pages} {136259} (\bibinfo {year} {2021})},\ \Eprint
  {http://arxiv.org/abs/2101.12304} {arXiv:2101.12304 [hep-ex]} \BibitemShut
  {NoStop}%
\bibitem [{\citenamefont {Cooper-Sarkar}\ \emph
  {et~al.}(1985{\natexlab{b}})\citenamefont {Cooper-Sarkar} \emph
  {et~al.}}]{WA66:1985mfx}%
  \BibitemOpen
  \bibfield  {author} {\bibinfo {author} {\bibfnamefont {A.~M.}\ \bibnamefont
  {Cooper-Sarkar}} \emph {et~al.} (\bibinfo {collaboration} {WA66}),\ }\href
  {\doibase 10.1016/0370-2693(85)91493-5} {\bibfield  {journal} {\bibinfo
  {journal} {Phys. Lett. B}\ }\textbf {\bibinfo {volume} {160}},\ \bibinfo
  {pages} {207} (\bibinfo {year} {1985}{\natexlab{b}})}\BibitemShut {NoStop}%
\bibitem [{\citenamefont {Vaitaitis}\ \emph {et~al.}(1999)\citenamefont
  {Vaitaitis} \emph {et~al.}}]{Vaitaitis:1999wq}%
  \BibitemOpen
  \bibfield  {author} {\bibinfo {author} {\bibfnamefont {A.}~\bibnamefont
  {Vaitaitis}} \emph {et~al.} (\bibinfo {collaboration} {NuTeV, E815}),\ }\href
  {\doibase 10.1103/PhysRevLett.83.4943} {\bibfield  {journal} {\bibinfo
  {journal} {Phys. Rev. Lett.}\ }\textbf {\bibinfo {volume} {83}},\ \bibinfo
  {pages} {4943} (\bibinfo {year} {1999})},\ \Eprint
  {http://arxiv.org/abs/hep-ex/9908011} {arXiv:hep-ex/9908011} \BibitemShut
  {NoStop}%
\bibitem [{\citenamefont {Plestid}(2021)}]{Plestid:2020ssy}%
  \BibitemOpen
  \bibfield  {author} {\bibinfo {author} {\bibfnamefont {R.}~\bibnamefont
  {Plestid}},\ }\href {\doibase 10.1103/PhysRevD.104.075028} {\bibfield
  {journal} {\bibinfo  {journal} {Phys. Rev. D}\ }\textbf {\bibinfo {volume}
  {104}},\ \bibinfo {pages} {075028} (\bibinfo {year} {2021})},\ \bibinfo
  {note} {[Erratum: Phys.Rev.D 105, 099901 (2022), Erratum: Phys.Rev.D 105,
  099901 (2022)]},\ \Eprint {http://arxiv.org/abs/2010.09523} {arXiv:2010.09523
  [hep-ph]} \BibitemShut {NoStop}%
\bibitem [{\citenamefont {Gustafson}\ \emph {et~al.}(2022)\citenamefont
  {Gustafson}, \citenamefont {Plestid},\ and\ \citenamefont
  {Shoemaker}}]{Gustafson:2022rsz}%
  \BibitemOpen
  \bibfield  {author} {\bibinfo {author} {\bibfnamefont {R.~A.}\ \bibnamefont
  {Gustafson}}, \bibinfo {author} {\bibfnamefont {R.}~\bibnamefont {Plestid}},
  \ and\ \bibinfo {author} {\bibfnamefont {I.~M.}\ \bibnamefont {Shoemaker}},\
  }\href {\doibase 10.1103/PhysRevD.106.095037} {\bibfield  {journal} {\bibinfo
   {journal} {Phys. Rev. D}\ }\textbf {\bibinfo {volume} {106}},\ \bibinfo
  {pages} {095037} (\bibinfo {year} {2022})},\ \Eprint
  {http://arxiv.org/abs/2205.02234} {arXiv:2205.02234 [hep-ph]} \BibitemShut
  {NoStop}%
\bibitem [{\citenamefont {Dentler}\ \emph {et~al.}(2018)\citenamefont
  {Dentler}, \citenamefont {Hern\'andez-Cabezudo}, \citenamefont {Kopp},
  \citenamefont {Machado}, \citenamefont {Maltoni}, \citenamefont
  {Martinez-Soler},\ and\ \citenamefont {Schwetz}}]{Dentler:2018sju}%
  \BibitemOpen
  \bibfield  {author} {\bibinfo {author} {\bibfnamefont {M.}~\bibnamefont
  {Dentler}}, \bibinfo {author} {\bibfnamefont {A.}~\bibnamefont
  {Hern\'andez-Cabezudo}}, \bibinfo {author} {\bibfnamefont {J.}~\bibnamefont
  {Kopp}}, \bibinfo {author} {\bibfnamefont {P.~A.~N.}\ \bibnamefont
  {Machado}}, \bibinfo {author} {\bibfnamefont {M.}~\bibnamefont {Maltoni}},
  \bibinfo {author} {\bibfnamefont {I.}~\bibnamefont {Martinez-Soler}}, \ and\
  \bibinfo {author} {\bibfnamefont {T.}~\bibnamefont {Schwetz}},\ }\href
  {\doibase 10.1007/JHEP08(2018)010} {\bibfield  {journal} {\bibinfo  {journal}
  {JHEP}\ }\textbf {\bibinfo {volume} {08}},\ \bibinfo {pages} {010} (\bibinfo
  {year} {2018})},\ \Eprint {http://arxiv.org/abs/1803.10661} {arXiv:1803.10661
  [hep-ph]} \BibitemShut {NoStop}%
\bibitem [{\citenamefont {Arg\"uelles}\ \emph {et~al.}(2023)\citenamefont
  {Arg\"uelles} \emph {et~al.}}]{Arguelles:2022tki}%
  \BibitemOpen
  \bibfield  {author} {\bibinfo {author} {\bibfnamefont {C.~A.}\ \bibnamefont
  {Arg\"uelles}} \emph {et~al.},\ }\href {\doibase
  10.1140/epjc/s10052-022-11049-7} {\bibfield  {journal} {\bibinfo  {journal}
  {Eur. Phys. J. C}\ }\textbf {\bibinfo {volume} {83}},\ \bibinfo {pages} {15}
  (\bibinfo {year} {2023})},\ \Eprint {http://arxiv.org/abs/2203.10811}
  {arXiv:2203.10811 [hep-ph]} \BibitemShut {NoStop}%
\bibitem [{\citenamefont {Blennow}\ \emph {et~al.}(2017)\citenamefont
  {Blennow}, \citenamefont {Coloma}, \citenamefont {Fernandez-Martinez},
  \citenamefont {Hernandez-Garcia},\ and\ \citenamefont
  {Lopez-Pavon}}]{Blennow:2016jkn}%
  \BibitemOpen
  \bibfield  {author} {\bibinfo {author} {\bibfnamefont {M.}~\bibnamefont
  {Blennow}}, \bibinfo {author} {\bibfnamefont {P.}~\bibnamefont {Coloma}},
  \bibinfo {author} {\bibfnamefont {E.}~\bibnamefont {Fernandez-Martinez}},
  \bibinfo {author} {\bibfnamefont {J.}~\bibnamefont {Hernandez-Garcia}}, \
  and\ \bibinfo {author} {\bibfnamefont {J.}~\bibnamefont {Lopez-Pavon}},\
  }\href {\doibase 10.1007/JHEP04(2017)153} {\bibfield  {journal} {\bibinfo
  {journal} {JHEP}\ }\textbf {\bibinfo {volume} {04}},\ \bibinfo {pages} {153}
  (\bibinfo {year} {2017})},\ \Eprint {http://arxiv.org/abs/1609.08637}
  {arXiv:1609.08637 [hep-ph]} \BibitemShut {NoStop}%
\bibitem [{\citenamefont {Orloff}\ \emph {et~al.}(2002)\citenamefont {Orloff},
  \citenamefont {Rozanov},\ and\ \citenamefont {Santoni}}]{Orloff:2002de}%
  \BibitemOpen
  \bibfield  {author} {\bibinfo {author} {\bibfnamefont {J.}~\bibnamefont
  {Orloff}}, \bibinfo {author} {\bibfnamefont {A.~N.}\ \bibnamefont {Rozanov}},
  \ and\ \bibinfo {author} {\bibfnamefont {C.}~\bibnamefont {Santoni}},\ }\href
  {\doibase 10.1016/S0370-2693(02)02769-7} {\bibfield  {journal} {\bibinfo
  {journal} {Phys. Lett. B}\ }\textbf {\bibinfo {volume} {550}},\ \bibinfo
  {pages} {8} (\bibinfo {year} {2002})},\ \Eprint
  {http://arxiv.org/abs/hep-ph/0208075} {arXiv:hep-ph/0208075} \BibitemShut
  {NoStop}%
\bibitem [{\citenamefont {Boiarska}\ \emph {et~al.}(2021)\citenamefont
  {Boiarska}, \citenamefont {Boyarsky}, \citenamefont {Mikulenko},\ and\
  \citenamefont {Ovchynnikov}}]{Boiarska:2021yho}%
  \BibitemOpen
  \bibfield  {author} {\bibinfo {author} {\bibfnamefont {I.}~\bibnamefont
  {Boiarska}}, \bibinfo {author} {\bibfnamefont {A.}~\bibnamefont {Boyarsky}},
  \bibinfo {author} {\bibfnamefont {O.}~\bibnamefont {Mikulenko}}, \ and\
  \bibinfo {author} {\bibfnamefont {M.}~\bibnamefont {Ovchynnikov}},\ }\href
  {\doibase 10.1103/PhysRevD.104.095019} {\bibfield  {journal} {\bibinfo
  {journal} {Phys. Rev. D}\ }\textbf {\bibinfo {volume} {104}},\ \bibinfo
  {pages} {095019} (\bibinfo {year} {2021})},\ \Eprint
  {http://arxiv.org/abs/2107.14685} {arXiv:2107.14685 [hep-ph]} \BibitemShut
  {NoStop}%
\bibitem [{\citenamefont {Acciarri}\ \emph {et~al.}(2021)\citenamefont
  {Acciarri} \emph {et~al.}}]{ArgoNeuT:2021clc}%
  \BibitemOpen
  \bibfield  {author} {\bibinfo {author} {\bibfnamefont {R.}~\bibnamefont
  {Acciarri}} \emph {et~al.} (\bibinfo {collaboration} {ArgoNeuT}),\ }\href
  {\doibase 10.1103/PhysRevLett.127.121801} {\bibfield  {journal} {\bibinfo
  {journal} {Phys. Rev. Lett.}\ }\textbf {\bibinfo {volume} {127}},\ \bibinfo
  {pages} {121801} (\bibinfo {year} {2021})},\ \Eprint
  {http://arxiv.org/abs/2106.13684} {arXiv:2106.13684 [hep-ex]} \BibitemShut
  {NoStop}%
\bibitem [{\citenamefont {Lees}\ \emph {et~al.}(2022)\citenamefont {Lees} \emph
  {et~al.}}]{BaBar:2022cqj}%
  \BibitemOpen
  \bibfield  {author} {\bibinfo {author} {\bibfnamefont {J.~P.}\ \bibnamefont
  {Lees}} \emph {et~al.} (\bibinfo {collaboration} {BaBar}),\ }\href {\doibase
  10.1103/PhysRevD.107.052009} {\bibfield  {journal} {\bibinfo  {journal}
  {Phys. Rev. D}\ }\textbf {\bibinfo {volume} {107}},\ \bibinfo {pages}
  {052009} (\bibinfo {year} {2022})},\ \Eprint
  {http://arxiv.org/abs/2207.09575} {arXiv:2207.09575 [hep-ex]} \BibitemShut
  {NoStop}%
\bibitem [{\citenamefont {Vincent}\ \emph {et~al.}(2015)\citenamefont
  {Vincent}, \citenamefont {Martinez}, \citenamefont {Hern\'andez},
  \citenamefont {Lattanzi},\ and\ \citenamefont {Mena}}]{Vincent:2014rja}%
  \BibitemOpen
  \bibfield  {author} {\bibinfo {author} {\bibfnamefont {A.~C.}\ \bibnamefont
  {Vincent}}, \bibinfo {author} {\bibfnamefont {E.~F.}\ \bibnamefont
  {Martinez}}, \bibinfo {author} {\bibfnamefont {P.}~\bibnamefont
  {Hern\'andez}}, \bibinfo {author} {\bibfnamefont {M.}~\bibnamefont
  {Lattanzi}}, \ and\ \bibinfo {author} {\bibfnamefont {O.}~\bibnamefont
  {Mena}},\ }\href {\doibase 10.1088/1475-7516/2015/04/006} {\bibfield
  {journal} {\bibinfo  {journal} {JCAP}\ }\textbf {\bibinfo {volume} {04}},\
  \bibinfo {pages} {006} (\bibinfo {year} {2015})},\ \Eprint
  {http://arxiv.org/abs/1408.1956} {arXiv:1408.1956 [astro-ph.CO]} \BibitemShut
  {NoStop}%
\bibitem [{\citenamefont {Langhoff}\ \emph {et~al.}(2022)\citenamefont
  {Langhoff}, \citenamefont {Outmezguine},\ and\ \citenamefont
  {Rodd}}]{Langhoff:2022bij}%
  \BibitemOpen
  \bibfield  {author} {\bibinfo {author} {\bibfnamefont {K.}~\bibnamefont
  {Langhoff}}, \bibinfo {author} {\bibfnamefont {N.~J.}\ \bibnamefont
  {Outmezguine}}, \ and\ \bibinfo {author} {\bibfnamefont {N.~L.}\ \bibnamefont
  {Rodd}},\ }\href {\doibase 10.1103/PhysRevLett.129.241101} {\bibfield
  {journal} {\bibinfo  {journal} {Phys. Rev. Lett.}\ }\textbf {\bibinfo
  {volume} {129}},\ \bibinfo {pages} {241101} (\bibinfo {year} {2022})},\
  \Eprint {http://arxiv.org/abs/2209.06216} {arXiv:2209.06216 [hep-ph]}
  \BibitemShut {NoStop}%
\bibitem [{\citenamefont {Dolgov}\ \emph {et~al.}(2000)\citenamefont {Dolgov},
  \citenamefont {Hansen}, \citenamefont {Raffelt},\ and\ \citenamefont
  {Semikoz}}]{Dolgov:2000pj}%
  \BibitemOpen
  \bibfield  {author} {\bibinfo {author} {\bibfnamefont {A.~D.}\ \bibnamefont
  {Dolgov}}, \bibinfo {author} {\bibfnamefont {S.~H.}\ \bibnamefont {Hansen}},
  \bibinfo {author} {\bibfnamefont {G.}~\bibnamefont {Raffelt}}, \ and\
  \bibinfo {author} {\bibfnamefont {D.~V.}\ \bibnamefont {Semikoz}},\ }\href
  {\doibase 10.1016/S0550-3213(00)00203-0} {\bibfield  {journal} {\bibinfo
  {journal} {Nucl. Phys. B}\ }\textbf {\bibinfo {volume} {580}},\ \bibinfo
  {pages} {331} (\bibinfo {year} {2000})},\ \Eprint
  {http://arxiv.org/abs/hep-ph/0002223} {arXiv:hep-ph/0002223} \BibitemShut
  {NoStop}%
\bibitem [{\citenamefont {Ruchayskiy}\ and\ \citenamefont
  {Ivashko}(2012{\natexlab{b}})}]{Ruchayskiy:2012si}%
  \BibitemOpen
  \bibfield  {author} {\bibinfo {author} {\bibfnamefont {O.}~\bibnamefont
  {Ruchayskiy}}\ and\ \bibinfo {author} {\bibfnamefont {A.}~\bibnamefont
  {Ivashko}},\ }\href {\doibase 10.1088/1475-7516/2012/10/014} {\bibfield
  {journal} {\bibinfo  {journal} {JCAP}\ }\textbf {\bibinfo {volume} {10}},\
  \bibinfo {pages} {014} (\bibinfo {year} {2012}{\natexlab{b}})},\ \Eprint
  {http://arxiv.org/abs/1202.2841} {arXiv:1202.2841 [hep-ph]} \BibitemShut
  {NoStop}%
\bibitem [{\citenamefont {Gelmini}\ \emph {et~al.}(2020)\citenamefont
  {Gelmini}, \citenamefont {Kawasaki}, \citenamefont {Kusenko}, \citenamefont
  {Murai},\ and\ \citenamefont {Takhistov}}]{Gelmini:2020ekg}%
  \BibitemOpen
  \bibfield  {author} {\bibinfo {author} {\bibfnamefont {G.~B.}\ \bibnamefont
  {Gelmini}}, \bibinfo {author} {\bibfnamefont {M.}~\bibnamefont {Kawasaki}},
  \bibinfo {author} {\bibfnamefont {A.}~\bibnamefont {Kusenko}}, \bibinfo
  {author} {\bibfnamefont {K.}~\bibnamefont {Murai}}, \ and\ \bibinfo {author}
  {\bibfnamefont {V.}~\bibnamefont {Takhistov}},\ }\href {\doibase
  10.1088/1475-7516/2020/09/051} {\bibfield  {journal} {\bibinfo  {journal}
  {JCAP}\ }\textbf {\bibinfo {volume} {09}},\ \bibinfo {pages} {051} (\bibinfo
  {year} {2020})},\ \Eprint {http://arxiv.org/abs/2005.06721} {arXiv:2005.06721
  [hep-ph]} \BibitemShut {NoStop}%
\bibitem [{\citenamefont {Sabti}\ \emph {et~al.}(2020)\citenamefont {Sabti},
  \citenamefont {Magalich},\ and\ \citenamefont {Filimonova}}]{Sabti:2020yrt}%
  \BibitemOpen
  \bibfield  {author} {\bibinfo {author} {\bibfnamefont {N.}~\bibnamefont
  {Sabti}}, \bibinfo {author} {\bibfnamefont {A.}~\bibnamefont {Magalich}}, \
  and\ \bibinfo {author} {\bibfnamefont {A.}~\bibnamefont {Filimonova}},\
  }\href {\doibase 10.1088/1475-7516/2020/11/056} {\bibfield  {journal}
  {\bibinfo  {journal} {JCAP}\ }\textbf {\bibinfo {volume} {11}},\ \bibinfo
  {pages} {056} (\bibinfo {year} {2020})},\ \Eprint
  {http://arxiv.org/abs/2006.07387} {arXiv:2006.07387 [hep-ph]} \BibitemShut
  {NoStop}%
\bibitem [{\citenamefont {Boyarsky}\ \emph {et~al.}(2021)\citenamefont
  {Boyarsky}, \citenamefont {Ovchynnikov}, \citenamefont {Ruchayskiy},\ and\
  \citenamefont {Syvolap}}]{Boyarsky:2020dzc}%
  \BibitemOpen
  \bibfield  {author} {\bibinfo {author} {\bibfnamefont {A.}~\bibnamefont
  {Boyarsky}}, \bibinfo {author} {\bibfnamefont {M.}~\bibnamefont
  {Ovchynnikov}}, \bibinfo {author} {\bibfnamefont {O.}~\bibnamefont
  {Ruchayskiy}}, \ and\ \bibinfo {author} {\bibfnamefont {V.}~\bibnamefont
  {Syvolap}},\ }\href {\doibase 10.1103/PhysRevD.104.023517} {\bibfield
  {journal} {\bibinfo  {journal} {Phys. Rev. D}\ }\textbf {\bibinfo {volume}
  {104}},\ \bibinfo {pages} {023517} (\bibinfo {year} {2021})},\ \Eprint
  {http://arxiv.org/abs/2008.00749} {arXiv:2008.00749 [hep-ph]} \BibitemShut
  {NoStop}%
\bibitem [{\citenamefont {ATLAS}(2022)}]{ATLAS:2022vkf}%
  \BibitemOpen
  \bibfield  {author} {\bibinfo {author} {\bibnamefont {ATLAS}},\ }\href
  {\doibase 10.1038/s41586-022-04893-w} {\bibfield  {journal} {\bibinfo
  {journal} {Nature}\ }\textbf {\bibinfo {volume} {607}},\ \bibinfo {pages}
  {52} (\bibinfo {year} {2022})},\ \bibinfo {note} {[Erratum: Nature 612, E24
  (2022)]},\ \Eprint {http://arxiv.org/abs/2207.00092} {arXiv:2207.00092
  [hep-ex]} \BibitemShut {NoStop}%
\bibitem [{\citenamefont {Tumasyan}\ \emph
  {et~al.}(2022{\natexlab{b}})\citenamefont {Tumasyan} \emph
  {et~al.}}]{CMS:2022dwd}%
  \BibitemOpen
  \bibfield  {author} {\bibinfo {author} {\bibfnamefont {A.}~\bibnamefont
  {Tumasyan}} \emph {et~al.} (\bibinfo {collaboration} {CMS}),\ }\href
  {\doibase 10.1038/s41586-022-04892-x} {\bibfield  {journal} {\bibinfo
  {journal} {Nature}\ }\textbf {\bibinfo {volume} {607}},\ \bibinfo {pages}
  {60} (\bibinfo {year} {2022}{\natexlab{b}})},\ \Eprint
  {http://arxiv.org/abs/2207.00043} {arXiv:2207.00043 [hep-ex]} \BibitemShut
  {NoStop}%
\bibitem [{\citenamefont {Fern\'andez-Mart\'\i{}nez}\ \emph
  {et~al.}(2022)\citenamefont {Fern\'andez-Mart\'\i{}nez}, \citenamefont
  {L\'opez-Pav\'on}, \citenamefont {No}, \citenamefont {Ota},\ and\
  \citenamefont {Rosauro-Alcaraz}}]{Fernandez-Martinez:2022stj}%
  \BibitemOpen
  \bibfield  {author} {\bibinfo {author} {\bibfnamefont {E.}~\bibnamefont
  {Fern\'andez-Mart\'\i{}nez}}, \bibinfo {author} {\bibfnamefont
  {J.}~\bibnamefont {L\'opez-Pav\'on}}, \bibinfo {author} {\bibfnamefont
  {J.~M.}\ \bibnamefont {No}}, \bibinfo {author} {\bibfnamefont
  {T.}~\bibnamefont {Ota}}, \ and\ \bibinfo {author} {\bibfnamefont
  {S.}~\bibnamefont {Rosauro-Alcaraz}},\ }\href@noop {} {\  (\bibinfo {year}
  {2022})},\ \Eprint {http://arxiv.org/abs/2210.16279} {arXiv:2210.16279
  [hep-ph]} \BibitemShut {NoStop}%
\bibitem [{\citenamefont {Feldman}\ and\ \citenamefont
  {Cousins}(1998)}]{Feldman:1997qc}%
  \BibitemOpen
  \bibfield  {author} {\bibinfo {author} {\bibfnamefont {G.~J.}\ \bibnamefont
  {Feldman}}\ and\ \bibinfo {author} {\bibfnamefont {R.~D.}\ \bibnamefont
  {Cousins}},\ }\href {\doibase 10.1103/PhysRevD.57.3873} {\bibfield  {journal}
  {\bibinfo  {journal} {Phys. Rev. D}\ }\textbf {\bibinfo {volume} {57}},\
  \bibinfo {pages} {3873} (\bibinfo {year} {1998})},\ \Eprint
  {http://arxiv.org/abs/physics/9711021} {arXiv:physics/9711021} \BibitemShut
  {NoStop}%
\bibitem [{\citenamefont {Workman}\ \emph {et~al.}(2022)\citenamefont {Workman}
  \emph {et~al.}}]{ParticleDataGroup:2022pth}%
  \BibitemOpen
  \bibfield  {author} {\bibinfo {author} {\bibfnamefont {R.~L.}\ \bibnamefont
  {Workman}} \emph {et~al.} (\bibinfo {collaboration} {Particle Data Group}),\
  }\href {\doibase 10.1093/ptep/ptac097} {\bibfield  {journal} {\bibinfo
  {journal} {PTEP}\ }\textbf {\bibinfo {volume} {2022}},\ \bibinfo {pages}
  {083C01} (\bibinfo {year} {2022})}\BibitemShut {NoStop}%
\bibitem [{\citenamefont {Janot}\ and\ \citenamefont
  {Jadach}(2020)}]{Janot:2019oyi}%
  \BibitemOpen
  \bibfield  {author} {\bibinfo {author} {\bibfnamefont {P.}~\bibnamefont
  {Janot}}\ and\ \bibinfo {author} {\bibfnamefont {S.}~\bibnamefont {Jadach}},\
  }\href {\doibase 10.1016/j.physletb.2020.135319} {\bibfield  {journal}
  {\bibinfo  {journal} {Phys. Lett. B}\ }\textbf {\bibinfo {volume} {803}},\
  \bibinfo {pages} {135319} (\bibinfo {year} {2020})},\ \Eprint
  {http://arxiv.org/abs/1912.02067} {arXiv:1912.02067 [hep-ph]} \BibitemShut
  {NoStop}%
\bibitem [{\citenamefont {Marciano}\ and\ \citenamefont
  {Parsa}(1996)}]{Marciano:1996wy}%
  \BibitemOpen
  \bibfield  {author} {\bibinfo {author} {\bibfnamefont {W.~J.}\ \bibnamefont
  {Marciano}}\ and\ \bibinfo {author} {\bibfnamefont {Z.}~\bibnamefont
  {Parsa}},\ }\href {\doibase 10.1103/PhysRevD.53.R1} {\bibfield  {journal}
  {\bibinfo  {journal} {Phys. Rev. D}\ }\textbf {\bibinfo {volume} {53}},\
  \bibinfo {pages} {1} (\bibinfo {year} {1996})}\BibitemShut {NoStop}%
\bibitem [{\citenamefont {Chang}\ \emph {et~al.}(1998)\citenamefont {Chang},
  \citenamefont {Lebedev},\ and\ \citenamefont {Ng}}]{Chang:1997tq}%
  \BibitemOpen
  \bibfield  {author} {\bibinfo {author} {\bibfnamefont {L.~N.}\ \bibnamefont
  {Chang}}, \bibinfo {author} {\bibfnamefont {O.}~\bibnamefont {Lebedev}}, \
  and\ \bibinfo {author} {\bibfnamefont {J.~N.}\ \bibnamefont {Ng}},\ }\href
  {\doibase 10.1016/S0370-2693(98)01147-2} {\bibfield  {journal} {\bibinfo
  {journal} {Phys. Lett. B}\ }\textbf {\bibinfo {volume} {441}},\ \bibinfo
  {pages} {419} (\bibinfo {year} {1998})},\ \Eprint
  {http://arxiv.org/abs/hep-ph/9806487} {arXiv:hep-ph/9806487} \BibitemShut
  {NoStop}%
\bibitem [{\citenamefont {Cortina~Gil}\ \emph
  {et~al.}(2020{\natexlab{b}})\citenamefont {Cortina~Gil} \emph
  {et~al.}}]{CortinaGil:2020zwa}%
  \BibitemOpen
  \bibfield  {author} {\bibinfo {author} {\bibfnamefont {E.}~\bibnamefont
  {Cortina~Gil}} \emph {et~al.} (\bibinfo {collaboration} {NA62}),\ }\href
  {\doibase 10.1007/JHEP02(2021)201} {\bibfield  {journal} {\bibinfo  {journal}
  {JHEP}\ }\textbf {\bibinfo {volume} {02}},\ \bibinfo {pages} {201} (\bibinfo
  {year} {2020}{\natexlab{b}})},\ \Eprint {http://arxiv.org/abs/2010.07644}
  {arXiv:2010.07644 [hep-ex]} \BibitemShut {NoStop}%
\bibitem [{\citenamefont {Aubert}\ \emph {et~al.}(2009)\citenamefont {Aubert}
  \emph {et~al.}}]{BaBar:2009gco}%
  \BibitemOpen
  \bibfield  {author} {\bibinfo {author} {\bibfnamefont {B.}~\bibnamefont
  {Aubert}} \emph {et~al.} (\bibinfo {collaboration} {BaBar}),\ }\href
  {\doibase 10.1103/PhysRevLett.103.251801} {\bibfield  {journal} {\bibinfo
  {journal} {Phys. Rev. Lett.}\ }\textbf {\bibinfo {volume} {103}},\ \bibinfo
  {pages} {251801} (\bibinfo {year} {2009})},\ \Eprint
  {http://arxiv.org/abs/0908.2840} {arXiv:0908.2840 [hep-ex]} \BibitemShut
  {NoStop}%
\bibitem [{\citenamefont {Natale}(1991)}]{Natale:1990yx}%
  \BibitemOpen
  \bibfield  {author} {\bibinfo {author} {\bibfnamefont {A.~A.}\ \bibnamefont
  {Natale}},\ }\href {\doibase 10.1016/0370-2693(91)91237-P} {\bibfield
  {journal} {\bibinfo  {journal} {Phys. Lett. B}\ }\textbf {\bibinfo {volume}
  {258}},\ \bibinfo {pages} {227} (\bibinfo {year} {1991})}\BibitemShut
  {NoStop}%
\bibitem [{\citenamefont {Abdallah}\ \emph {et~al.}(2005)\citenamefont
  {Abdallah} \emph {et~al.}}]{DELPHI:2003dlq}%
  \BibitemOpen
  \bibfield  {author} {\bibinfo {author} {\bibfnamefont {J.}~\bibnamefont
  {Abdallah}} \emph {et~al.} (\bibinfo {collaboration} {DELPHI}),\ }\href
  {\doibase 10.1140/epjc/s2004-02051-8} {\bibfield  {journal} {\bibinfo
  {journal} {Eur. Phys. J. C}\ }\textbf {\bibinfo {volume} {38}},\ \bibinfo
  {pages} {395} (\bibinfo {year} {2005})},\ \Eprint
  {http://arxiv.org/abs/hep-ex/0406019} {arXiv:hep-ex/0406019} \BibitemShut
  {NoStop}%
\bibitem [{\citenamefont {Fox}\ \emph {et~al.}(2011)\citenamefont {Fox},
  \citenamefont {Harnik}, \citenamefont {Kopp},\ and\ \citenamefont
  {Tsai}}]{Fox:2011fx}%
  \BibitemOpen
  \bibfield  {author} {\bibinfo {author} {\bibfnamefont {P.~J.}\ \bibnamefont
  {Fox}}, \bibinfo {author} {\bibfnamefont {R.}~\bibnamefont {Harnik}},
  \bibinfo {author} {\bibfnamefont {J.}~\bibnamefont {Kopp}}, \ and\ \bibinfo
  {author} {\bibfnamefont {Y.}~\bibnamefont {Tsai}},\ }\href {\doibase
  10.1103/PhysRevD.84.014028} {\bibfield  {journal} {\bibinfo  {journal} {Phys.
  Rev. D}\ }\textbf {\bibinfo {volume} {84}},\ \bibinfo {pages} {014028}
  (\bibinfo {year} {2011})},\ \Eprint {http://arxiv.org/abs/1103.0240}
  {arXiv:1103.0240 [hep-ph]} \BibitemShut {NoStop}%
\bibitem [{\citenamefont {Hirata}\ \emph {et~al.}(1987)\citenamefont {Hirata}
  \emph {et~al.}}]{Hirata:1987hu}%
  \BibitemOpen
  \bibfield  {author} {\bibinfo {author} {\bibfnamefont {K.}~\bibnamefont
  {Hirata}} \emph {et~al.} (\bibinfo {collaboration} {Kamiokande-II}),\ }\href
  {\doibase 10.1103/PhysRevLett.58.1490} {\bibfield  {journal} {\bibinfo
  {journal} {Phys. Rev. Lett.}\ }\textbf {\bibinfo {volume} {58}},\ \bibinfo
  {pages} {1490} (\bibinfo {year} {1987})}\BibitemShut {NoStop}%
\bibitem [{\citenamefont {Bionta}\ \emph {et~al.}(1987)\citenamefont {Bionta}
  \emph {et~al.}}]{Bionta:1987qt}%
  \BibitemOpen
  \bibfield  {author} {\bibinfo {author} {\bibfnamefont {R.~M.}\ \bibnamefont
  {Bionta}} \emph {et~al.},\ }\href {\doibase 10.1103/PhysRevLett.58.1494}
  {\bibfield  {journal} {\bibinfo  {journal} {Phys. Rev. Lett.}\ }\textbf
  {\bibinfo {volume} {58}},\ \bibinfo {pages} {1494} (\bibinfo {year}
  {1987})}\BibitemShut {NoStop}%
\bibitem [{\citenamefont {Raffelt}\ and\ \citenamefont
  {Seckel}(1988)}]{Raffelt:1987yt}%
  \BibitemOpen
  \bibfield  {author} {\bibinfo {author} {\bibfnamefont {G.}~\bibnamefont
  {Raffelt}}\ and\ \bibinfo {author} {\bibfnamefont {D.}~\bibnamefont
  {Seckel}},\ }\href {\doibase 10.1103/PhysRevLett.60.1793} {\bibfield
  {journal} {\bibinfo  {journal} {Phys. Rev. Lett.}\ }\textbf {\bibinfo
  {volume} {60}},\ \bibinfo {pages} {1793} (\bibinfo {year}
  {1988})}\BibitemShut {NoStop}%
\bibitem [{\citenamefont {DeRocco}\ \emph {et~al.}(2019)\citenamefont
  {DeRocco}, \citenamefont {Graham}, \citenamefont {Kasen}, \citenamefont
  {Marques-Tavares},\ and\ \citenamefont {Rajendran}}]{DeRocco:2019jti}%
  \BibitemOpen
  \bibfield  {author} {\bibinfo {author} {\bibfnamefont {W.}~\bibnamefont
  {DeRocco}}, \bibinfo {author} {\bibfnamefont {P.~W.}\ \bibnamefont {Graham}},
  \bibinfo {author} {\bibfnamefont {D.}~\bibnamefont {Kasen}}, \bibinfo
  {author} {\bibfnamefont {G.}~\bibnamefont {Marques-Tavares}}, \ and\ \bibinfo
  {author} {\bibfnamefont {S.}~\bibnamefont {Rajendran}},\ }\href {\doibase
  10.1103/PhysRevD.100.075018} {\bibfield  {journal} {\bibinfo  {journal}
  {Phys. Rev. D}\ }\textbf {\bibinfo {volume} {100}},\ \bibinfo {pages}
  {075018} (\bibinfo {year} {2019})},\ \Eprint
  {http://arxiv.org/abs/1905.09284} {arXiv:1905.09284 [hep-ph]} \BibitemShut
  {NoStop}%
\bibitem [{\citenamefont {Magill}\ \emph {et~al.}(2018)\citenamefont {Magill},
  \citenamefont {Plestid}, \citenamefont {Pospelov},\ and\ \citenamefont
  {Tsai}}]{Magill:2018jla}%
  \BibitemOpen
  \bibfield  {author} {\bibinfo {author} {\bibfnamefont {G.}~\bibnamefont
  {Magill}}, \bibinfo {author} {\bibfnamefont {R.}~\bibnamefont {Plestid}},
  \bibinfo {author} {\bibfnamefont {M.}~\bibnamefont {Pospelov}}, \ and\
  \bibinfo {author} {\bibfnamefont {Y.-D.}\ \bibnamefont {Tsai}},\ }\href
  {\doibase 10.1103/PhysRevD.98.115015} {\bibfield  {journal} {\bibinfo
  {journal} {Phys. Rev. D}\ }\textbf {\bibinfo {volume} {98}},\ \bibinfo
  {pages} {115015} (\bibinfo {year} {2018})},\ \Eprint
  {http://arxiv.org/abs/1803.03262} {arXiv:1803.03262 [hep-ph]} \BibitemShut
  {NoStop}%
\bibitem [{\citenamefont {Brdar}\ \emph {et~al.}(2020)\citenamefont {Brdar},
  \citenamefont {Greljo}, \citenamefont {Kopp},\ and\ \citenamefont
  {Opferkuch}}]{Brdar:2020quo}%
  \BibitemOpen
  \bibfield  {author} {\bibinfo {author} {\bibfnamefont {V.}~\bibnamefont
  {Brdar}}, \bibinfo {author} {\bibfnamefont {A.}~\bibnamefont {Greljo}},
  \bibinfo {author} {\bibfnamefont {J.}~\bibnamefont {Kopp}}, \ and\ \bibinfo
  {author} {\bibfnamefont {T.}~\bibnamefont {Opferkuch}},\ }\href {\doibase
  10.1088/1475-7516/2021/01/039} {\bibfield  {journal} {\bibinfo  {journal}
  {JCAP}\ }\textbf {\bibinfo {volume} {01}},\ \bibinfo {pages} {039} (\bibinfo
  {year} {2020})},\ \Eprint {http://arxiv.org/abs/2007.15563} {arXiv:2007.15563
  [hep-ph]} \BibitemShut {NoStop}%
\bibitem [{\citenamefont {Coloma}\ \emph {et~al.}(2017)\citenamefont {Coloma},
  \citenamefont {Machado}, \citenamefont {Martinez-Soler},\ and\ \citenamefont
  {Shoemaker}}]{Coloma:2017ppo}%
  \BibitemOpen
  \bibfield  {author} {\bibinfo {author} {\bibfnamefont {P.}~\bibnamefont
  {Coloma}}, \bibinfo {author} {\bibfnamefont {P.~A.~N.}\ \bibnamefont
  {Machado}}, \bibinfo {author} {\bibfnamefont {I.}~\bibnamefont
  {Martinez-Soler}}, \ and\ \bibinfo {author} {\bibfnamefont {I.~M.}\
  \bibnamefont {Shoemaker}},\ }\href {\doibase 10.1103/PhysRevLett.119.201804}
  {\bibfield  {journal} {\bibinfo  {journal} {Phys. Rev. Lett.}\ }\textbf
  {\bibinfo {volume} {119}},\ \bibinfo {pages} {201804} (\bibinfo {year}
  {2017})},\ \Eprint {http://arxiv.org/abs/1707.08573} {arXiv:1707.08573
  [hep-ph]} \BibitemShut {NoStop}%
\bibitem [{\citenamefont {Gninenko}\ and\ \citenamefont
  {Krasnikov}(1999)}]{Gninenko:1998nn}%
  \BibitemOpen
  \bibfield  {author} {\bibinfo {author} {\bibfnamefont {S.~N.}\ \bibnamefont
  {Gninenko}}\ and\ \bibinfo {author} {\bibfnamefont {N.~V.}\ \bibnamefont
  {Krasnikov}},\ }\href {\doibase 10.1016/S0370-2693(99)00130-6} {\bibfield
  {journal} {\bibinfo  {journal} {Phys.Lett. B}\ }\textbf {\bibinfo {volume}
  {450}},\ \bibinfo {pages} {165} (\bibinfo {year} {1999})},\ \Eprint
  {http://arxiv.org/abs/hep-ph/9808370} {arXiv:hep-ph/9808370} \BibitemShut
  {NoStop}%
\bibitem [{\citenamefont {Schwienhorst}\ \emph {et~al.}(2001)\citenamefont
  {Schwienhorst} \emph {et~al.}}]{DONUT:2001zvi}%
  \BibitemOpen
  \bibfield  {author} {\bibinfo {author} {\bibfnamefont {R.}~\bibnamefont
  {Schwienhorst}} \emph {et~al.} (\bibinfo {collaboration} {DONUT}),\ }\href
  {\doibase 10.1016/S0370-2693(01)00746-8} {\bibfield  {journal} {\bibinfo
  {journal} {Phys. Lett. B}\ }\textbf {\bibinfo {volume} {513}},\ \bibinfo
  {pages} {23} (\bibinfo {year} {2001})},\ \Eprint
  {http://arxiv.org/abs/hep-ex/0102026} {arXiv:hep-ex/0102026} \BibitemShut
  {NoStop}%
\bibitem [{\citenamefont {Schwetz}\ \emph {et~al.}(2020)\citenamefont
  {Schwetz}, \citenamefont {Zhou},\ and\ \citenamefont
  {Zhu}}]{Schwetz:2020xra}%
  \BibitemOpen
  \bibfield  {author} {\bibinfo {author} {\bibfnamefont {T.}~\bibnamefont
  {Schwetz}}, \bibinfo {author} {\bibfnamefont {A.}~\bibnamefont {Zhou}}, \
  and\ \bibinfo {author} {\bibfnamefont {J.-Y.}\ \bibnamefont {Zhu}},\ }\href
  {\doibase 10.1007/JHEP07(2021)200} {\bibfield  {journal} {\bibinfo  {journal}
  {JHEP}\ }\textbf {\bibinfo {volume} {21}},\ \bibinfo {pages} {200} (\bibinfo
  {year} {2020})},\ \Eprint {http://arxiv.org/abs/2105.09699} {arXiv:2105.09699
  [hep-ph]} \BibitemShut {NoStop}%
\bibitem [{\citenamefont {Chivukula}\ and\ \citenamefont
  {Georgi}(1987)}]{Chivukula:1987py}%
  \BibitemOpen
  \bibfield  {author} {\bibinfo {author} {\bibfnamefont {R.~S.}\ \bibnamefont
  {Chivukula}}\ and\ \bibinfo {author} {\bibfnamefont {H.}~\bibnamefont
  {Georgi}},\ }\href {\doibase 10.1016/0370-2693(87)90713-1} {\bibfield
  {journal} {\bibinfo  {journal} {Phys. Lett. B}\ }\textbf {\bibinfo {volume}
  {188}},\ \bibinfo {pages} {99} (\bibinfo {year} {1987})}\BibitemShut
  {NoStop}%
\bibitem [{\citenamefont {D'Ambrosio}\ \emph {et~al.}(2002)\citenamefont
  {D'Ambrosio}, \citenamefont {Giudice}, \citenamefont {Isidori},\ and\
  \citenamefont {Strumia}}]{DAmbrosio:2002vsn}%
  \BibitemOpen
  \bibfield  {author} {\bibinfo {author} {\bibfnamefont {G.}~\bibnamefont
  {D'Ambrosio}}, \bibinfo {author} {\bibfnamefont {G.~F.}\ \bibnamefont
  {Giudice}}, \bibinfo {author} {\bibfnamefont {G.}~\bibnamefont {Isidori}}, \
  and\ \bibinfo {author} {\bibfnamefont {A.}~\bibnamefont {Strumia}},\ }\href
  {\doibase 10.1016/S0550-3213(02)00836-2} {\bibfield  {journal} {\bibinfo
  {journal} {Nucl. Phys. B}\ }\textbf {\bibinfo {volume} {645}},\ \bibinfo
  {pages} {155} (\bibinfo {year} {2002})},\ \Eprint
  {http://arxiv.org/abs/hep-ph/0207036} {arXiv:hep-ph/0207036} \BibitemShut
  {NoStop}%
\bibitem [{\citenamefont {Coloma}\ \emph {et~al.}(2021)\citenamefont {Coloma},
  \citenamefont {Fern\'andez-Mart\'\i{}nez}, \citenamefont
  {Gonz\'alez-L\'opez}, \citenamefont {Hern\'andez-Garc\'\i{}a},\ and\
  \citenamefont {Pavlovic}}]{Coloma:2020lgy}%
  \BibitemOpen
  \bibfield  {author} {\bibinfo {author} {\bibfnamefont {P.}~\bibnamefont
  {Coloma}}, \bibinfo {author} {\bibfnamefont {E.}~\bibnamefont
  {Fern\'andez-Mart\'\i{}nez}}, \bibinfo {author} {\bibfnamefont
  {M.}~\bibnamefont {Gonz\'alez-L\'opez}}, \bibinfo {author} {\bibfnamefont
  {J.}~\bibnamefont {Hern\'andez-Garc\'\i{}a}}, \ and\ \bibinfo {author}
  {\bibfnamefont {Z.}~\bibnamefont {Pavlovic}},\ }\href {\doibase
  10.1140/epjc/s10052-021-08861-y} {\bibfield  {journal} {\bibinfo  {journal}
  {Eur. Phys. J. C}\ }\textbf {\bibinfo {volume} {81}},\ \bibinfo {pages} {78}
  (\bibinfo {year} {2021})},\ \Eprint {http://arxiv.org/abs/2007.03701}
  {arXiv:2007.03701 [hep-ph]} \BibitemShut {NoStop}%
\bibitem [{\citenamefont {Bondarenko}\ \emph {et~al.}(2018)\citenamefont
  {Bondarenko}, \citenamefont {Boyarsky}, \citenamefont {Gorbunov},\ and\
  \citenamefont {Ruchayskiy}}]{Bondarenko:2018ptm}%
  \BibitemOpen
  \bibfield  {author} {\bibinfo {author} {\bibfnamefont {K.}~\bibnamefont
  {Bondarenko}}, \bibinfo {author} {\bibfnamefont {A.}~\bibnamefont
  {Boyarsky}}, \bibinfo {author} {\bibfnamefont {D.}~\bibnamefont {Gorbunov}},
  \ and\ \bibinfo {author} {\bibfnamefont {O.}~\bibnamefont {Ruchayskiy}},\
  }\href {\doibase 10.1007/JHEP11(2018)032} {\bibfield  {journal} {\bibinfo
  {journal} {JHEP}\ }\textbf {\bibinfo {volume} {11}},\ \bibinfo {pages} {032}
  (\bibinfo {year} {2018})},\ \Eprint {http://arxiv.org/abs/1805.08567}
  {arXiv:1805.08567 [hep-ph]} \BibitemShut {NoStop}%
\bibitem [{\citenamefont {Hwang}\ and\ \citenamefont
  {Kim}(1997)}]{Hwang:1997ie}%
  \BibitemOpen
  \bibfield  {author} {\bibinfo {author} {\bibfnamefont {D.~S.}\ \bibnamefont
  {Hwang}}\ and\ \bibinfo {author} {\bibfnamefont {G.-H.}\ \bibnamefont
  {Kim}},\ }\href {\doibase 10.1007/s002880050533} {\bibfield  {journal}
  {\bibinfo  {journal} {Z. Phys. C}\ }\textbf {\bibinfo {volume} {76}},\
  \bibinfo {pages} {107} (\bibinfo {year} {1997})},\ \Eprint
  {http://arxiv.org/abs/hep-ph/9703364} {arXiv:hep-ph/9703364} \BibitemShut
  {NoStop}%
\bibitem [{\citenamefont {Merlo}\ \emph {et~al.}(2019)\citenamefont {Merlo},
  \citenamefont {Pobbe}, \citenamefont {Rigolin},\ and\ \citenamefont
  {Sumensari}}]{Merlo:2019anv}%
  \BibitemOpen
  \bibfield  {author} {\bibinfo {author} {\bibfnamefont {L.}~\bibnamefont
  {Merlo}}, \bibinfo {author} {\bibfnamefont {F.}~\bibnamefont {Pobbe}},
  \bibinfo {author} {\bibfnamefont {S.}~\bibnamefont {Rigolin}}, \ and\
  \bibinfo {author} {\bibfnamefont {O.}~\bibnamefont {Sumensari}},\ }\href
  {\doibase 10.1007/JHEP06(2019)091} {\bibfield  {journal} {\bibinfo  {journal}
  {JHEP}\ }\textbf {\bibinfo {volume} {06}},\ \bibinfo {pages} {091} (\bibinfo
  {year} {2019})},\ \Eprint {http://arxiv.org/abs/1905.03259} {arXiv:1905.03259
  [hep-ph]} \BibitemShut {NoStop}%
\end{thebibliography}

\end{document}